\documentclass[12pt]{article}
\usepackage{setspace}
\usepackage{epsfig}
\usepackage{amsmath}
\usepackage{amssymb}
\usepackage{graphicx}
\def\be{\begin{equation}}
\def\ee{\end{equation}}
\def\ba{\begin{array}}
\def\ea{\end{array}}
\def\beqn{\begin{eqnarray}}
\def\eeqn{\end{eqnarray}}

\def\bt{\begin{tabular}}
\def\et{\end{tabular}}
\def\bc{\begin{center}}
\def\ec{\end{center}}

\begin{document}
\title{Majorana neutrinos and non minimal lepton mass textures}
\author{Samandeep Sharma, Priyanka Fakay, Gulsheen Ahuja$^*$, Manmohan
Gupta\\ {\it Department of Physics, Centre of Advanced Study,
P.U.,
 Chandigarh, India.}\\
\\{\it $^*$gulsheenahuja@yahoo.co.in}}
\date{}
\maketitle

\renewcommand{\baselinestretch}{1.50}\normalsize

\begin{abstract}
In the light of the recent measurement of the leptonic mixing angle $\theta_{13}$, implications of the latest
mixing data have been investigated for non-minimal textures of lepton mass matrices assuming the 
neutrinos to be Majorana like. Large number of possible
 texture specific lepton mass matrices have been examined for their compatibility with the lepton 
 mixing data in the case of normal hierarchy, 
inverted hierarchy and degenerate scenario of neutrino masses. Specifically, apart from 
other phenomenological quantities the
implications of the lepton mixing angle $\theta_{13}$
have been investigated on the lightest neutrino mass as well as the effective Majorana mass $<m_{ee}>$.

\end{abstract}

\section{Introduction}
The last few years have witnessed spectacular advances in fixing the neutrino masses and mixing
parameters through various solar \cite{davis}-\cite{sno}, atmospheric \cite{super}, 
reactor \cite{kam},\cite{chooz} and accelerator \cite{minos}-\cite{min} neutrino experiments. In this context,
the recent observations \cite{db}, \cite{reno} regarding the unexpectedly `large' value of 
the mixing angle $\theta_{13}$ have led to an increased interest in investigating the
implications of the neutrino oscillations phenomenology. In particular, the nonzero value of
$\theta_{13}$ could lead to the possibilty of exsistence of CP violation in the leptonic sector, therefore
leading to an increased amount of activity in this area. These observations have also 
deepened the mystery of flavor mixings as the patterns of quark and lepton mixing angles now look to be significantly different. 
Noting that the mixing angles and CP violating phases are very much related to the corresponding mass matrices, one essentially 
has to formulate the fermion mass matrices to unravel some of the deeper aspects of flavor physics. In case
one assumes unification or quark-lepton 
complimentarity \cite{smirnov}, it becomes desirable to understand the quark and lepton mixings
from the same perspective as far as possible, thus making 
the formulation of viable fermion mass matrices all the more complicated. 
\par In the absence of any compelling theory of flavor dynamics, one usually resorts to phenomenological models. In this context, the 
concept of texture specific mass matrices, introduced implicitly by Weinberg \cite{wein} 
and explicitly by Fritzsch \cite{fritzsch}, has got a good deal of attention in 
the literature, for details we refer the reader to a recent review article \cite{singreview}. 
\par Considering neutrinos to be Majorana particles, after the recent measurement of $\theta_{13}$, a few analyses have been 
carried out for texture specific mass matrices in the non-flavor basis. In particular, Fukugita {\it{et al.}} \cite{fukugita}
have investigated the 
implications of angle $\theta_{13}$ on minimal texture mass matrices (Fritzsch-like texture six zero) for normal hierarchy of neutrino 
masses. This analysis has been extended further by Fakay {\it{et al.}} \cite{ourplb} wherein 
for all hierarchies of neutrino masses, Fritzsch-like 
texture zero and five zero mass matrices have been examined in detail. However, a detailed and comprehensive analysis for non-Fritzsch like
lepton mass matrices with five texture zeroes and beyond is yet to be carried out. In this 
context, it may be noted that Branco {\it{et al.}} \cite{branco}
have carried out a detailed analysis for all possible structures of texture four zero lepton mass matrices, however 
similar attempts have 
not been carried out after the recent measurements of $\theta_{13}$.
\par In the present paper, we have attempted to carry out detailed calculations pertaining to lepton mass matrices with
non-minimal textures for all the three possibilities of neutrino masses. In particular, the analysis has been carried out for 
Fritzsch-like texture two zero mass matrices as well as for all possibilities of texture four zero and five zero lepton mass matrices. The 
compatibility of these texture specific mass matrices has been examined by plotting the parameter space corresponding to any two mixing
angles. Further, the implications of mixing angles on the lightest neutrino mass as well as the effective Majorana mass $<m_{ee}>$
have also been investigated. 
\par The detailed plan of the paper is as follows. In Section
(2), we detail the essentials of the formalism regarding
the texture specific mass matrices. Inputs used in the present
analysis have been given in Section (3) and the
discussion of the calculations and results have been presented in
Section (4). Finally, Section (5) summarizes our
conclusions.

\section{Texture specific mass lepton mass matrices in the Standard Model and the PMNS matrix \label{form}}
Using the facility of weak basis (WB) transformations \cite{frxing}, it can be shown that the most general lepton mass matrices
within the framework of standard model (SM) can be expressed as
\be
 M_{l}=\left( \ba{ccc}
C_{l} & A _{l} & 0      \\
A_{l}^{*} & D_{l} &  B_{l}     \\
 0 &     B_{l}^{*}  &  E_{l} \ea \right), \qquad
M_{\nu D}=\left( \ba{ccc} C_{\nu } & A _{\nu } & 0\\
A_{\nu }^{*} & D_{\nu } & B_{\nu }     \\
 0 &     B_{\nu }^{*}  &  E_{\nu } \ea \right),
\label{t20}\ee
$M_{l}$ and $M_{\nu D}$  corresponding to 
charged lepton and Dirac-like neutrino mass matrices respectively. Both the matrices are
texture 1 zero type with $A_{l(\nu)}
=|A_{l(\nu)}|e^{i\alpha_{l(\nu)}}$
 and $B_{l(\nu)} = |B_{l(\nu)}|e^{i\beta_{l(\nu)}}$. Further, to facilitate the formulation
of phenomenological mass matrices, which perhaps are compatible with the GUT scale mass matrices, it 
has been suggested \cite{nmm} that in order to avoid fine tuning amongst the elements of the mass matrices, these should follow 
a `natural hierarchy' i.e.$ (1,1),(1,2),(1,3)\lesssim (2,2), (2,3)
\lesssim (3,3)$. 
%

For the complete diagonalizing matrix pertaining to structure presented in equation (\ref{t20}) we refer the
reader to \cite{fulldirac}, however to understand the relationship between diagonalizing transformations
for different hierarchies of neutrino masses and for the charged lepton case as detailed in \cite{ourplb}, we present here the 
first element of the diagonalizing transformation $O_k$, 
\be
O_k(11) =  \sqrt{\frac{(E_k -m_1)(D_k + E_k - m_1 -
m_2)(D_k + E_k -m_1 -m_3)}{(D_k +2 E_k -m_1 -m_2 -m_3) (m_1 -m_2)
(m_1-m_3)}}, 
\label{diat20}
\ee
 $m_1$, -$m_2$, $m_3$ being the eigen values of a general mass matrix $M_k$ with structure as 
 given in equation (\ref{t20}). In the case of charged leptons,
because of the hierarchy $m_e \ll m_{\mu} \ll m_{\tau}$, the mass
eigenstates can be approximated to the respective flavor
eigenstates, i.e. $m_{l1} \simeq m_e$,
$m_{l2} \simeq m_{\mu}$ and $m_{l3} \simeq m_{\tau}$, and thus one can
obtain the first element of the matrix $O_l$ from the equation (\ref{diat20}), by replacing $m_1$, $m_2$,
$m_3$ by $m_e$, $m_{\mu}$, $m_{\tau}$, e.g.,
\be
O_l(11)=\sqrt{\frac{(E_l -m_e)(D_l + E_l - m_e - m_\mu)(D_l + E_l -m_e -m_\tau)}{(D_l +2 E_l -m_e -m_\mu -m_\tau)(m_e -m_\mu) (m_e-m_\tau)}}
\ee

Equation (\ref{diat20}) can also be used to obtain the first
element of diagonalizing transformation for Majorana neutrinos,
assuming normal hierarchy, defined as $m_{\nu_1}<m_{\nu_2}\ll
m_{\nu_3}$, and also valid for the degenerate case defined as
$m_{\nu_1} \lesssim m_{\nu_2} \sim m_{\nu_3}$, by replacing $m_1$,
$m_2$, $m_3$ by $\sqrt{m_{\nu 1} m_R}$, $\sqrt{m_{\nu 2} m_R}$,
$\sqrt{m_{\nu 3} m_R}$, e.g., 
\be O_{\nu}(11) =  \sqrt{\frac{(E_\nu -\sqrt{m_{\nu_1}})(D_\nu + E_\nu - \sqrt{m_{\nu_ 1}} -
\sqrt{m_{\nu _2}})(D_\nu + E_\nu -\sqrt{m_{\nu_ 1}} -\sqrt{m_{\nu_ 3}})}{(D_\nu +2 E_\nu
-\sqrt{m_{\nu_ 1}} -\sqrt{m_{\nu_ 2}} -\sqrt{m_{\nu_ 3}}) (\sqrt{m_{\nu_ 1}} -\sqrt{m_{\nu_ 2}})
(\sqrt{m_{\nu_ 1}}-\sqrt{m_{\nu_ 3}})}},  
\label{omajnh}
\ee 

where $m_{\nu_1}$,
$m_{\nu_2}$ and $m_{\nu_3}$ are neutrino masses. 
In the same manner, one can obtain the elements of diagonalizing
transformation for the inverted hierarchy case, defined as
$m_{\nu_3} \ll m_{\nu_1} < m_{\nu_2}$, by replacing $m_1$, $m_2$,
$m_3$ in equation (\ref{diat20}) with $\sqrt{m_{\nu_1} m_R}$,
$-\sqrt{m_{\nu_2} m_R}$, $-\sqrt{m_{\nu_3} m_R}$, e.g., 
\be O_{\nu}(11) =  \sqrt{\frac{(E_\nu -\sqrt{m_{\nu_1}})(D_\nu + E_\nu - \sqrt{m_{\nu_ 1}} -
\sqrt{m_{\nu _2}})(D_\nu + E_\nu -\sqrt{m_{\nu_ 1}} +\sqrt{m_{\nu_ 3}})}{(D_\nu +2 E_\nu
-\sqrt{m_{\nu_ 1}} -\sqrt{m_{\nu_ 2}} +\sqrt{m_{\nu_ 3}}) (\sqrt{m_{\nu_ 1}} -\sqrt{m_{\nu_ 2}})
(\sqrt{m_{\nu_ 1}}+\sqrt{m_{\nu_ 3}})}} 
\label{omajih}.
\ee
The other elements of
diagonalizing transformations in the case of neutrinos as well as
charged leptons can similarly be found. It can be shown that using the seesaw mechanism,
the effective neutrino mass matrix can be expressed as,
\be
M_{\nu}= P_{\nu D} O_{\nu D} \frac{(M_{\nu D}^{diag})^2}{(m_R)}
O_{\nu D}^T P_{\nu D},
\label{mnu} 
\ee
$m_R$ being the right handed neutrino mass scale. Further, the lepton mixing matrix 
can be expressed as
\be
 U = O_l^{\dagger} Q_l P_{\nu D} O_{\nu D},
 \ee
where $Q_l P_{\nu D}$, without loss of generality, can be taken as
$(e^{i\phi_1},\,1,\,e^{i\phi_2})$, $\phi_1$ and $\phi_2$ being
related to the phases of mass matrices and can be treated as free
parameters. 
\section{Inputs used for the analysis}
In the present analysis,
we have made use of the results of a latest global three neutrino oscillation analysis \cite{fogli2012}, in 
table (\ref{data}) we present the $ 1 \sigma$ and $ 3 \sigma$ ranges of the neutrino oscillation parameters.

\begin{table}[ht]
\centering
\begin{tabular}{c c c }
\hline
Parameter & $1 \sigma$ range & $3 \sigma$ range  \\ [0.5ex]
\hline

$\Delta m_{sol}^2$ $[ 10^{-5} eV^2]$ & (7.32-7.80) & (6.99-8.18) \\
\hline
$\Delta m_{atm}^2$ $[ 10^{-3} eV^2]$ & (2.33-2.49)(NH); (2.31-2.49) (IH) & (2.19-2.62)(NH); (2.17-2.61)(IH)  \\
\hline
$sin^2 \theta_{13}$ $[10^{-2}]$ & (2.16-2.66)(NH); (2.19-2.67)(IH) & (1.69-3.13)(NH); (1.71-3.15) (IH)  \\
\hline
$sin^2 \theta_{12}$ $[10^{-1}]$ & (2.91-3.25) & (2.59-3.59)  \\
\hline
$sin^2 \theta_{23}$ $[10^{-1}]$ & (3.65-4.10)(NH);(3.70-4.31)(IH)  & (3.31-6.37)(NH);(3.35-6.63)(IH)  \\
\hline
\end{tabular}
\caption{The $1\sigma$ and $3\sigma$ ranges of neutrino oscillation parameters presented in \cite{fogli2012}.}
\label{data}
\end{table}

While carrying out the analysis, the lightest neutrino mass, $m_1$ for the case of NH and $m_3$ for the case of IH, 
is considered as a free
parameter, which is explored within the range $10^{-8}\,\rm{eV}-10^{-1}\,\rm{eV}$ for all the three possible mass
hierarchies of neutrinos i.e. normal, inverted and degenerate scenario.
In the absence of any constraint on the phases, $\phi_1$ and $\phi_2$
have been given full variation from 0 to $2\pi$. The parameters $D_{l,
\nu}$ and $C_{l, \nu}$ have been considered as free parameters, however, they have been constrained
such that diagonalizing transformations $O_l$ and $O_{\nu}$ always
remain real.
\section{Results and discussions}
\subsection{Texture two zero lepton mass matrices} \label{tex20}
To examine the compatibility of texture two zero lepton mass matrices given in equation (\ref{t20}) with the recent
mixing data, we carry out a detailed analysis pertaining to all three possible neutrino mass hierarchies. Firstly, 
we attempt to examine the compatibilty of these texture two zero matrices with the inverted hierarchy of neutrino masses.
To this end, in figure (1) we present the plot showing the  parameter space corresponding to the 
mixing angle $s_{12}$ along with  $s_{13}$. Giving full allowed 
variation to other parameters, figure (1) has been obtained by constraining the angle $s_{23}$ by its $3\sigma$ 
experimental bound. The blank rectangular regions in this plot shows the experimentally allowed $3\sigma$ regions for
$s_{12}$ and $s_{13}$. Interestingly, a general look at this plot reveals the viablity of inverted hierarchy of neutrino 
masses for the texture two zero mass matrices presented in eqn.(\ref{t20}) as can be seen from the significant
overlap between the parameter space allowed by this 
structure with the experimentally allowed $3\sigma$ region.
 
 \begin{figure}[hbt]
\bc
\includegraphics[width=2.in,angle=270]{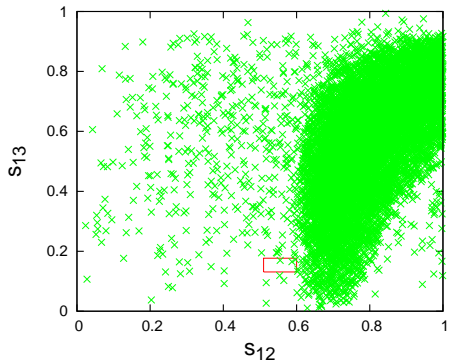}
\caption{Plot showing the parameter space corresponding to $s_{12}$ and $s_{13}$ for 
texture two zero mass matrices (inverted hierarchy).} 
 \ec
 \label{t2ih1}
\end{figure}

\begin{figure}
\begin{tabular}{cc}
  \includegraphics[width=0.2\paperwidth,height=0.2\paperheight,angle=-90]{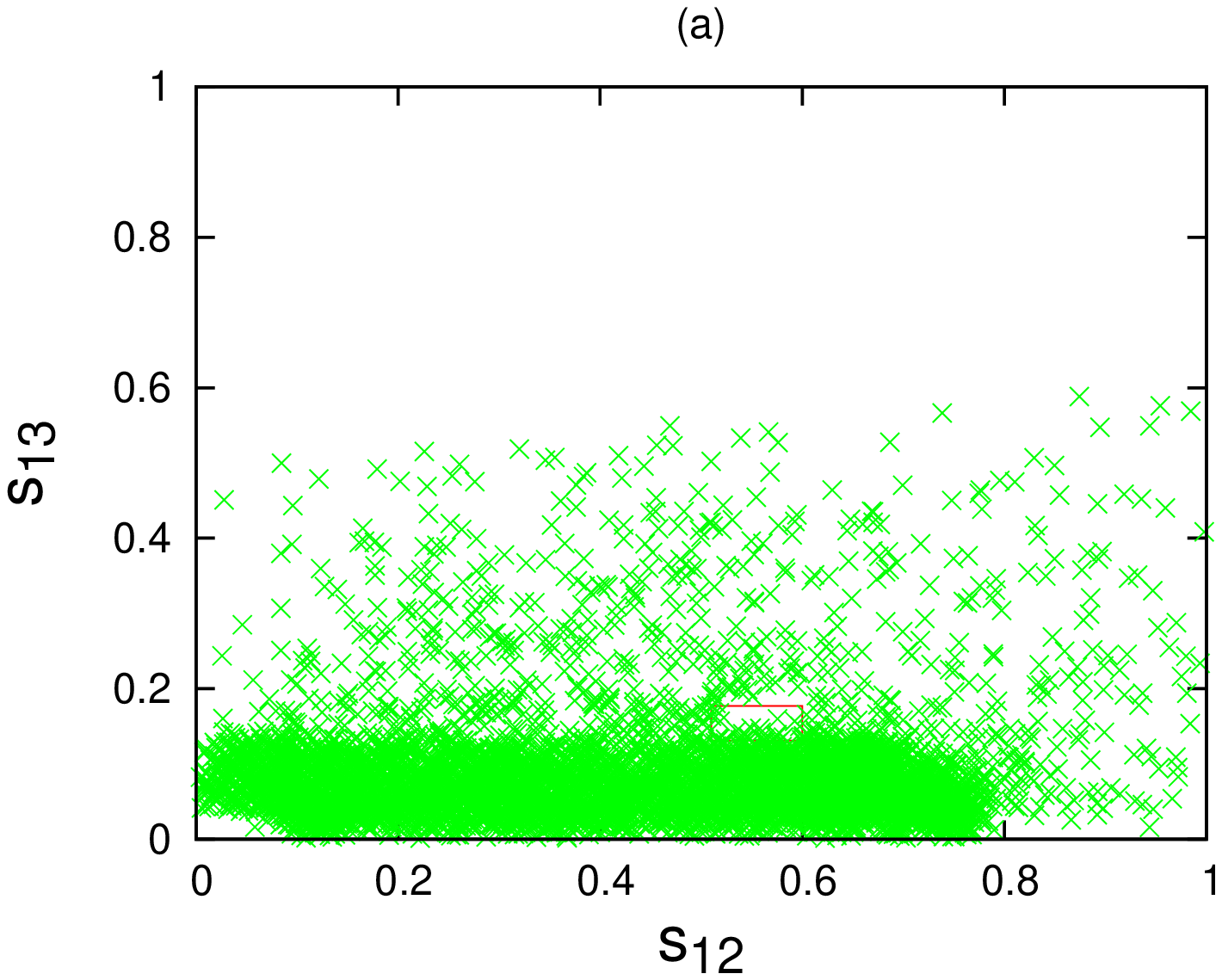}
  \includegraphics[width=0.2\paperwidth,height=0.2\paperheight,angle=-90]{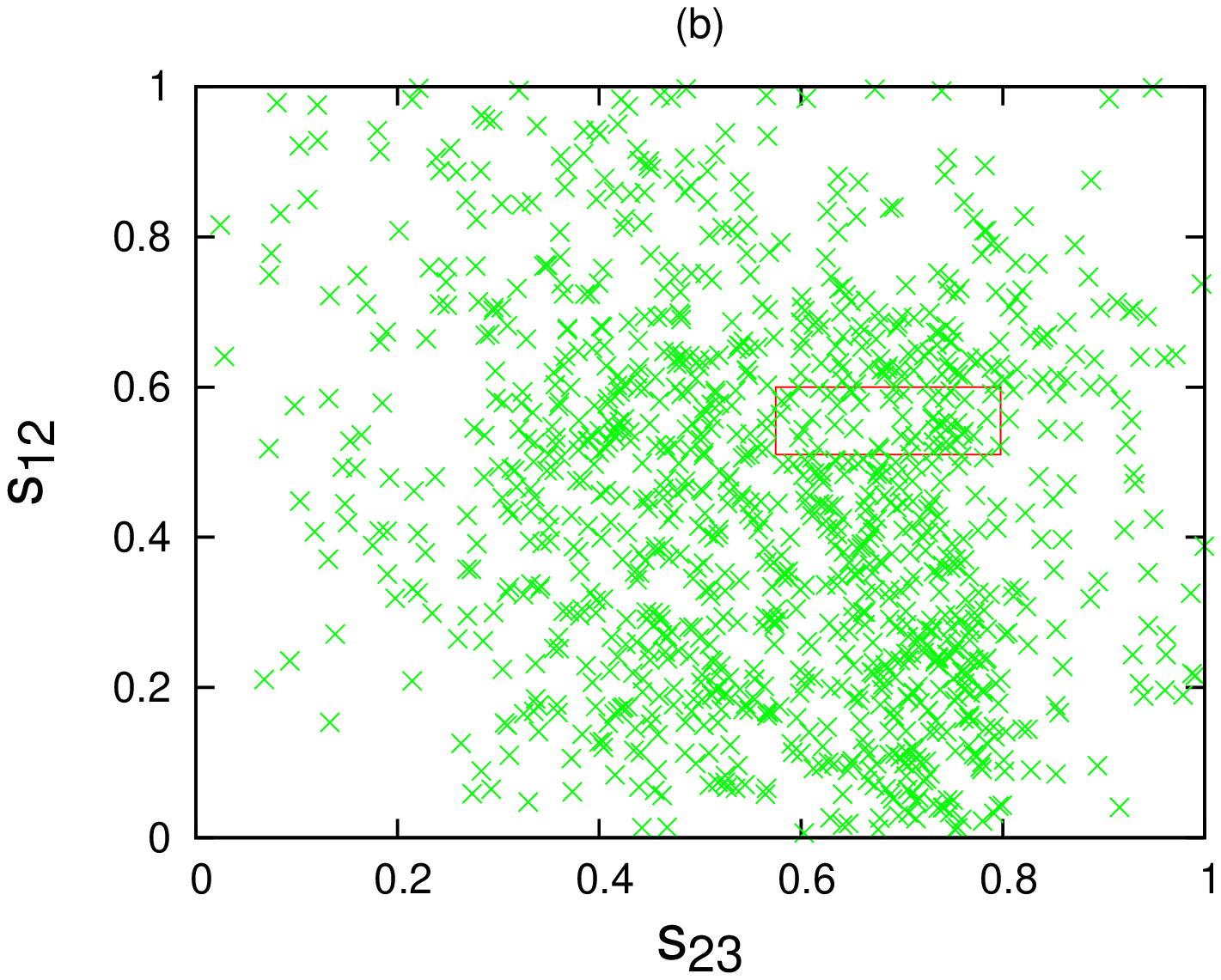}
  \includegraphics[width=0.2\paperwidth,height=0.2\paperheight,angle=-90]{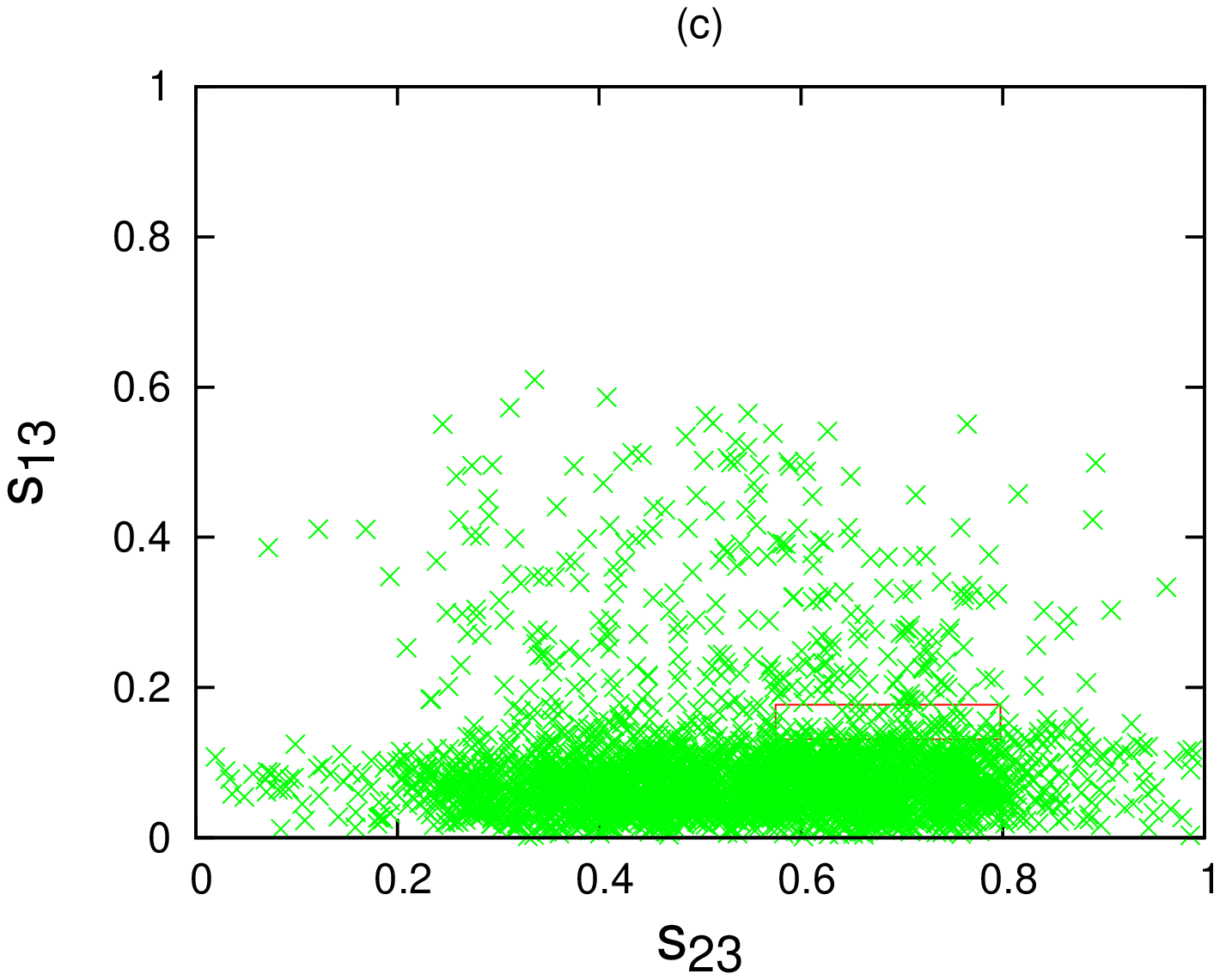}
\end{tabular}
\caption{Plots showing the parameter space for any two mixing angles for texture two zero mass matrices (normal hierarchy).}
\label{t2nh1}
\end{figure}

After examining the viability of inverted hierarchy of neutrino masses for 
the mass matrix structure given in equation (\ref{t20}), we now examine
the compatibility of these
matrices for the normal hierarchy case. To this end, in figure (\ref{t2nh1}) we present the plots showing the 
parameter space allowed for two mixing angles when the third one is constrained by its $1\sigma$ experimental bound for normal
neutrino mass hierarchy. The rectangular regions in these plots show the experimentally allowed $3\sigma$ regions of the 
plotted angles. A general look at the figure (\ref{t2nh1}) reveals that the structure (\ref{t20}) is compatible
with the normal neutrino mass hierarchy.

Further, in figures 3(a) and 3(b) we present the plots showing the variation of
the mixing angle $s_{13}$ with the parameters $C_l/m_e$ and $C_\nu/m_1$ respectively
for structure given in eqn.(\ref{t20}) pertaining to 
normal hierarchy of neutrino masses. While plotting these figures, the other
two mixing angles have been constrained by their $3\sigma$ experimental
bounds, while all the free parameters have been given full variation. The two parallel lines in these
figures show the $3 \sigma$ allowed range for the mixing angle $s_{13}$. Taking
a careful look at these
plots, one can note that
the leptonic mixing angles donot have much dependence on the parameters $C_l$ and $C_\nu$.
Further one can see that a fit for all the three mixing angles can be obtained for the values of parameters being 
$C_l \lesssim 0.7 m_e$ and $C_\nu \lesssim 0.8 m_1$.

\begin{figure}[hbt]
  \begin{minipage}{0.45\linewidth}   \centering
\includegraphics[width=2.in,angle=-90]{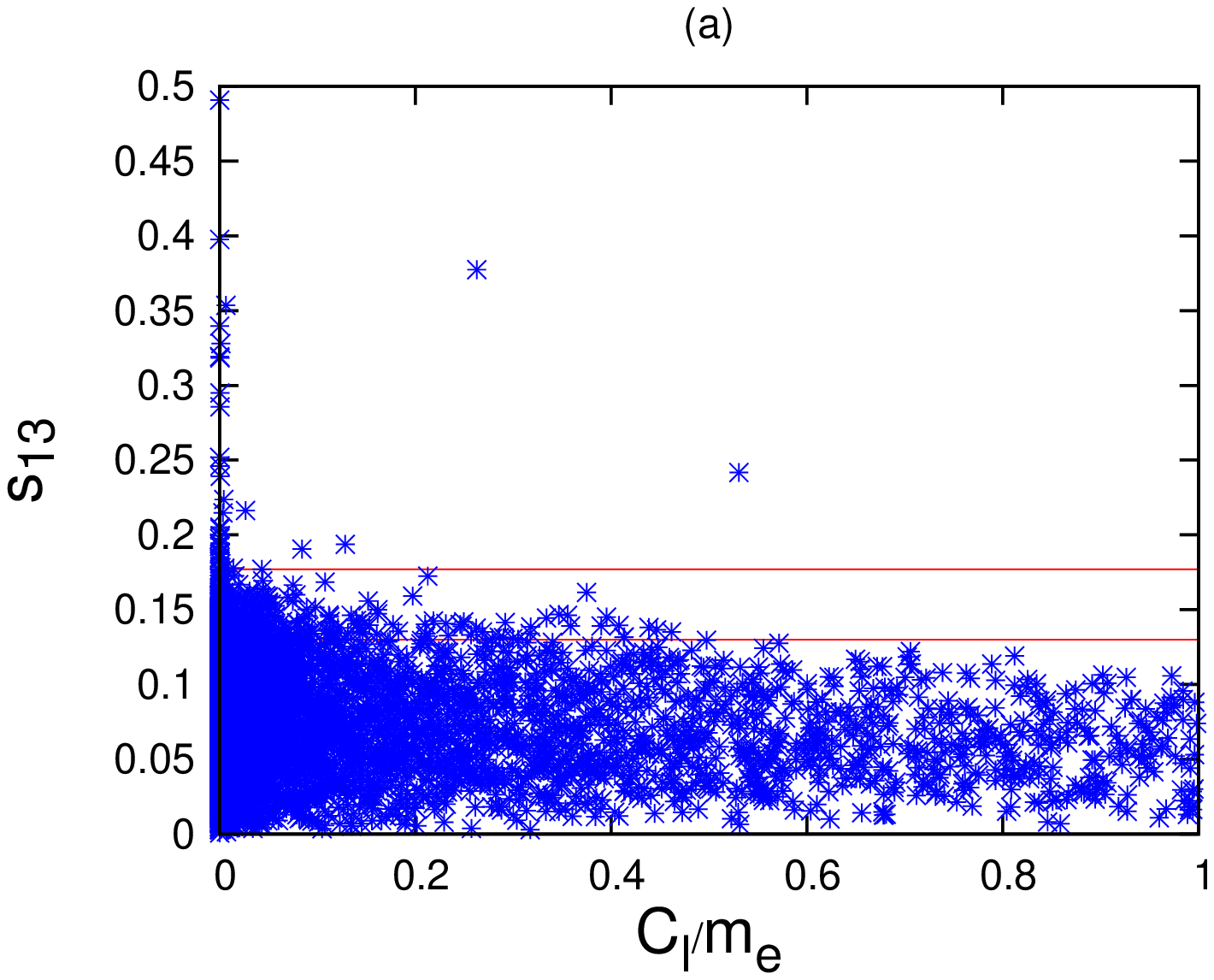}
    \end{minipage} \hspace{0.5cm}
\begin{minipage} {0.45\linewidth} \centering
\includegraphics[width=2.in,angle=-90]{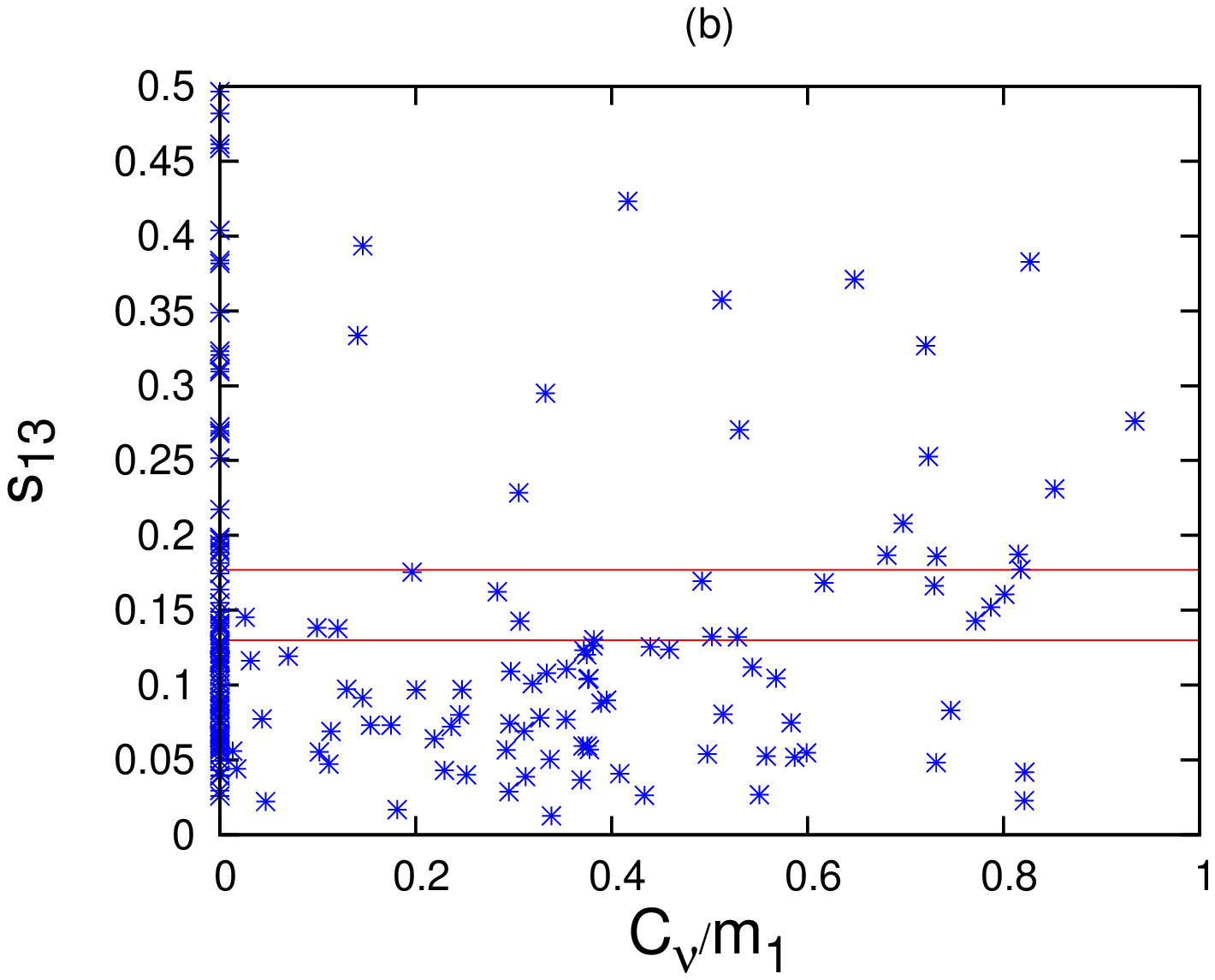}
\end{minipage}
\caption{Plots showing the variation of the mixing angle $s_{13}$ with the (1,1) element of the charged
lepton mass matrix and the neutrino mass matrix for texture two zero mass matrices (normal hierarchy).}
\label{t2nh2}
\end{figure}

Therefore, one can conclude that very small values of the parameters $C_l$ and $C_\nu$ are required to fit the latest experimental
data for the mass matrices pertaining to structure (\ref{t20}) and these 
 essentially reduce to
\be
 M_{l}=\left( \ba{ccc}
0 & A _{l} & 0      \\
A_{l}^{*} & D_{l} &  B_{l}     \\
 0 &     B_{l}^{*}  &  E_{l} \ea \right), \qquad
M_{\nu D}=\left( \ba{ccc} 0 & A _{\nu } & 0\\
A_{\nu }^{*} & D_{\nu } & B_{\nu }     \\
 0 &     B_{\nu }^{*}  &  E_{\nu } \ea \right).
\label{flt40}\ee
The structure (\ref{flt40}) is referred to as Fritzsch-like texture four zero structure and is studied extensively in
literature \cite{t40lep}. However, no such attempt has been made after the recent measurement 
of the mixing angle $s_{13}$. Therefore, it becomes
interesting to analyse this structure in detail for its compatibility with the latest lepton mixing data.

\subsection{Texture four zero lepton mass matrices}
As discussed in the previous section, starting with the most general lepton mass matrices having `natural
hierarchy' \cite{nmm} among their elements, one essentially arrives at the Fritzsch-like texture four zero
structure as given in eqn.(\ref{flt40}). In this context, it is interesting to
note that using the facility of WB transformation performed by  permutation matrices isomorphic
to $S_3$, all possible texture four zero mass matrices can essentially be classified into four distinct classes \cite{branco}. 
All the matrices belonging to a particular class have the same physical content. From the table (2) presented in \cite{fulldirac}
it is  clear that class IV is  not viable as all
the matrices in this class correspond to the scenario where one of
the generations gets decoupled from the other two.
Therefore, in the following subsection we attempt
to confront classes I, II, III  of lepton mass matrices
with the latest experimental data. 
\subsubsection{Class I ansatz}
To begin with, we carry out a detailed analysis for texture four zero mass matrices belonging to class I, i.e.,
\be
 M_{i}=\left( \ba{ccc}
0 & A _{i}e^{i\alpha_i} & 0      \\
A_{i}e^{-i\alpha_i} & D_{i} &  B_{i}e^{i\beta_i}     \\
 0 &     B_{i}e^{-i\beta_i} &  E_{i} \ea \right),
\label{cl1}
\ee
where $i=l,~\nu_D$ corresponds to the charged lepton and Dirac neutrino mass matrices
respectively. For the purpose of calculations, the elements
$D_l$, $D_\nu$ as well as the phases $\phi_1$ and $\phi_2$ have been considered as free parameters.
Following the methodology as discussed in section (\ref{form}), we attempt to carry out a detailed
study pertaining to normal, inverted as well as degenerate neutrino mass orderings. Firstly, we examine
the compatibility of mass matrices given in eqn.(\ref{flt40}) with the
inverted hierarchy of neutrino masses. For this purpose in figure
(4), we present the parameter space corresponding to the mixing angles $s_{12}$ and $s_{13}$ while
$s_{23}$ is being constrained by its $3\sigma$ range. The blank rectangular regions
in these figures represent the $3\sigma$ allowed ranges of the mixing angles $s_{12}$ and $s_{13}$. As can be
seen from this plot, the plotted parameter space does not show any overlap with the experimentally
allowed region. Therefore, we find that the class I ansatz of texture four zero lepton mass matrices
is ruled out for the inverted hierarchy scenario of neutrino masses.
\begin{figure}[hbt]
\bc
 \includegraphics[width=2.in,angle=270]{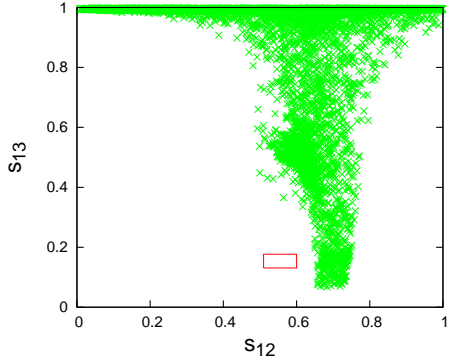}
 \caption{Plot showing the parameter space for $s_{12}$ and $s_{13}$ when $s_{23}$ is constrained by
its  $3 \sigma$ range for
 Class I ansatz of texture four zero mass matrices (inverted hierarchy).} 
 \ec
 \label{t4ih1}
\end{figure}

\begin{figure}
\begin{tabular}{cc}
  \includegraphics[width=0.2\paperwidth,height=0.2\paperheight,angle=-90]{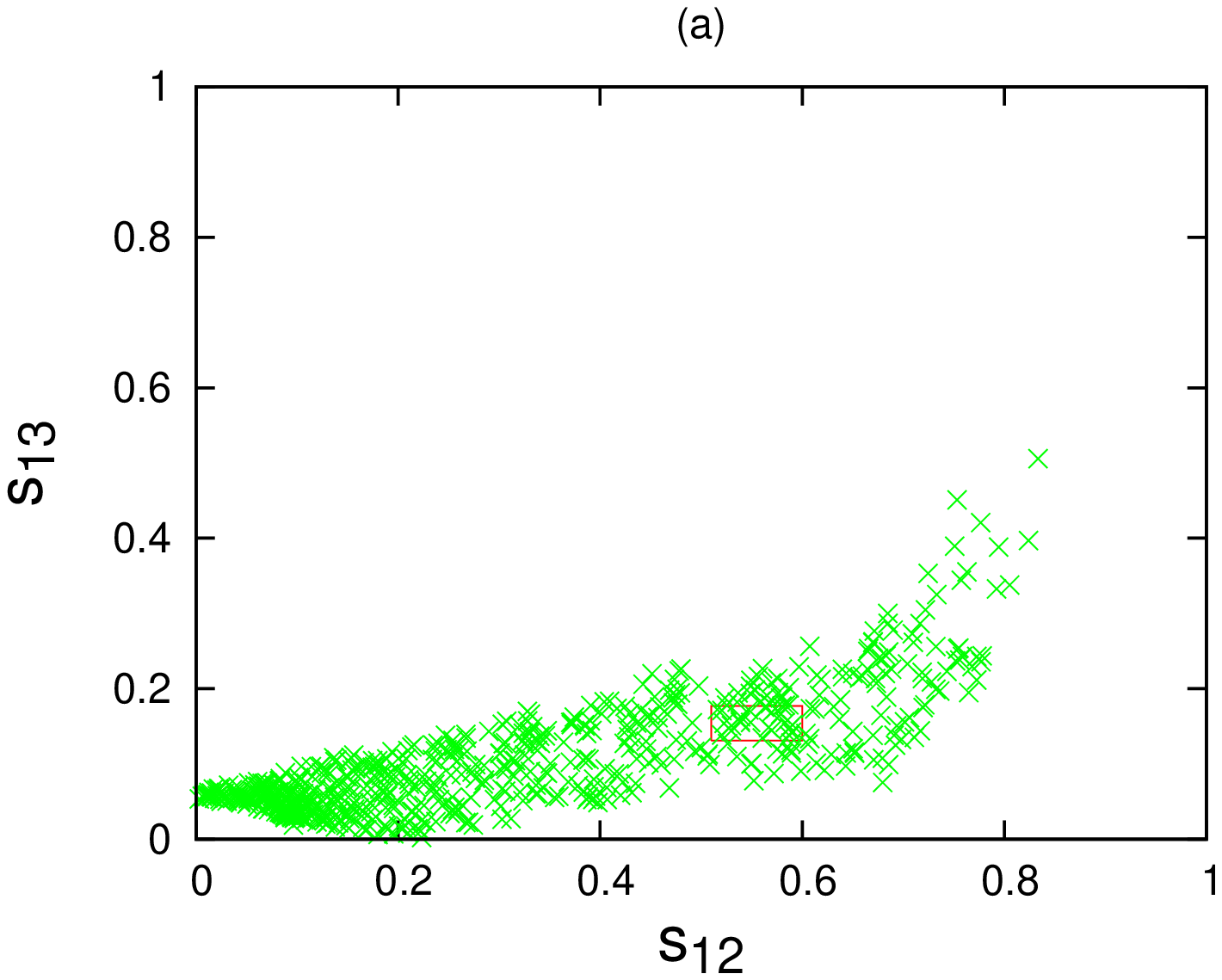}
  \includegraphics[width=0.2\paperwidth,height=0.2\paperheight,angle=-90]{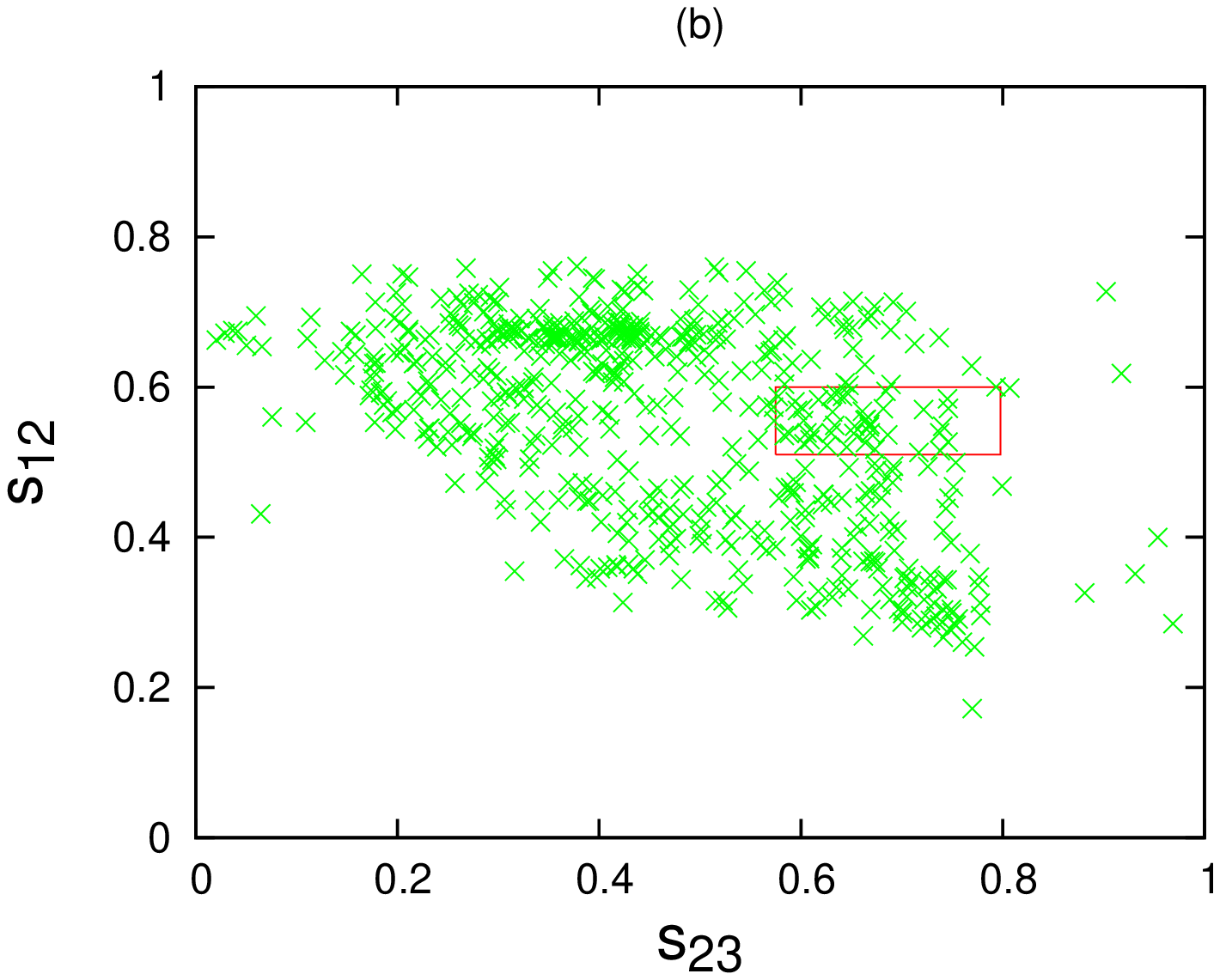}
  \includegraphics[width=0.2\paperwidth,height=0.2\paperheight,angle=-90]{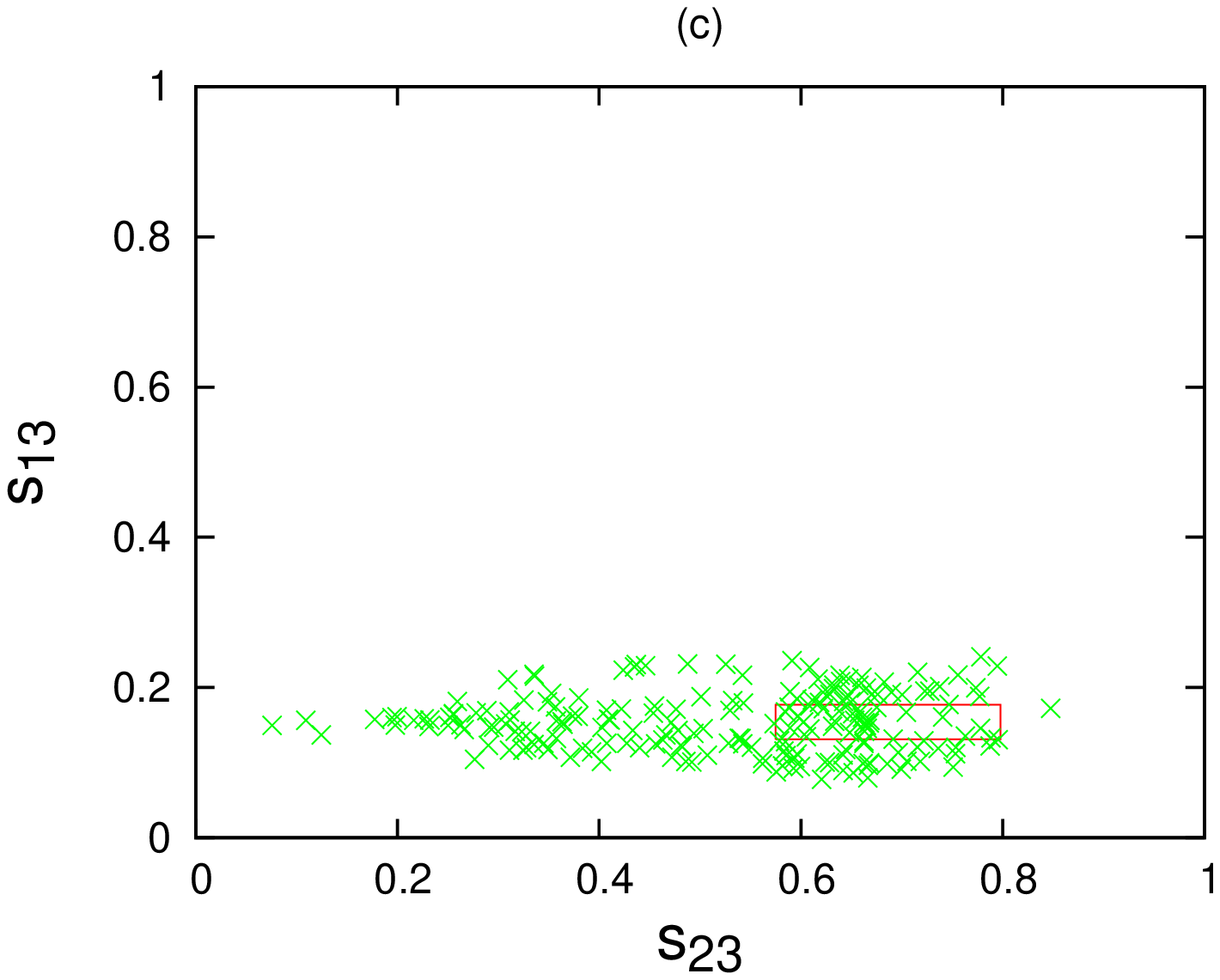}
\end{tabular}
\caption{Plots showing the parameter space for any two mixing angles when the third angle is constrained by
its  $1 \sigma$ range for 
Class I ansatz of texture four zero  mass matrices (normal hierarchy).}
\label{t4nh1}
\end{figure}

After ruling out the structure presented in eqn.(\ref{cl1}) for the inverted hierarchy, we now proceed to examine the compatibility
of these matrices for the normal hierarchy case. To this end, in figure (\ref{t4nh1}) we present the plots showing the
parameter space corresponding to any two mixing angles wherein the third one is constrained by its $1 \sigma$ range.
Interestingly, normal hierarchy seems to be viable in this case as can be seen from these plots,
wherein the parameter space shows significant overlap with the experimentally allowed
$3 \sigma$ region shown by the rectangular boxes in each plot.

\begin{figure}[hbt]
  \begin{minipage}{0.45\linewidth}   \centering
\includegraphics[width=2.in,angle=-90]{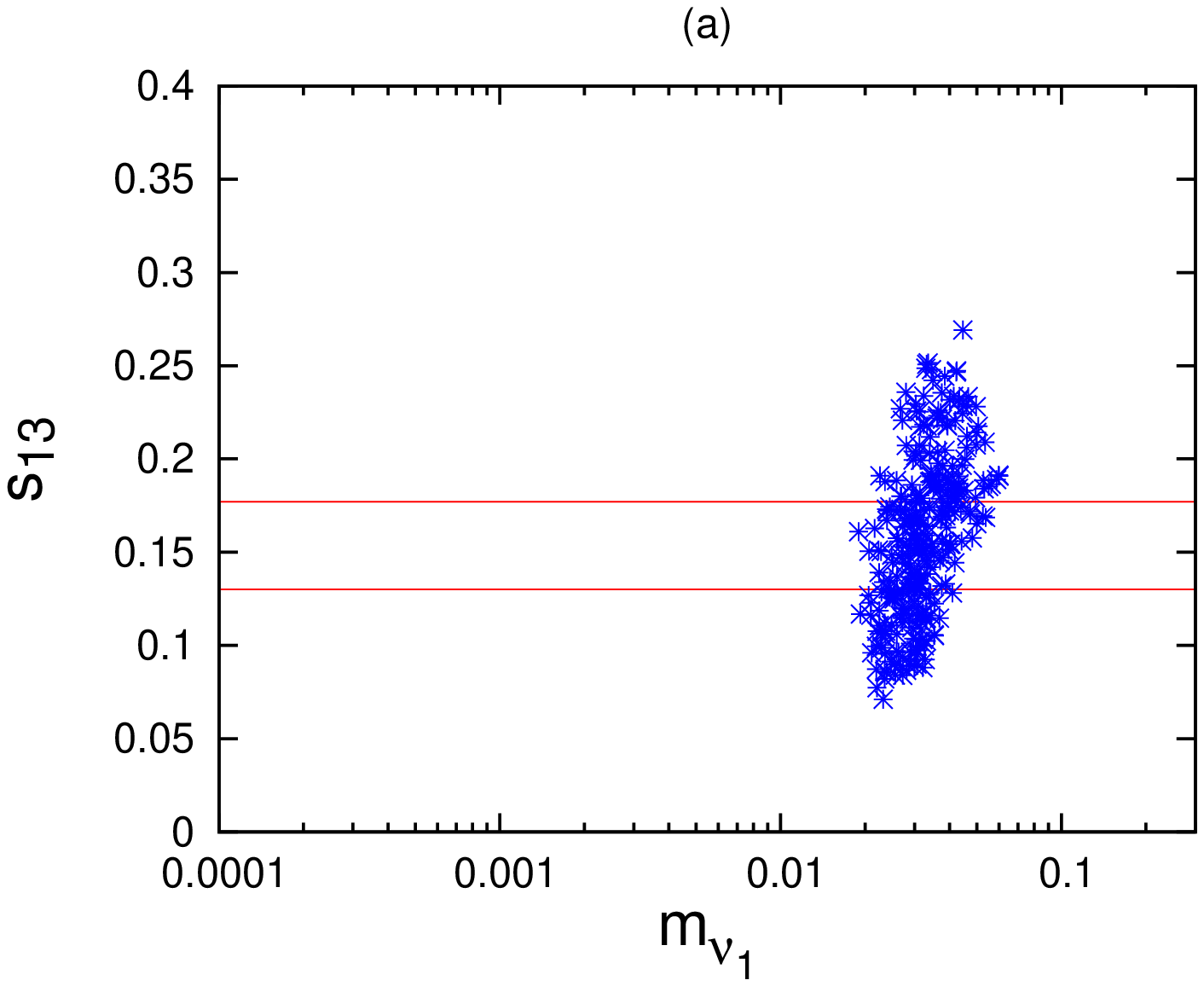}
    \end{minipage} \hspace{0.5cm}
\begin{minipage} {0.45\linewidth} \centering
\includegraphics[width=2.in,angle=-90]{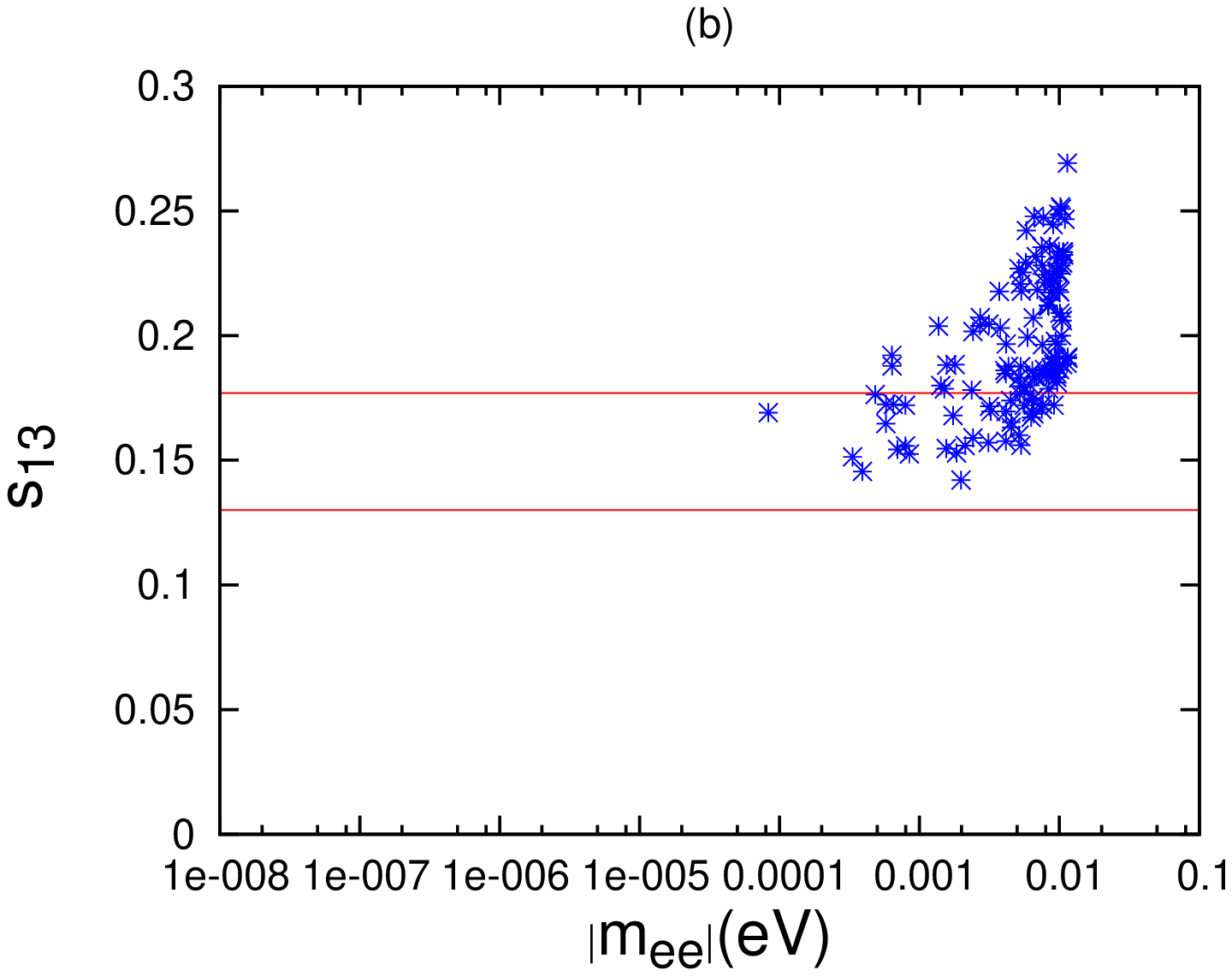}
\end{minipage}
\caption{Plots showing the dependence of mixing angle $s_{13}$ on (a) the lightest neutrino mass and 
(b) the effective Majorana mass, when the
other two angles are constrained by their $3 \sigma$ ranges 
for Class I ansatz of texture four zero  mass matrices (normal hierarchy).}
\label{t4nh2}
\end{figure}

Further, in figure (\ref{t4nh2}) we present the graphs showing the variation
of the lightest neutrino mass and the effective Majorana mass with the mixing angle $s_{13}$
for normal hierarchy, keeping the
other two mixing angles constrained by their $3\sigma$ bounds. The parallel lines in each plot
show the $3\sigma$ range of the mixing angle $s_{13}$. Taking a careful look at figure \ref{t4nh2}(a),
one can find upper and lower bounds on the lightest neutrino mass to be $0.01eV \lesssim m_{\nu 1} \lesssim 0.1 eV$ approximately.
Similarly, from the 
numerical results shown in figure \ref{t4nh2}(b), one can obtain an approximate
lower bound on the effective Majorana mass $|m_{ee}|$, viz. $|m_{ee}| \gtrsim 10^{-5} eV$. 

\par The degenerate scenario of the neutrino masses can be characterized by either $m_{\nu 1} \lesssim m_{\nu 2} \sim m_{\nu3} \sim 0.1 eV$
or $m_{\nu 3} \sim m_{\nu 1} \lesssim m_{\nu2} \sim~0.1eV$, corresponding to normal hierarchy and inverted hierarchy respectively. Since
while carrying out the calculations pertaining to both the normal as well as inverted hierarchy cases, the range of lightest neutrino
mass is taken to be $10^{-8} -10^{-1}$ eV, which includes the neutrino masses corresponding to the degenerate scenario; therefore, by
discussion similar to the one given for ruling out inverted hierarchy, degenerate scenario corresponding to inverted hierarchy
of neutrino masses seems to be ruled out for class I ansatz of texture four zero mass matrices. However, 
from figure (\ref{t4nh2}) the value of the lightest neutrino mass pertaining to the degenerate scenario,
$m_{\nu 1} \sim 0.1 eV$ seems to be included in the experimentally allowed region, 
thereby showing the viablity of Class I ansatz for degenerate scenario pertaining to normal hierarchy.

\subsubsection{Class II ansatz}
\begin{figure}
\begin{tabular}{cc}
  \includegraphics[width=0.2\paperwidth,height=0.2\paperheight,angle=-90]{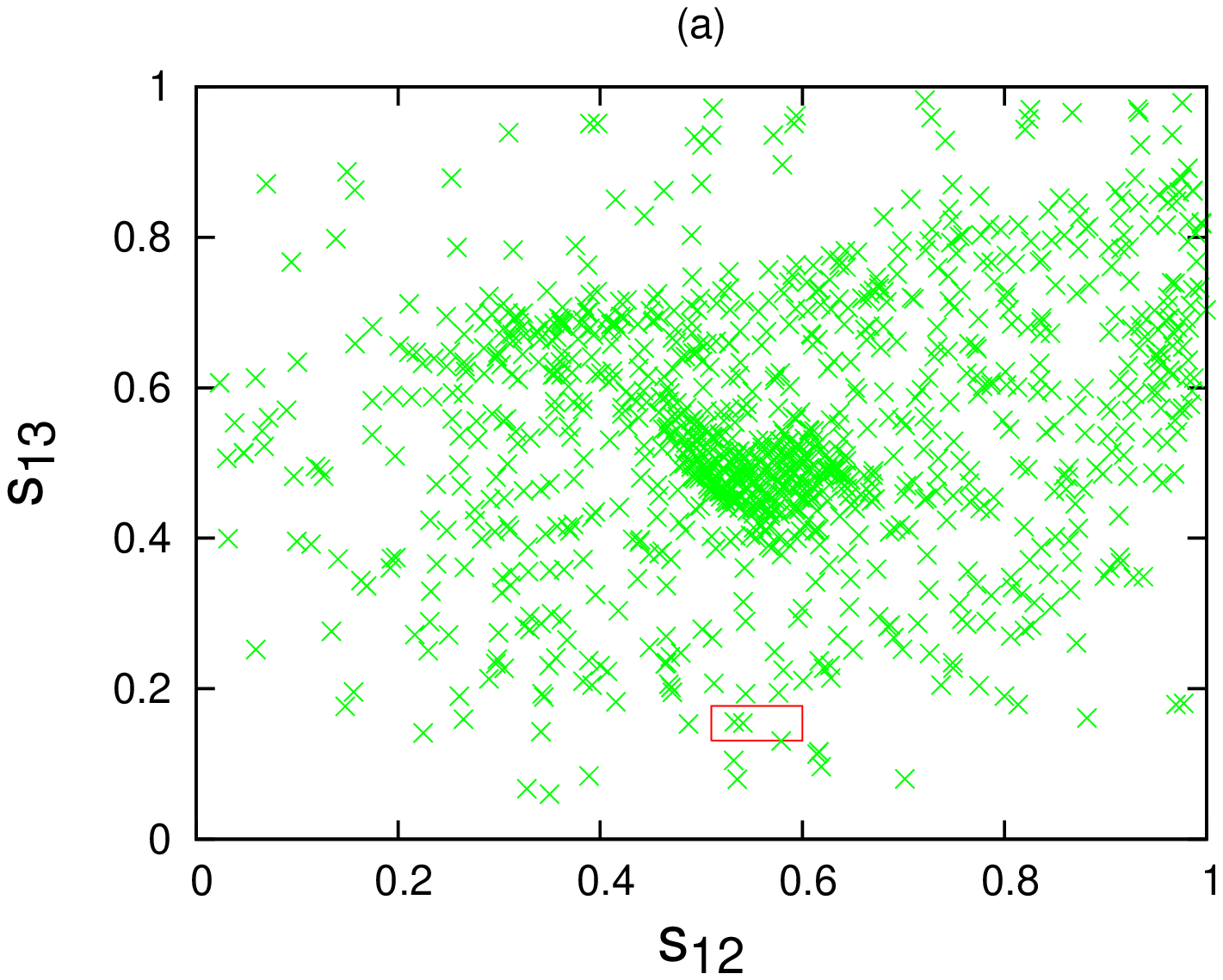}
  \includegraphics[width=0.2\paperwidth,height=0.2\paperheight,angle=-90]{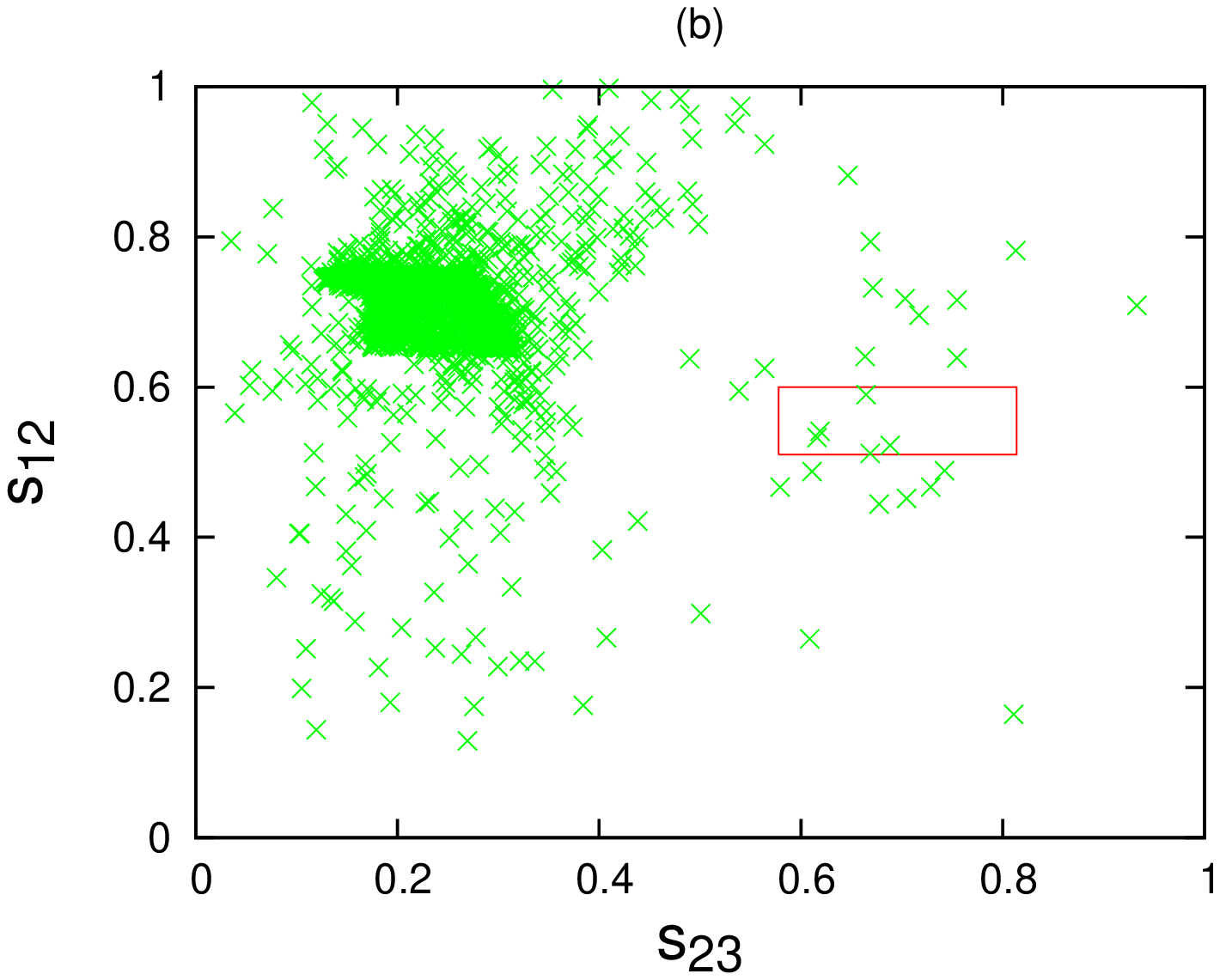}
  \includegraphics[width=0.2\paperwidth,height=0.2\paperheight,angle=-90]{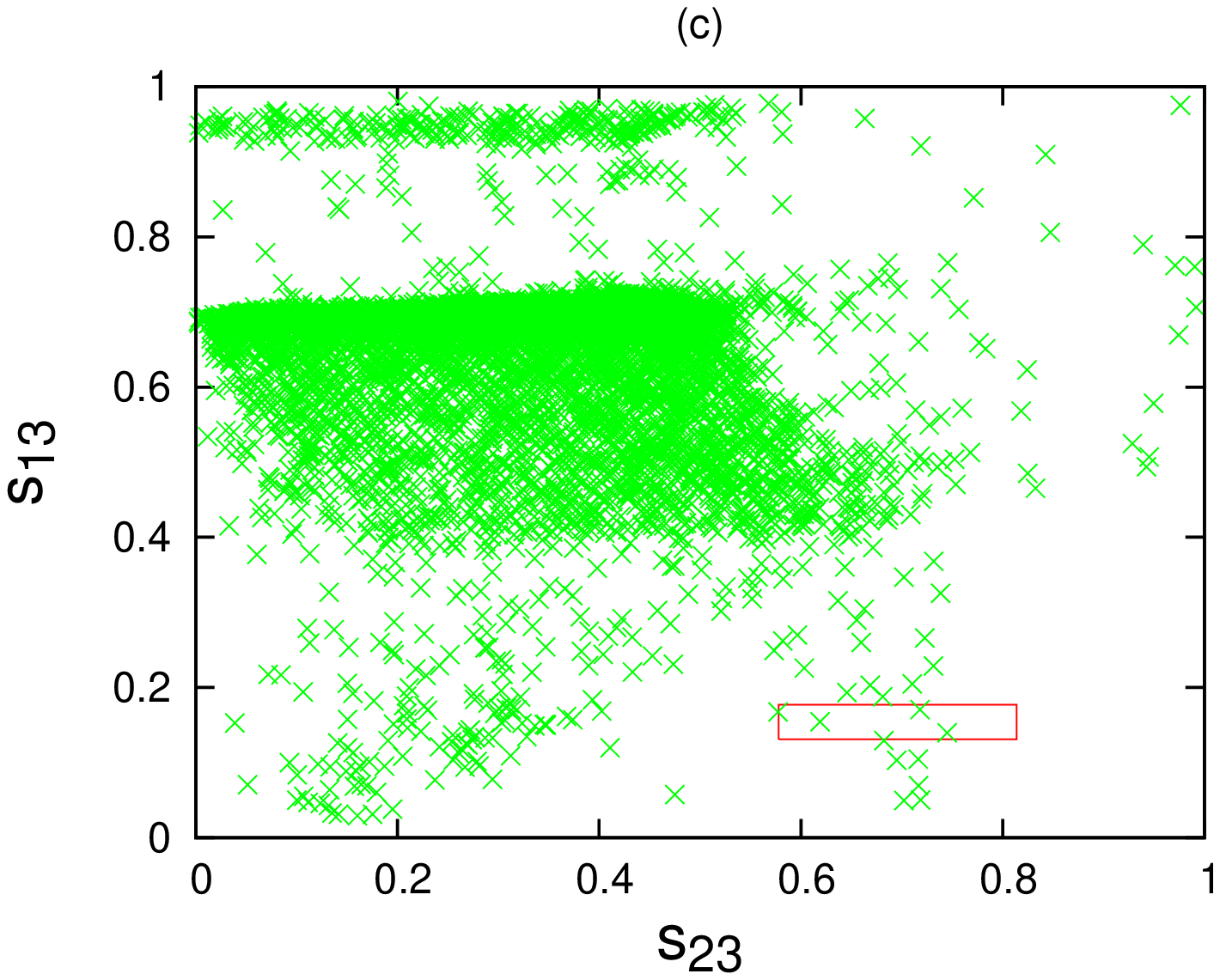}
\end{tabular}
\caption{Plots showing the parameter space for any two mixing angles when the third angle is constrained by
its  $1 \sigma$ range for Class II ansatz of texture four zero mass matrices (inverted hierarchy).}
\label{4aih1}
\end{figure}

To analyse this class we follow the same procedure as for class I Ans\"{a}tze. The
charged lepton and neutrino mass matrices which we choose to analyse for this class
can be given as,
\be
 M_{i}=\left( \ba{ccc}
D_i & A _{i}e^{i\alpha_i} & 0     \\
A_{i}e^{-i\alpha_i} & 0 &  B_{i}e^{i\beta_i}     \\
 0 & B_{i}e^{-i\beta_i} &  E_{i} \ea \right),
\label{cl2}
\ee
where $i=l,~\nu_D$ corresponds to the charged lepton and Dirac neutrino mass matrices
respectively. As for class I, the elements $D_l$ and $D_\nu$ as well
as the phases $\phi_1$ and $\phi_2$ are considered to be the free parameters. Firstly, we examine 
the viability of inverted hierarchy for the mass matrices presented in eqn.(\ref{cl2}). To this end in figure (\ref{4aih1}),
we present the plots showing the parameter space for two mixing angles wherein the third angle is constrained by its $1\sigma$ 
range. The rectangular boxes in these plots show the $3\sigma$ ranges for the
two mixing angles being considered. It is interesting to note that the inverted hierarchy seems to be compatible with
the $1\sigma$ 
ranges of the present lepton mixing data.
\par
\begin{figure}
\begin{tabular}{cc}
  \includegraphics[width=0.2\paperwidth,height=0.2\paperheight,angle=-90]{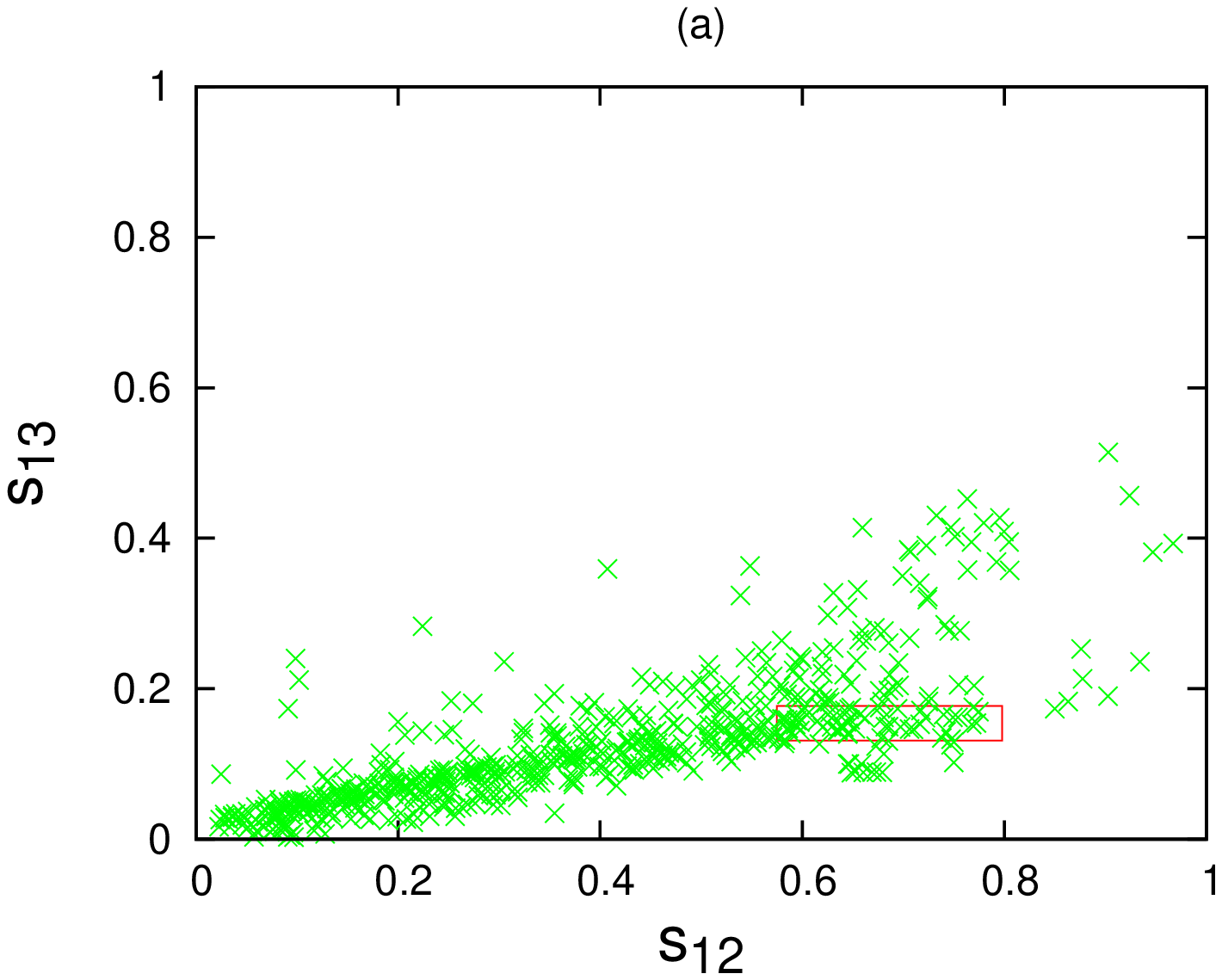}
  \includegraphics[width=0.2\paperwidth,height=0.2\paperheight,angle=-90]{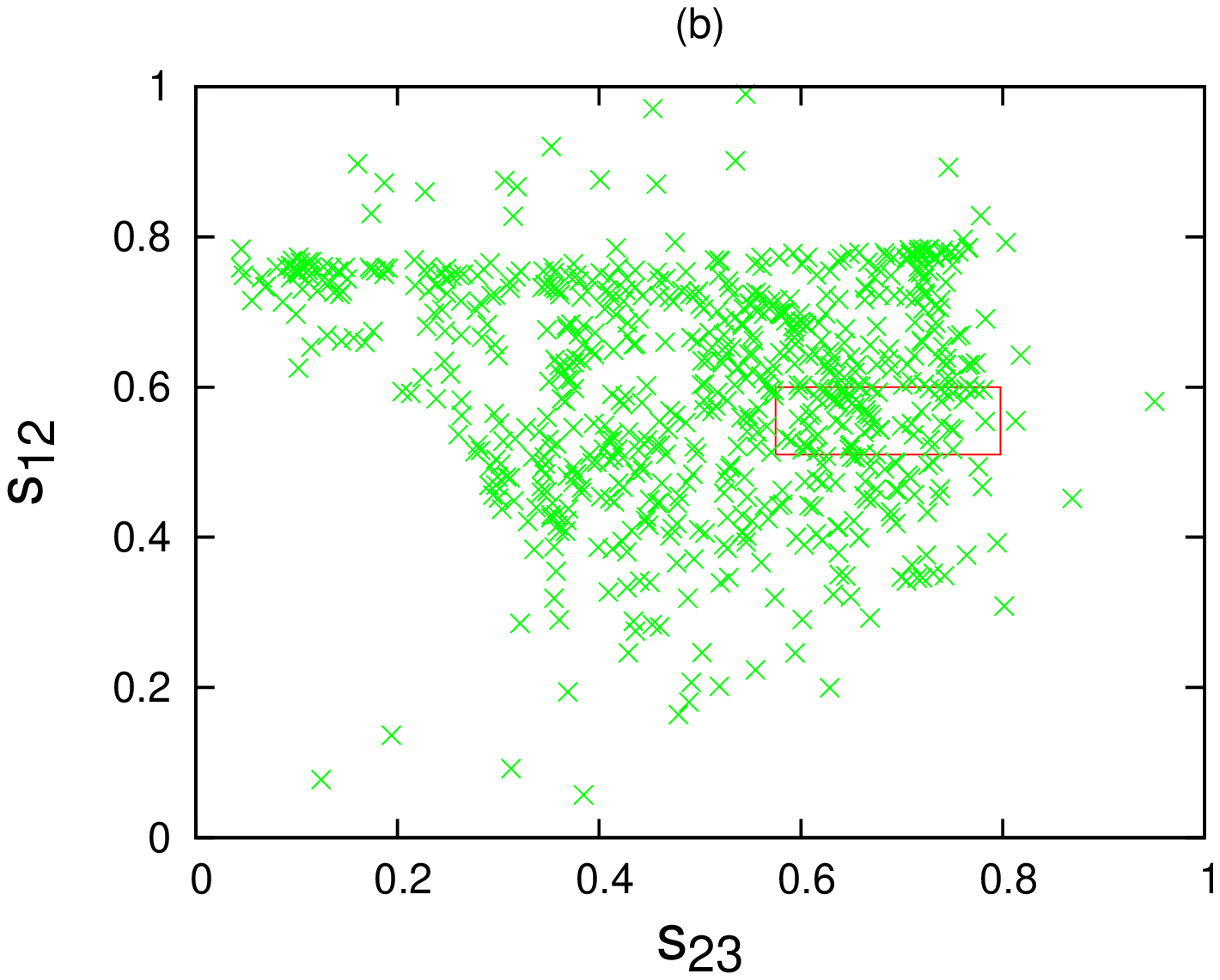}
  \includegraphics[width=0.2\paperwidth,height=0.2\paperheight,angle=-90]{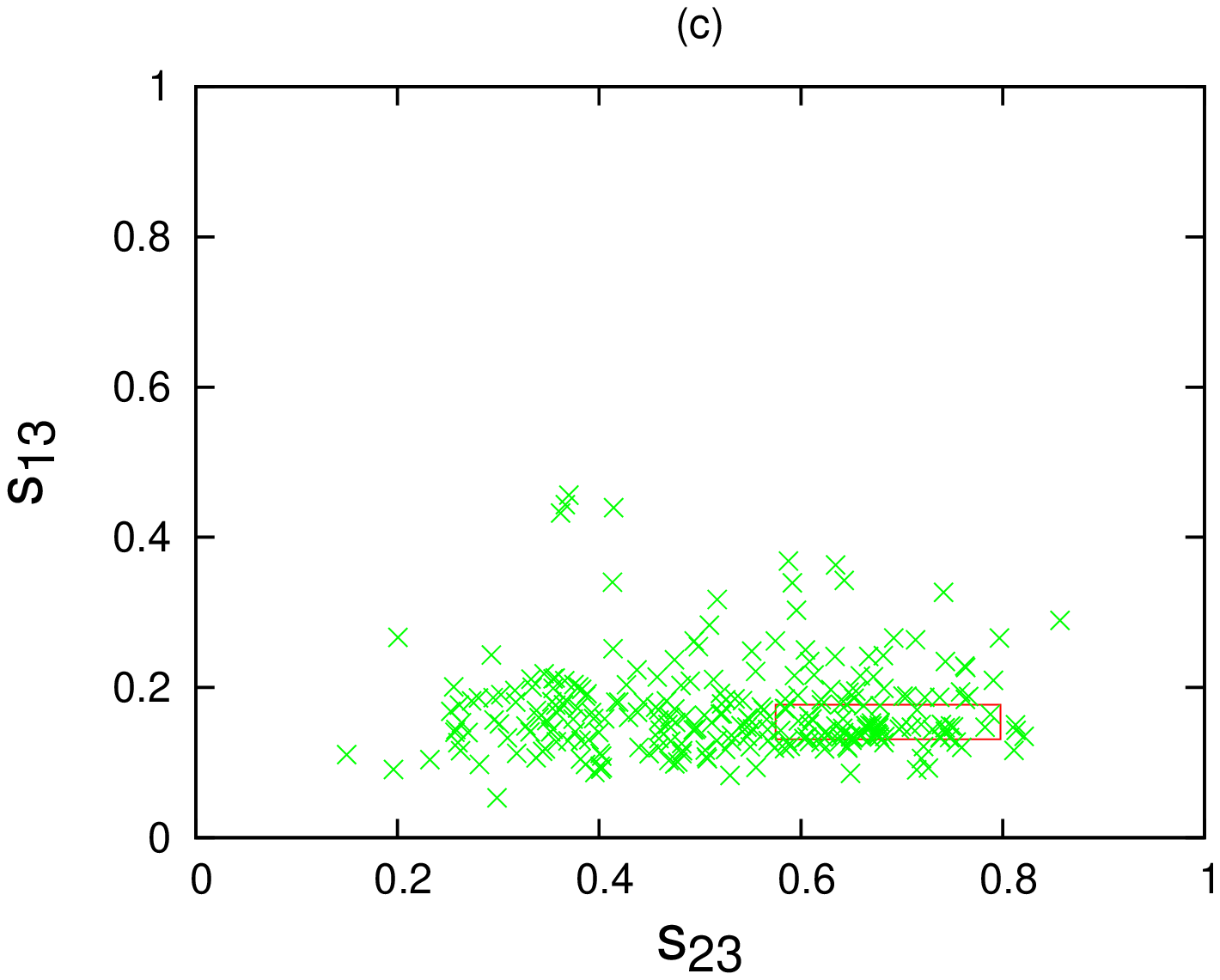}
\end{tabular}
\caption{Plots showing the parameter space for any two mixing angles when the third angle is constrained by
its  $1 \sigma$ range for Class II ansatz of texture four zero mass matrices (normal hierarchy).}
\label{4anh1}
\end{figure}

 
Next, in order to examine the compatibility of structure given in eqn.(\ref{cl2}) with the normal hierarchy, in figure
(\ref{4anh1}) we present the plots showing the parameter space for 
two mixing angles wherein the third angle is constrained by its $1\sigma$ experimental range. A general look at figure
(\ref{4anh1}) reveals that the mass matrices presented in eqn.(\ref{cl2}) are compatible with the normal hierarchy
scenario of neutrino masses.

\begin{figure}[hbt]
  \begin{minipage}{0.45\linewidth}   \centering
\includegraphics[width=2.in,angle=-90]{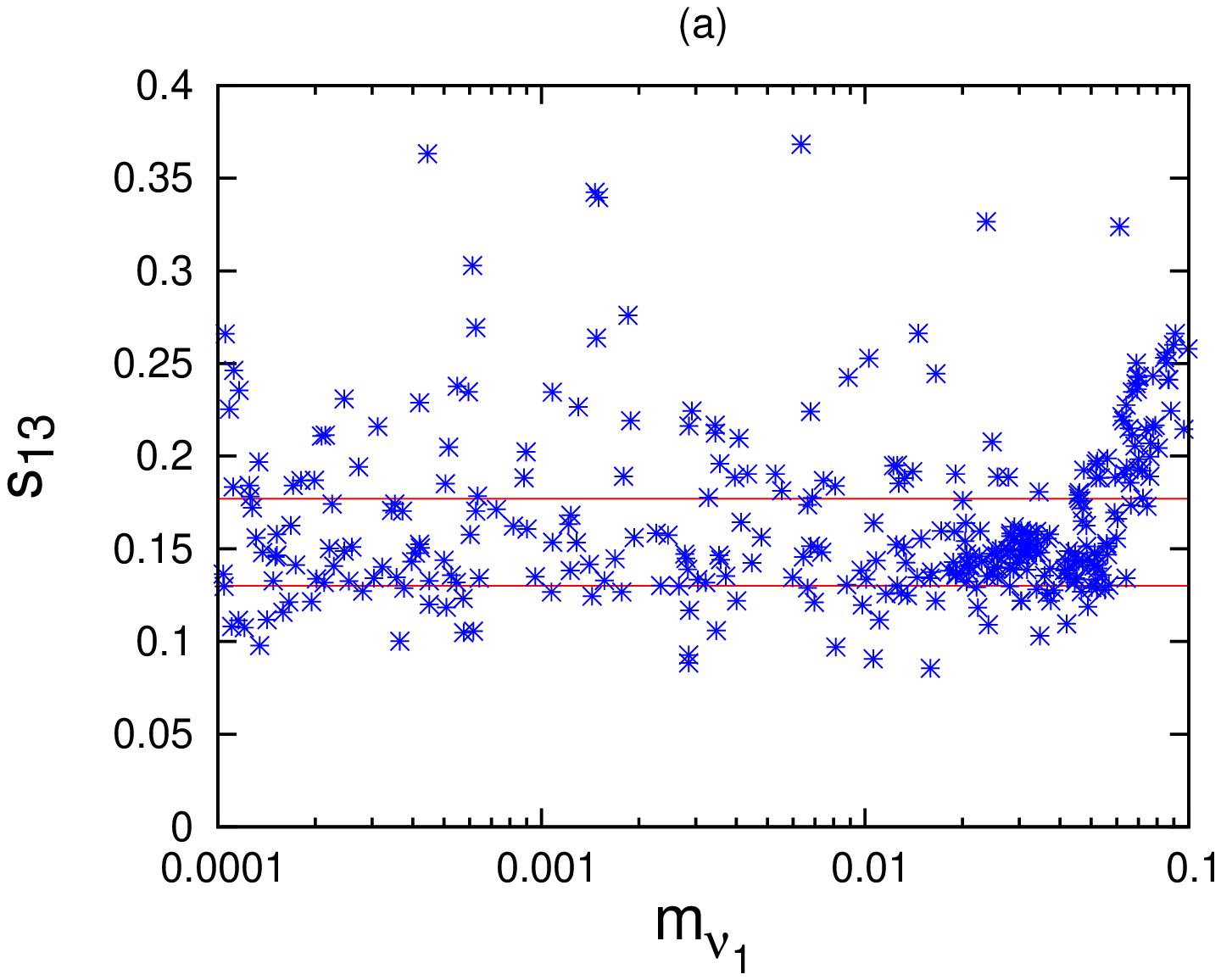}
    \end{minipage} \hspace{0.5cm}
\begin{minipage} {0.45\linewidth} \centering
\includegraphics[width=2.in,angle=-90]{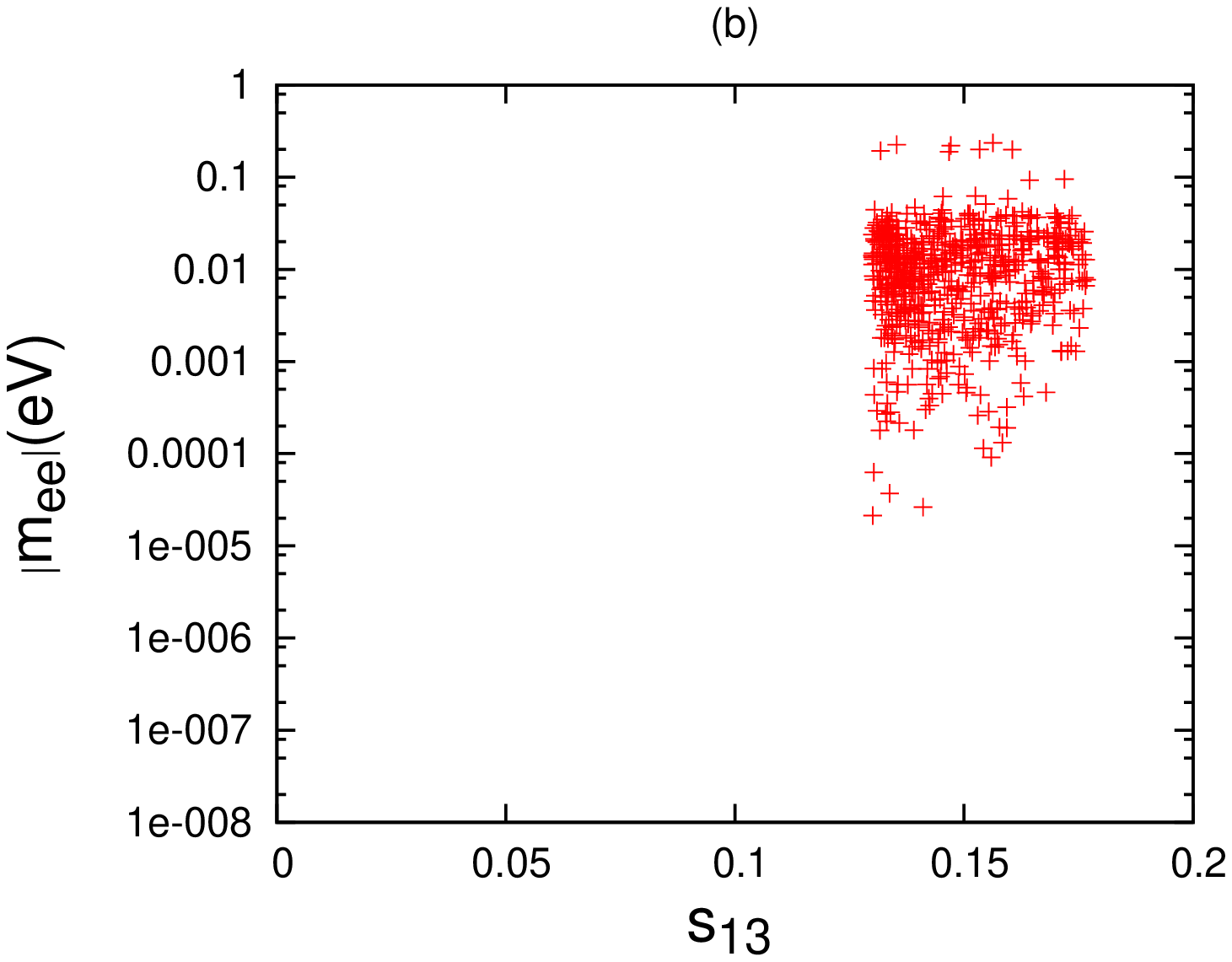}
\end{minipage}
\caption{Plots showing the dependence of mixing angle $s_{13}$ on (a) the lightest neutrino mass and 
(b) the effective Majorana mass,
for Class II ansatz of texture four zero  mass matrices (normal hierarchy).}
\label{4anh2}
\end{figure}

\begin{figure}[hbt]
  \begin{minipage}{0.45\linewidth}   \centering
\includegraphics[width=2.in,angle=-90]{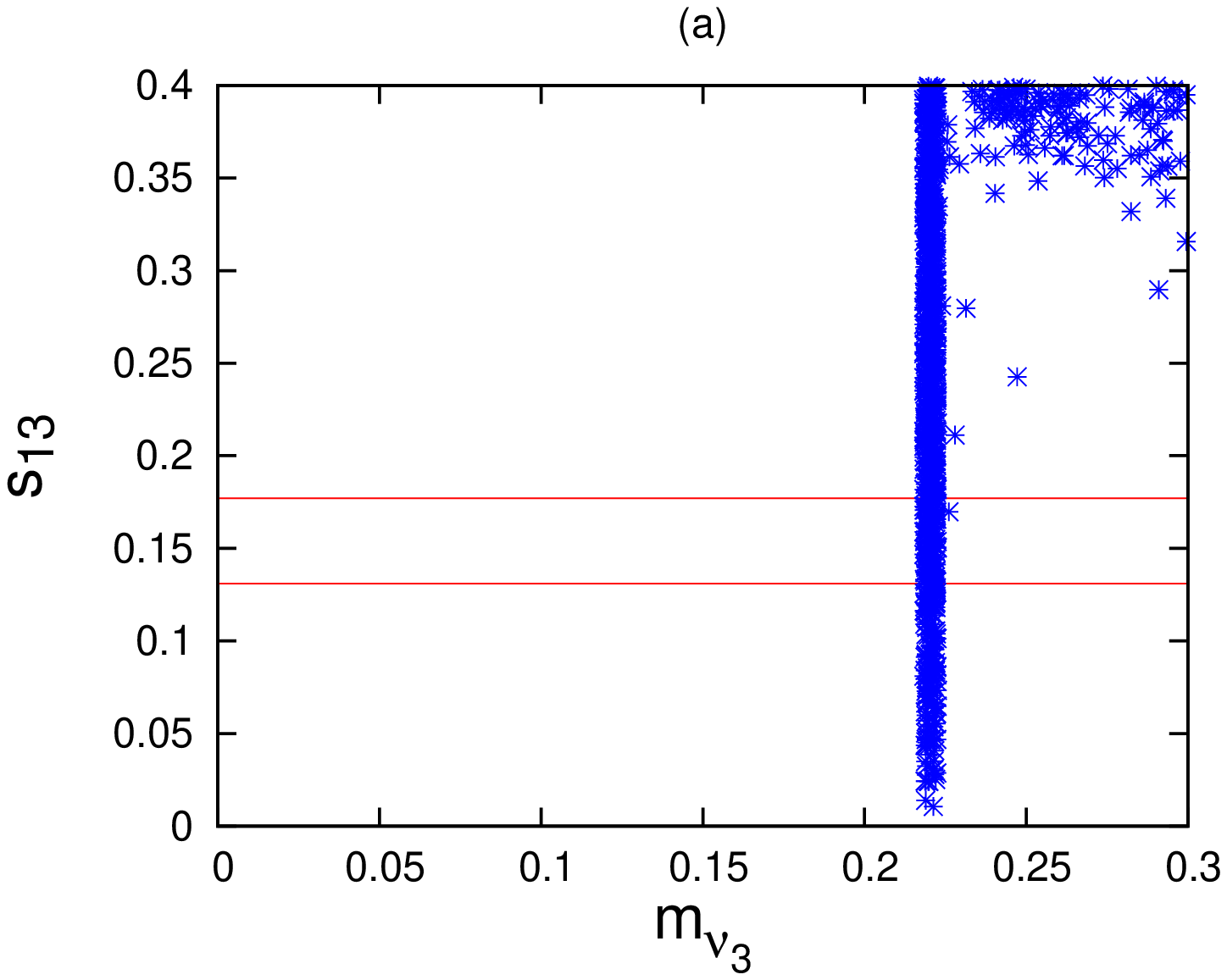}
    \end{minipage} \hspace{0.5cm}
\begin{minipage} {0.45\linewidth} \centering
\includegraphics[width=2.in,angle=-90]{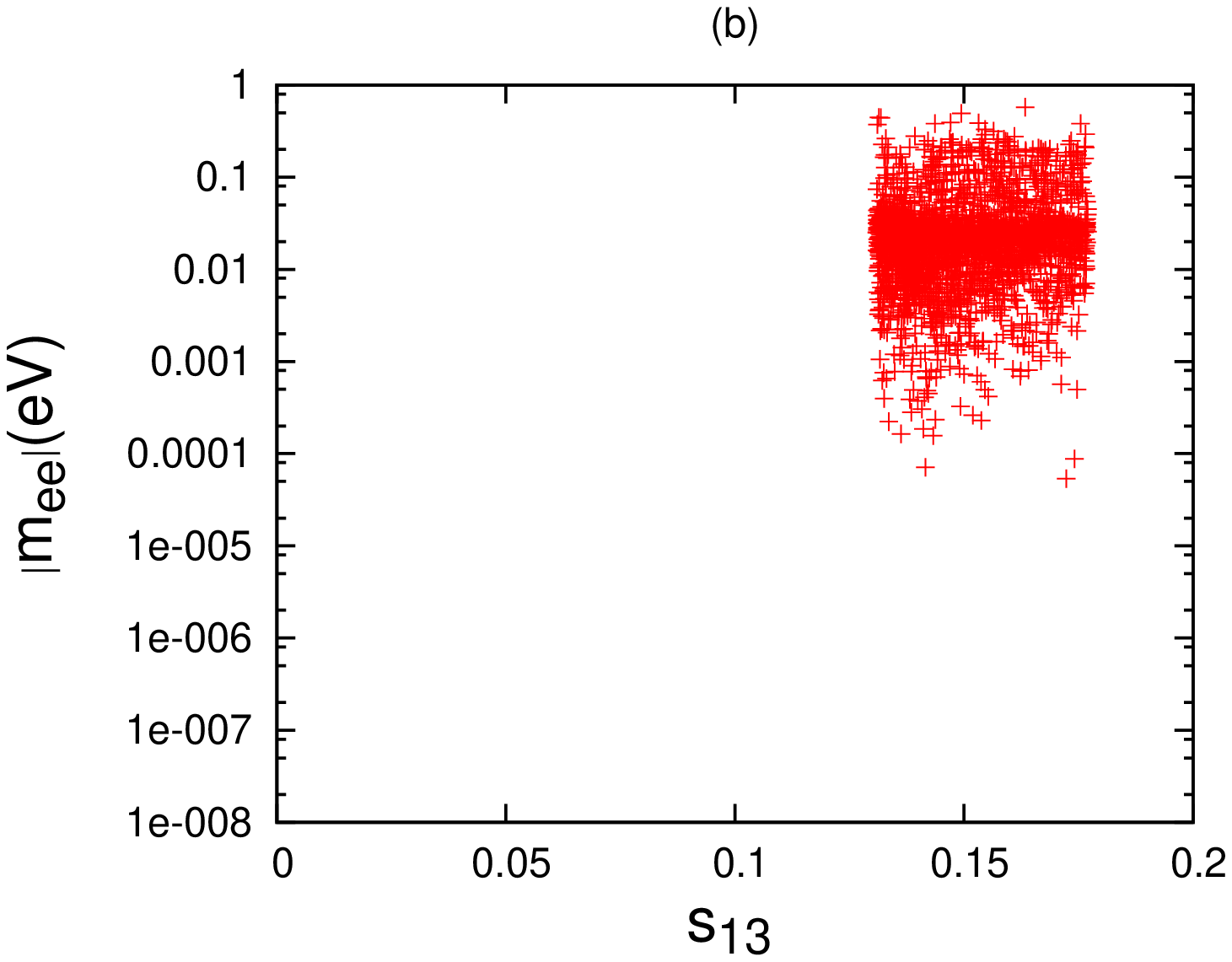}
\end{minipage}
\caption{Plots showing the dependence of mixing angle $s_{13}$ on (a) the lightest neutrino mass and 
(b) the effective Majorana mass, 
for Class II ansatz of texture four zero  mass matrices (inverted hierarchy).}
\label{4aih2}
\end{figure}
Thus, we find that  for texture four zero matrices pertaining to class II,
both the normal as well as inverted neutrino mass hierarchies are compatible
the lepton mixing data. As a next step, in figure
(\ref{4anh2}) we attempt to study the
implications of the $3\sigma$ ranges of the leptonic mixing angle $s_{13}$ on the lightest
neutrino mass and the effective Majorana mass respectively for normal hierarchy scenario of neutrino masses.
Analogous plots for inverted  hierarchy scenario have been presented in figure (\ref{4aih2}). While plotting 
figures \ref{4anh2}(a) and \ref{4aih2}(a), the other two mixing angles have been constrained by their
$3\sigma$ experimental bounds, whereas figures \ref{4anh2}(b) and \ref{4aih2}(b) have been obtained by 
constraining $s_{13}$ by its $3\sigma$ experimental bound. Interestingly, one finds that 
the present $3\sigma$ ranges of the mixing
angles provide no bound on the lightest neutrino mass for the normal hierarchy respectively, whereas for the 
inverted hierarchy case one gets an extremely narrow range for the lightest neutrino mass which clearly points to the 
degenerate scenario of neutrino masses. Interestingly, a careful look at figures \ref{4anh2}(b) and \ref{4aih2}(b)
reveals that the $3\sigma$ range for $s_{13}$
provides an approximate
lower bound on the effective Majorana mass $|m_{ee}|$, viz. $|m_{ee}| \gtrsim 10^{-5} eV$ and $|m_{ee}| \gtrsim 10^{-4} eV$
for the normal and inverted hierarchy scenarios respectively.

\subsubsection{Class III ansatz}\label{bbb}
\begin{figure}
\begin{tabular}{cc}
  \includegraphics[width=0.2\paperwidth,height=0.2\paperheight,angle=-90]{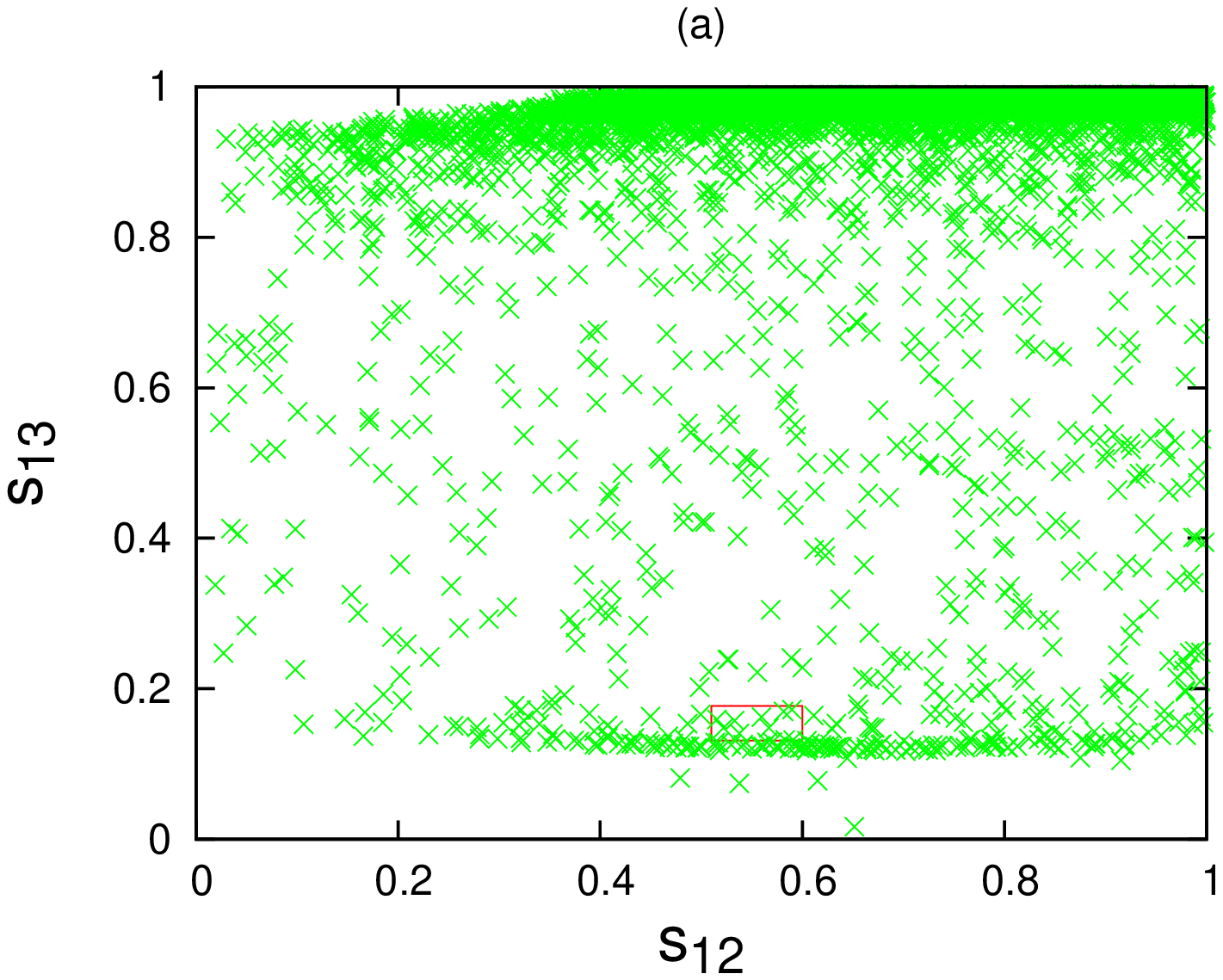}
  \includegraphics[width=0.2\paperwidth,height=0.2\paperheight,angle=-90]{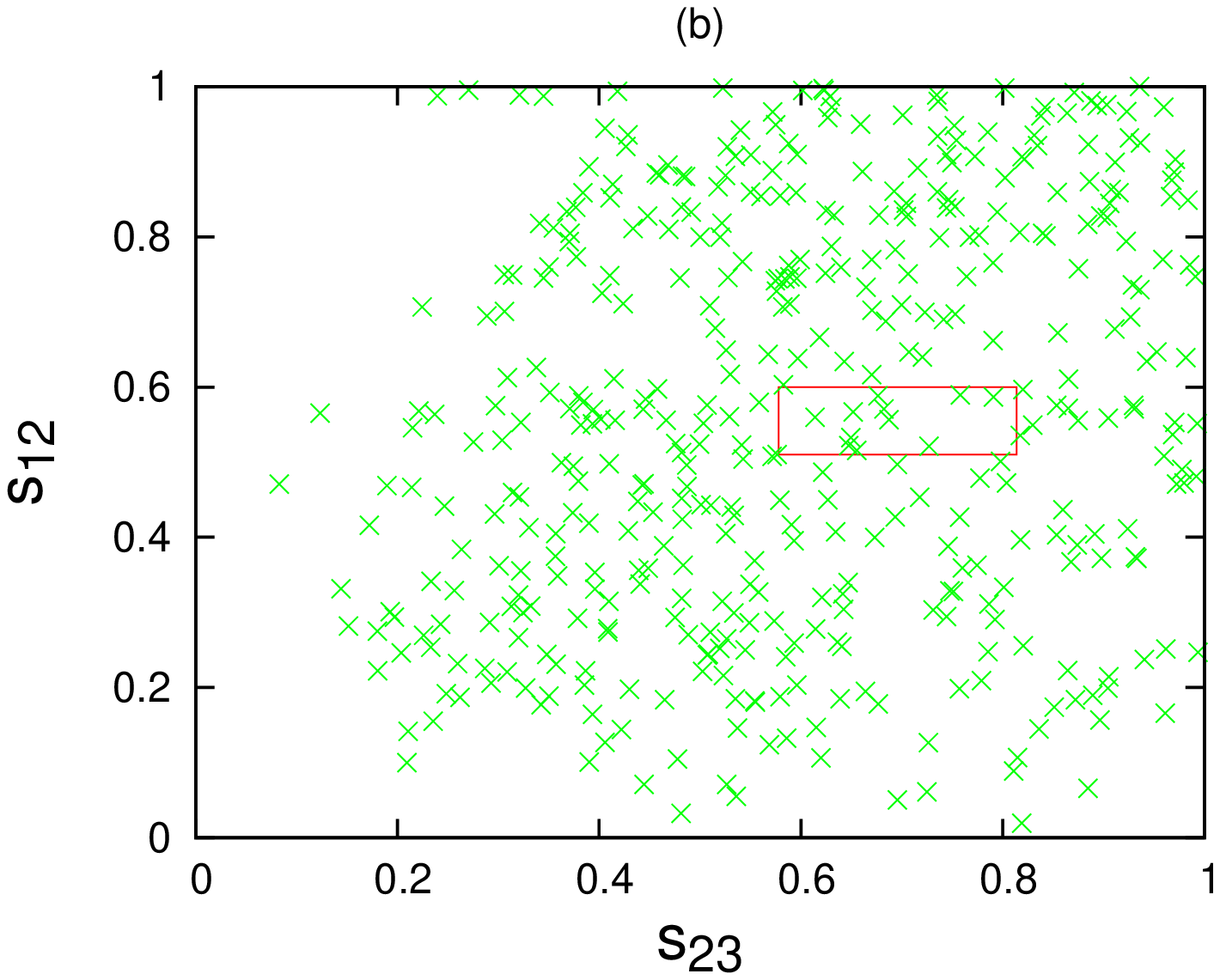}
  \includegraphics[width=0.2\paperwidth,height=0.2\paperheight,angle=-90]{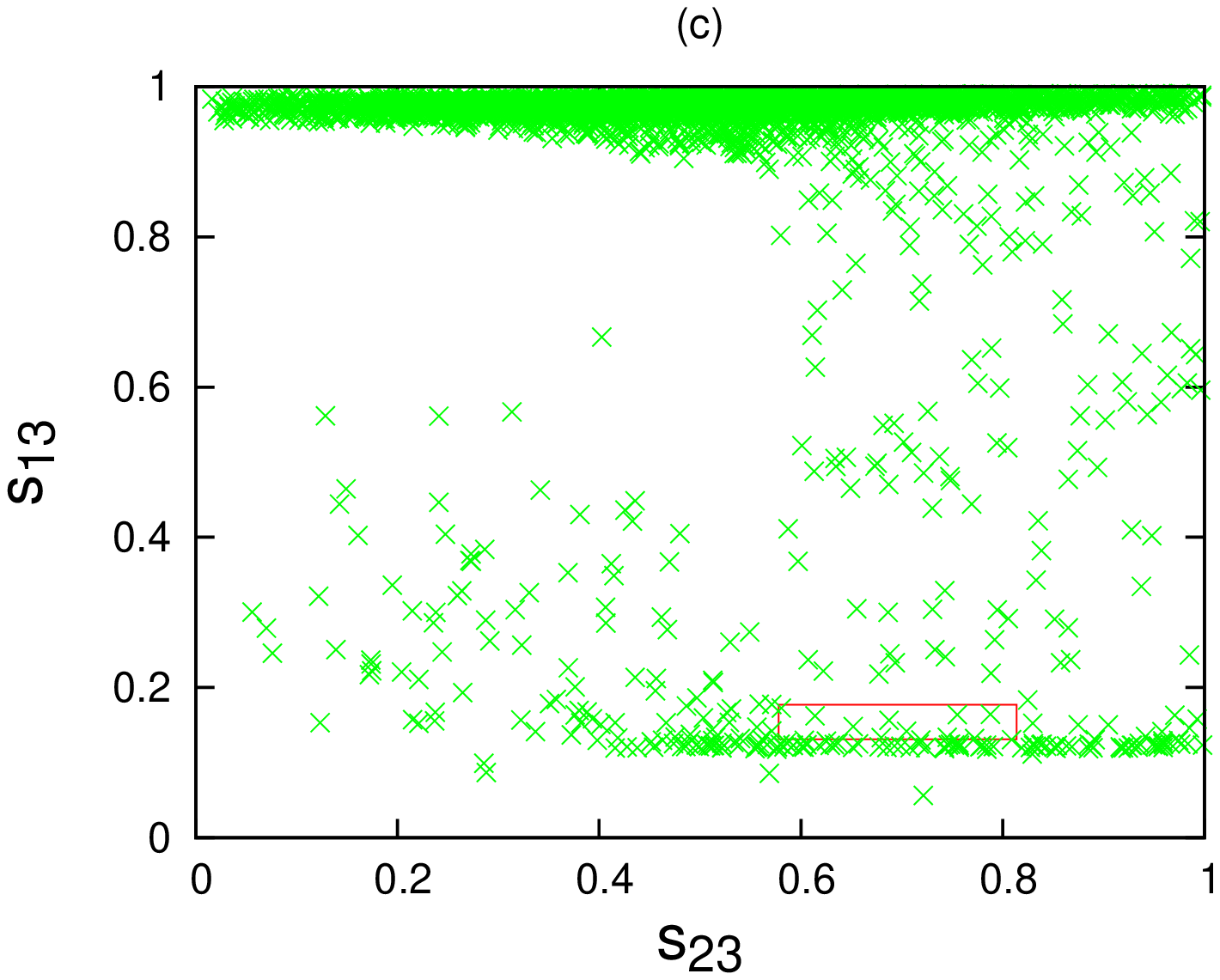}
\end{tabular}
\caption{Plots showing the parameter space for any two mixing angles when the third angle is constrained by
its  $1 \sigma$ range for Class III ansatz of texture four zero mass matrices (inverted hierarchy).}
\label{5aih1}
\end{figure}

To study the lepton mass matrices for this class, we choose to analyse the following
structure,

\be
 M_{i}=\left( \ba{ccc}
0 & A _{i}e^{i\alpha_i} & B_{i}e^{i\gamma_i}     \\
A_{i}e^{-i\alpha_i} & 0 &   D_{i}e^{i\beta_i}      \\
 B_{i}e^{-i\gamma_i} & D_{i}e^{-i\beta_i}  &  E_{i} \ea \right),
\label{cl3}
\ee
where $i=l,~\nu_D$ corresponds to the charged lepton and Dirac neutrino mass matrices
respectively. In the case of factorizable phases, the lepton mass matrices belonging to this
class can be analysed following a methodology similar to that for class I and class II ansatz.
For the purpose of calculations, the (2,3) element in each sector, $D_l$ and $D_\nu$, as well
as the phases $\phi_1$ and $\phi_2$ have been considered as free parameters. Firstly, we examine the texture 
four zero mass matrices in this class for their 
compatibility with the inverted hierarchy scenario. For this purpose, in figure (\ref{5aih1}) we present the plots 
showing the parameter space for any two mixing angles keeping the third one constrained by its $1\sigma$ experimental bound.
A general look at figure (\ref{5aih1}) shows that the  inverted mass neutrino mass ordering scenario is 
viable as can be seen
from significant overlap with the experimentally allowed $3\sigma$ regions shown by rectangular boxes in each plot.

After examining the viability of inverted hierarchy scenario for the texture four zero mass matrices pertaining to 
class III, we now proceed to examine their compatibility with the normal hierarchy of neutrino masses. To this end, 
in figure (12) we present the plot showing the parameter space corresponding to the mixing angles 
$s_{23}$ and $s_{12}$. While plotting this figure, the mixing angle $s_{13}$ has been constrained by its $1\sigma$ 
experimental bound. A general look at figure (12) reveals that normal hierarchy is ruled out by the $1\sigma$
ranges of the latest lepton mixing data. This can be understood by noting that the plotted parameter space of the 
two angles has no overlap with their experimentally allowed $1\sigma$ region shown by the rectangular box in the figure.

\begin{figure}[hbt]
\bc
 \includegraphics[width=2.in,angle=270]{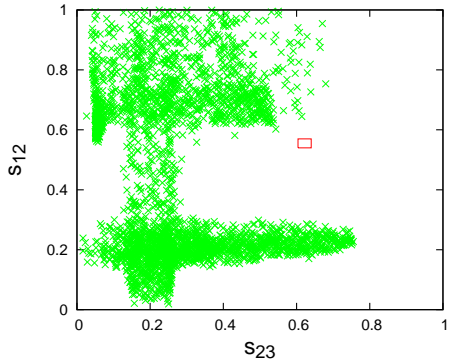}
 \caption{Plot showing the parameter space for $s_{23}$ and $s_{12}$ when $s_{13}$ is constrained by
its  $1 \sigma$ range for
 Class III ansatz of texture four zero mass matrices (normal hierarchy).} 
 \ec
 \label{aaaa}
\end{figure}

\begin{figure}
\begin{tabular}{cc}
  \includegraphics[width=0.2\paperwidth,height=0.2\paperheight,angle=-90]{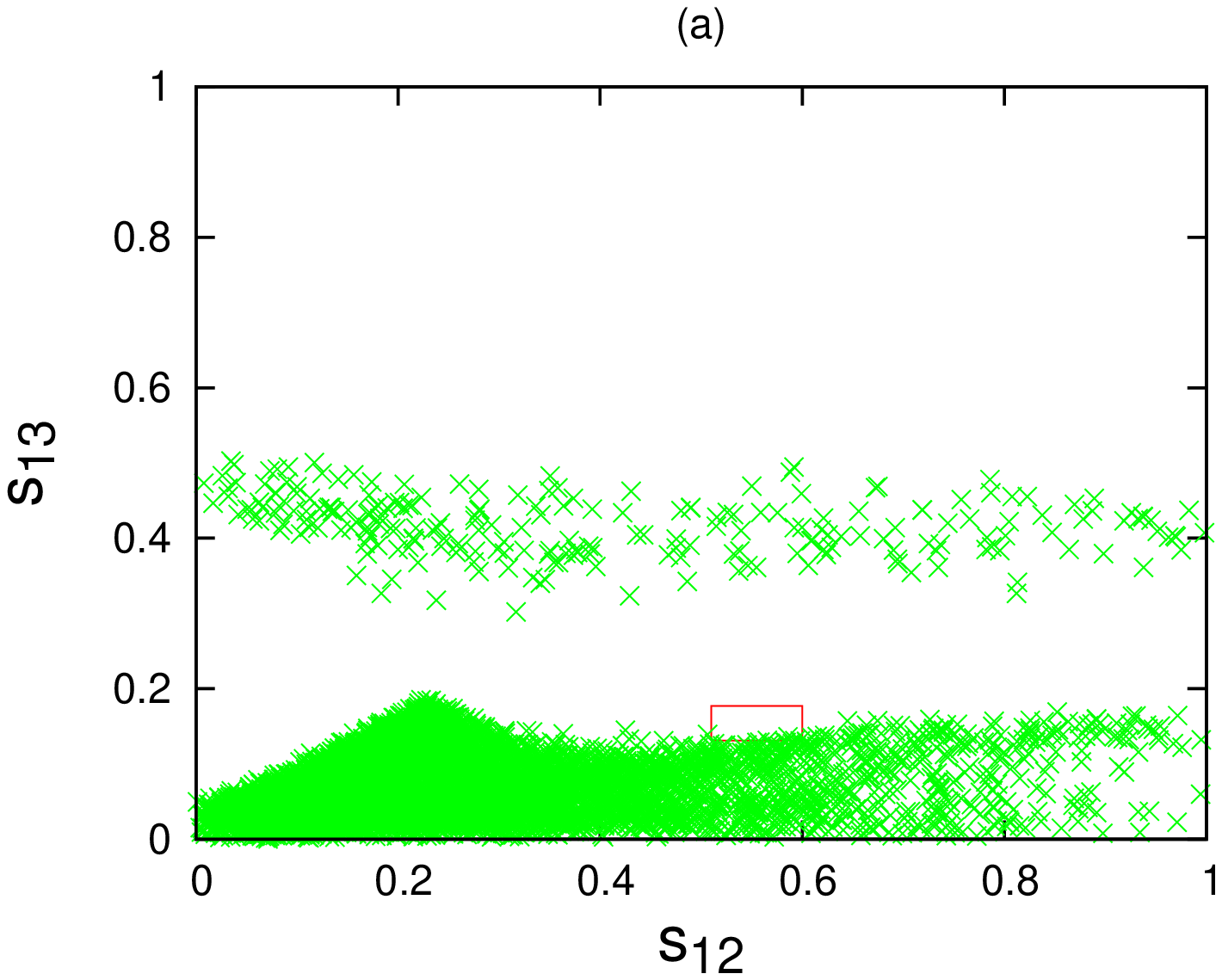}
  \includegraphics[width=0.2\paperwidth,height=0.2\paperheight,angle=-90]{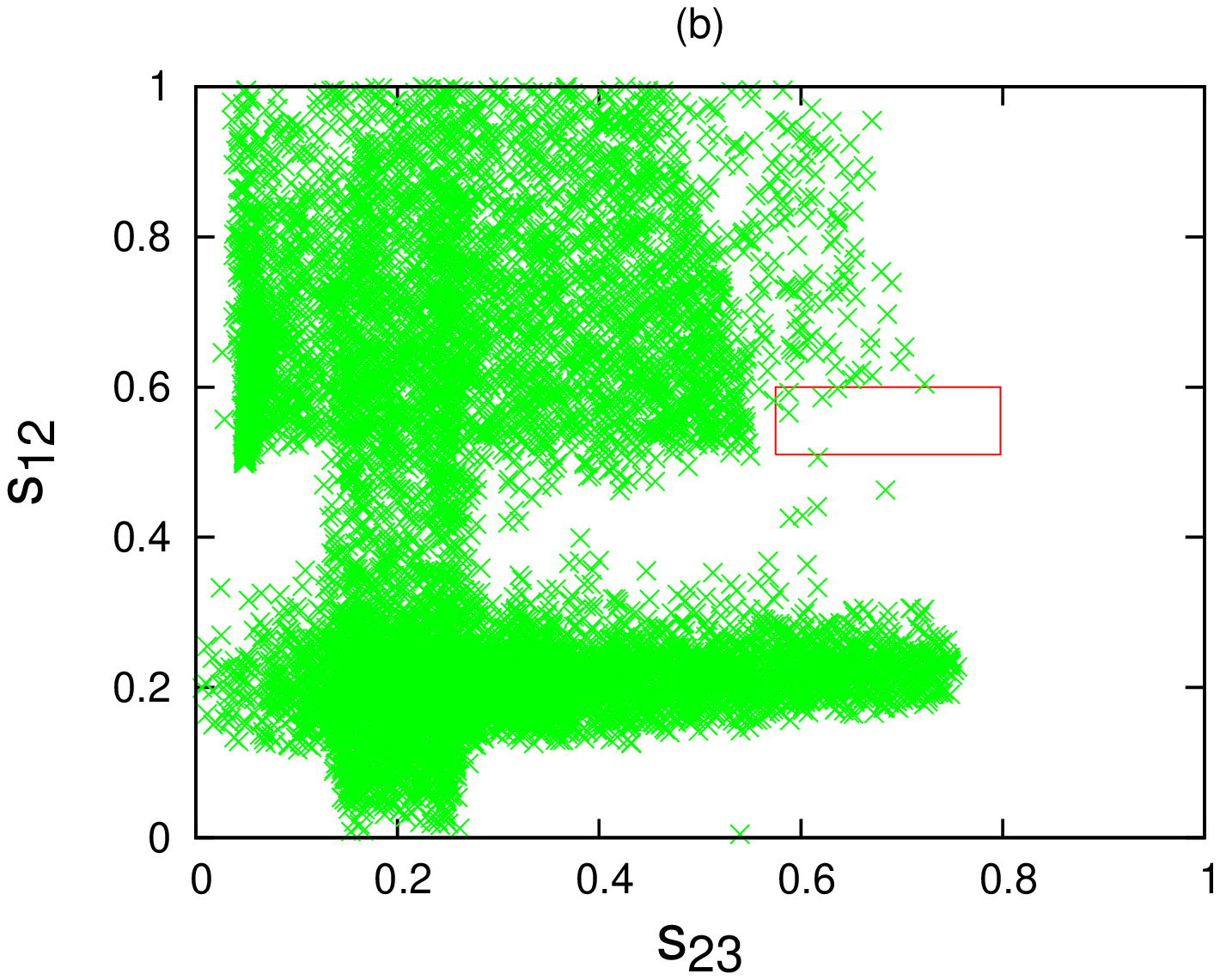}
  \includegraphics[width=0.2\paperwidth,height=0.2\paperheight,angle=-90]{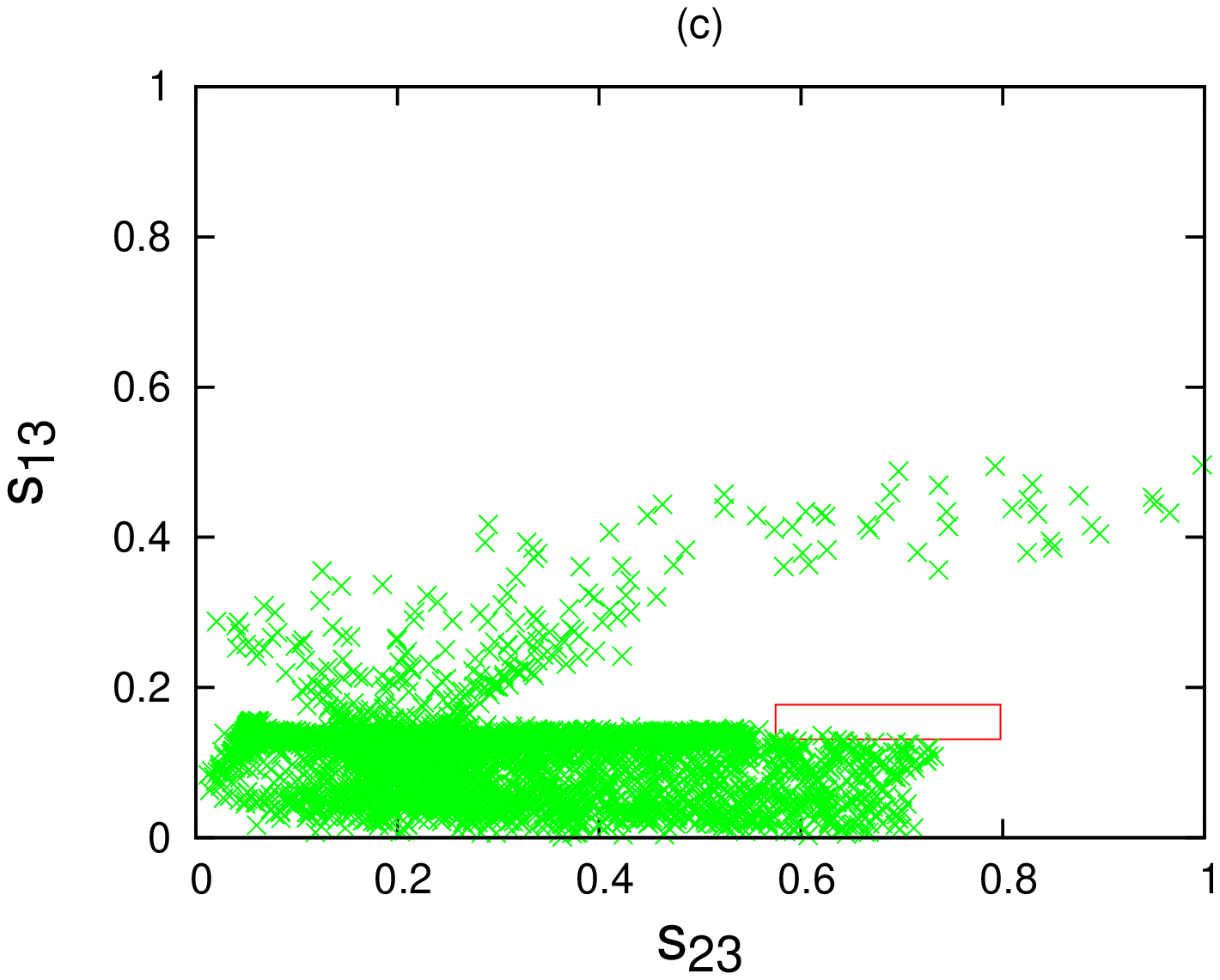}
\end{tabular}
\caption{Plots showing the parameter space for any two mixing angles when the third angle is constrained by
its  $3 \sigma$ range for
Class III ansatz of texture four zero mass matrices (normal hierarchy).}
\label{5anh2}
\end{figure}

Further, we examine the compatibility of these mass matrices with the normal hierarchy scenario for 
the $3\sigma$ ranges of the present lepton mixing data. For
this purpose in figure
(\ref{5anh2}), we present the parameter space corresponding to any two mixing angles while
the third one being constrained by its $3\sigma$ range. The blank rectangular regions
in figure (\ref{5anh2}) represent the $3\sigma$ 
allowed ranges of the two mixing angles being considered respectively. Interestingly, we find that 
the class III ansatz is compatible with the $3\sigma$ ranges of the present lepton 
mixing data pertaining to normal hierarchy scenario.

\begin{figure}[hbt]
  \begin{minipage}{0.45\linewidth}   \centering
\includegraphics[width=2.in,angle=-90]{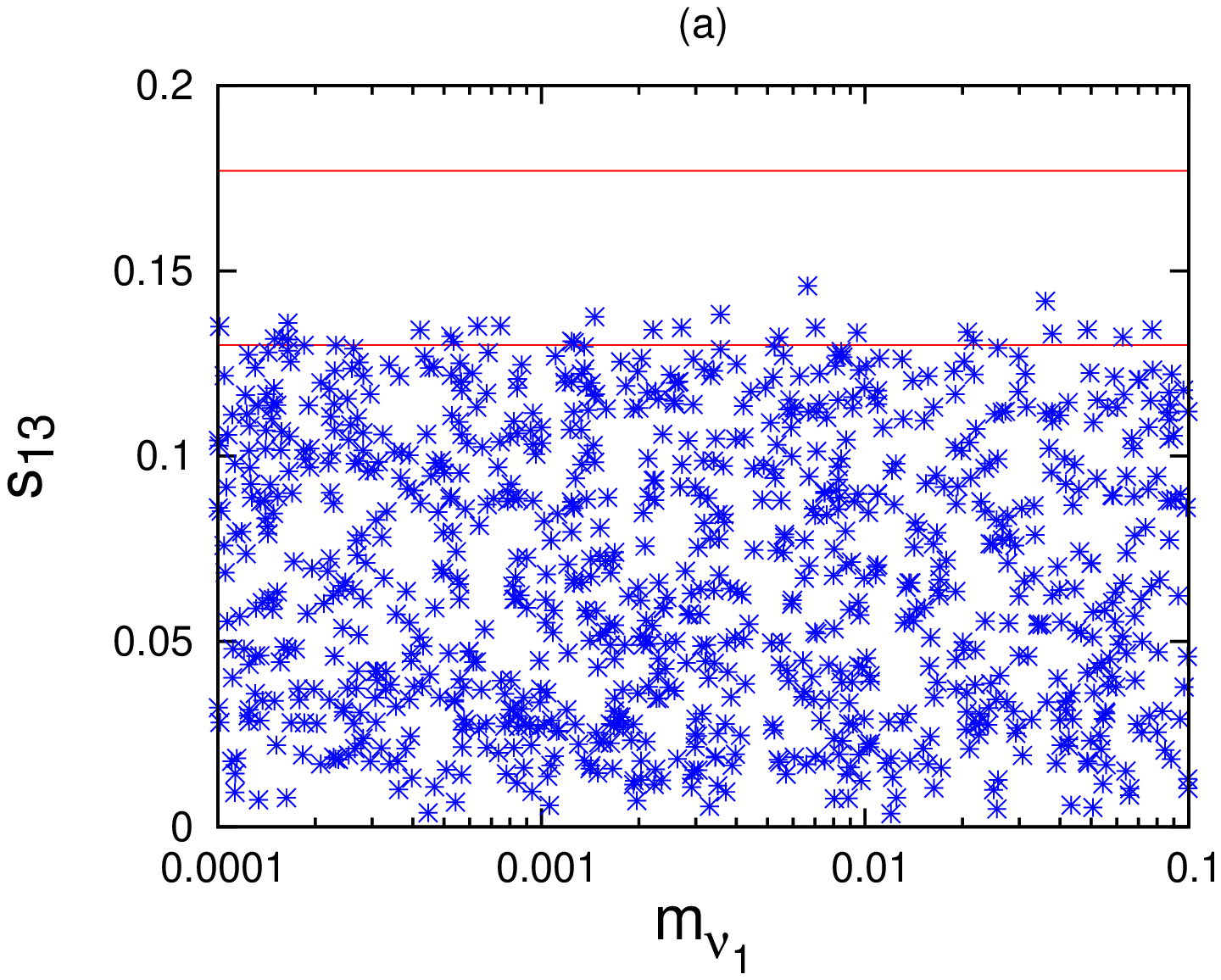}
    \end{minipage} \hspace{0.5cm}
\begin{minipage} {0.45\linewidth} \centering
\includegraphics[width=2.in,angle=-90]{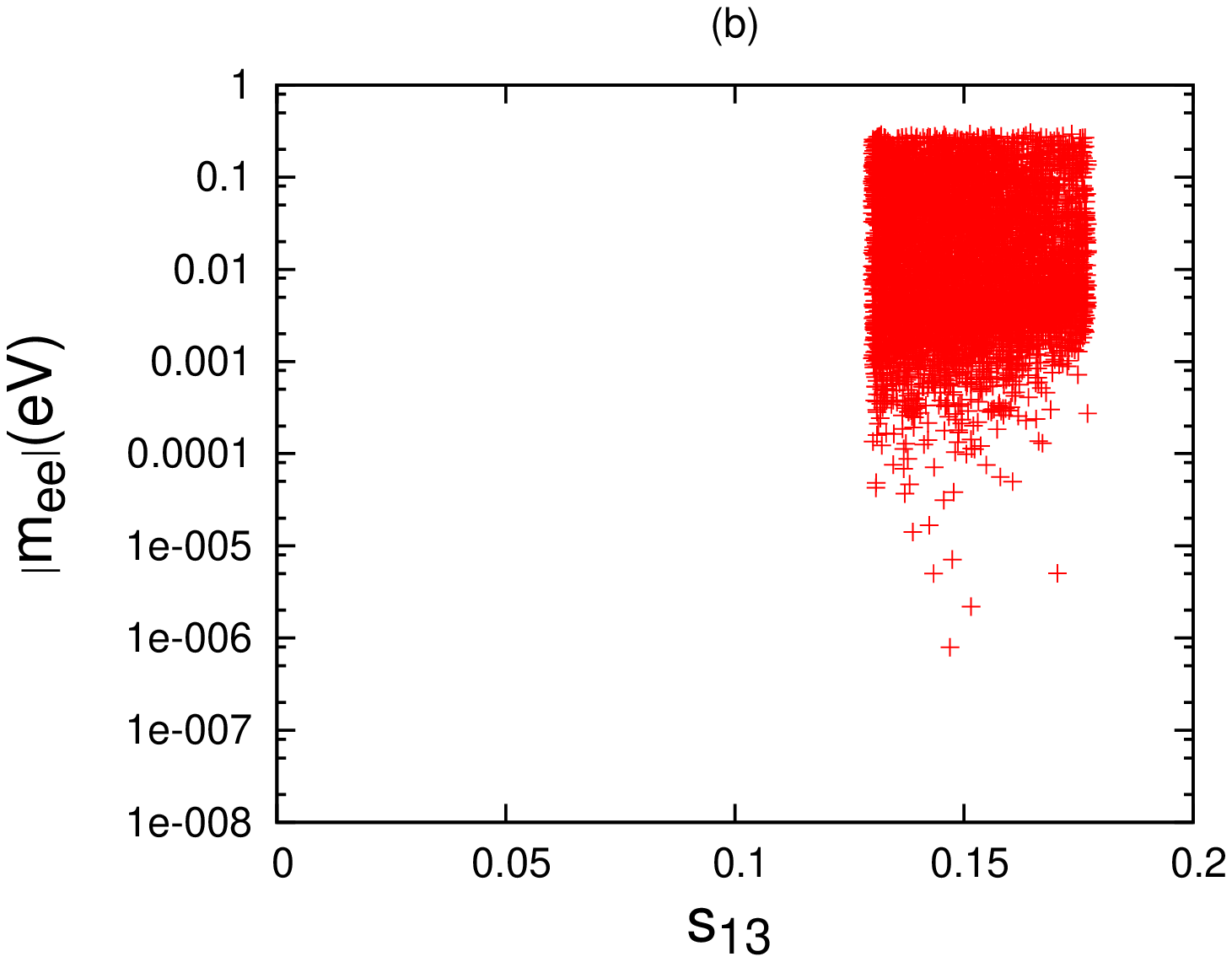}
\end{minipage}
\caption{Plots showing the dependence of mixing angle $s_{13}$ on (a) the lightest neutrino mass and 
(b) the effective Majorana mass,
for Class III ansatz of texture four zero  mass matrices (normal hierarchy).}
\label{5anh3}
\end{figure}

\begin{figure}[hbt]
  \begin{minipage}{0.45\linewidth}   \centering
\includegraphics[width=2.in,angle=-90]{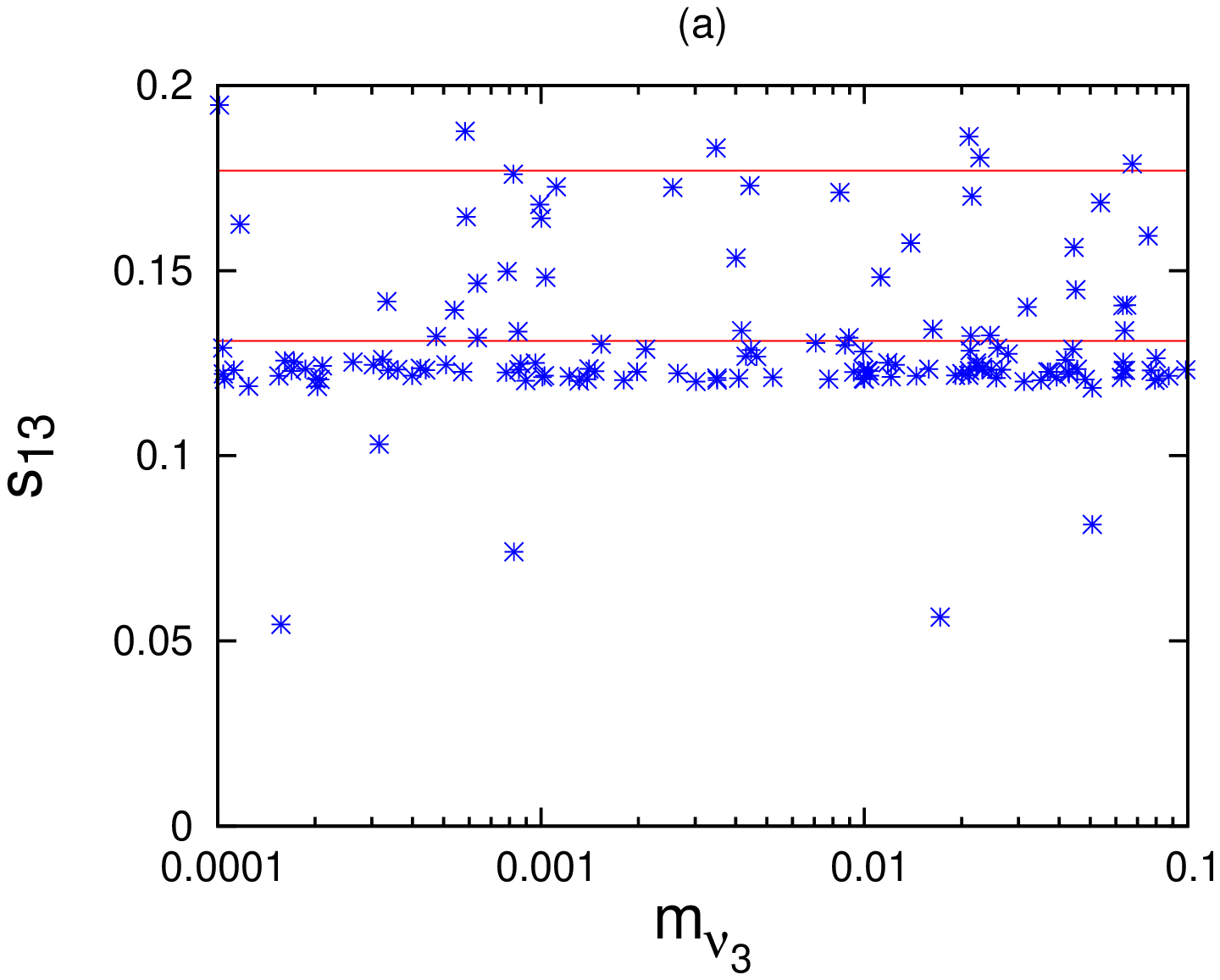}
    \end{minipage} \hspace{0.5cm}
\begin{minipage} {0.45\linewidth} \centering
\includegraphics[width=2.in,angle=-90]{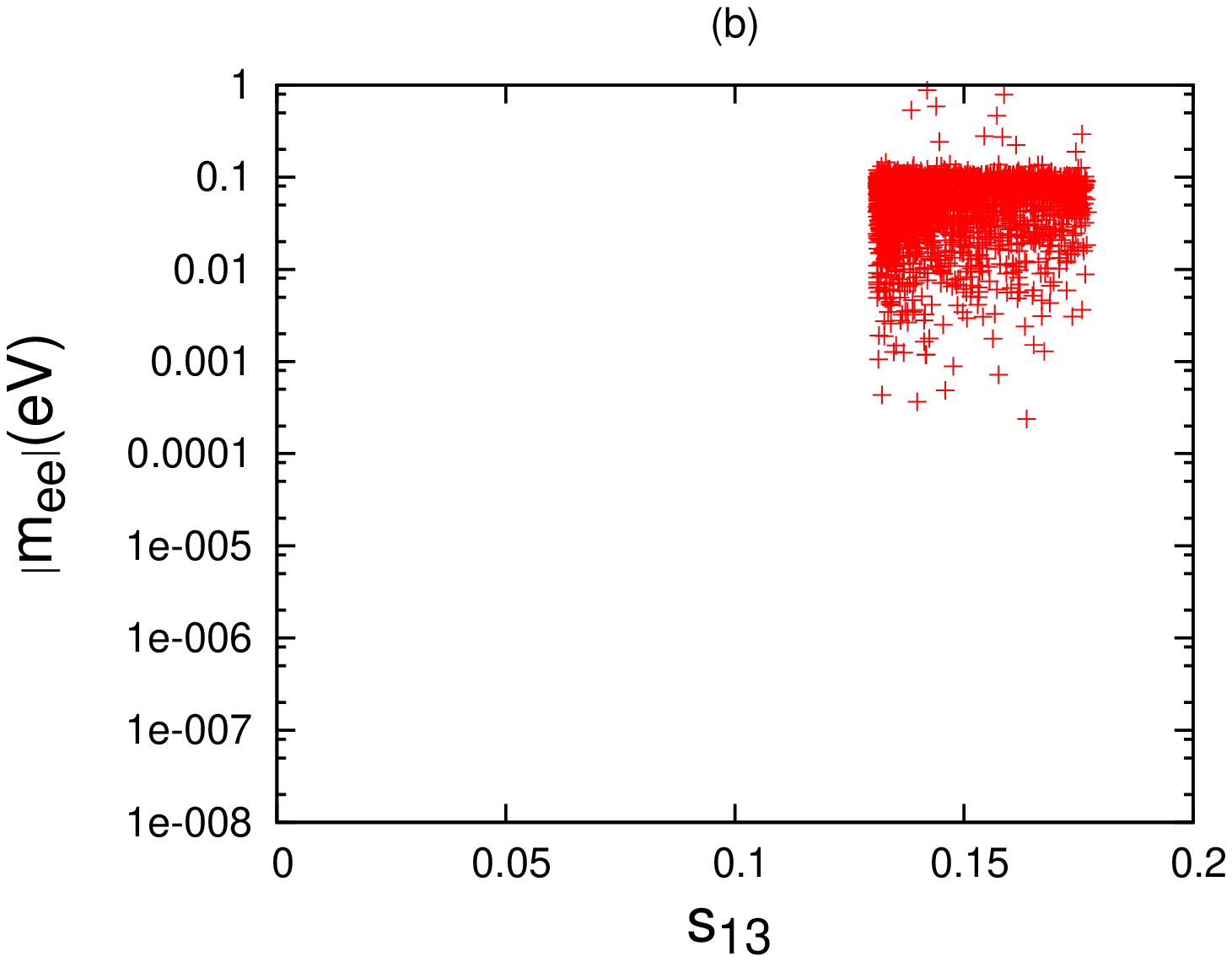}
\end{minipage}
\caption{Plots showing the dependence of mixing angle $s_{13}$ on (a) the lightest neutrino mass and 
(b) the effective Majorana mass, for
Class III ansatz of texture four zero  mass matrices (inverted hierarchy).}
\label{5aih2}
\end{figure}
As a next step, in figure
\ref{5anh3}(a) and \ref{5anh3}(b) we attempt to study the
implications of the $3\sigma$ ranges of the leptonic mixing angle $s_{13}$ on the lightest
neutrino mass and the effective Majorana mass respectively for normal hierarchy scenario of neutrino masses.
Analogous plots for inverted  hierarchy scenario have been presented in figure (\ref{5aih2}). While plotting 
figures \ref{5anh3}(a) and \ref{5aih2}(a), the other two mixing angles have been constrained by their
$3\sigma$ experimental bounds, whereas figures \ref{5anh3}(b) and \ref{5aih2}(b) have been obtained by 
constraining $s_{13}$ by its $3\sigma$ experimental bound. Interestingly, one finds that 
the present $3\sigma$ range of the mixing
angle $s_{13}$ provides no bound on the lightest neutrino mass for both the normal as well as inverted hierarchy scenario.
Further, a careful look at figures \ref{5anh3}(b) and \ref{5aih2}(b)
reveals the lower bound on $|m_{ee}|$ for the case of normal hierarchy is quite broader as compared to that 
for the inverted hierarchy scenario, viz., $|m_{ee}| \gtrsim 10^{-6} eV$ and $|m_{ee}| \gtrsim 10^{-4} eV$ 
for normal and inverted hierarchy respectively. Further, since the lightest neutrino mass for both normal 
as well as inverted hierarchy is unrestricted, therefore possibilty of degenerate neutrino mass ordering can not be ruled 
out for class III ansatz of texture four zero mass matrices.

\subsection{Texture five zero lepton mass matrices}
After studying all possible texture four zero lepton mass matrices, it becomes
interesting to explore the parallel texture five zero structures for each class which can be derived by
substituting  either $D_l=0$, $D_\nu \neq 0$ or $D_l\neq 0$, $D_l = 0$ in the corresponding
texture four zero mass matrices. In this subsection, we carry out a detailed study pertaining to all classes of
texture five zero lepton mass matrices for both the
possibilities leading to texture five zero structures. A detailed analysis \cite{ourplb} for both the cases of
Fritzsch-like 
texture five zero mass matrices has recently been carried out. Therefore, in the present work we present the
analyses for class II and class III ansatz of texture five zero mass matrices only.
\subsubsection{Class II ansatz}
The two possibilities for texture five zero lepton mass matrices for this class can be given as,
\be
 M_{l}=\left( \ba{ccc}
0 & A _{l} & 0    \\
A_{l}^{*} & 0 &  B_l    \\
 0 &   B_{l}^{*}     &  E_{l} \ea \right), \qquad
 M_{\nu}=\left( \ba{ccc}
D_\nu & A _{\nu} & 0    \\
A_{\nu}^{*} & 0 &  B_{\nu}    \\
 0  &  B_{\nu}^{*}    &  E_{\nu} \ea \right),
\label{cl2t51}\ee
or
\be
  M_{l}=\left( \ba{ccc}
D_l & A _{l} & 0   \\
A_{l}^{*} & 0 &   B_{l}     \\
 0 &  B_{l}^{*}     &  E_{l} \ea \right), \qquad
 M_{\nu}=\left( \ba{ccc}
0 & A _{\nu} & 0   \\
A_{\nu}^{*} & 0 &  B_{\nu}      \\
 0  &  B_{\nu}^{*}    &  E_{\nu} \ea \right),
\label{cl2t52}\ee

\begin{figure}[hbt]
  \begin{minipage}{0.45\linewidth}   \centering
\includegraphics[width=2.in,angle=-90]{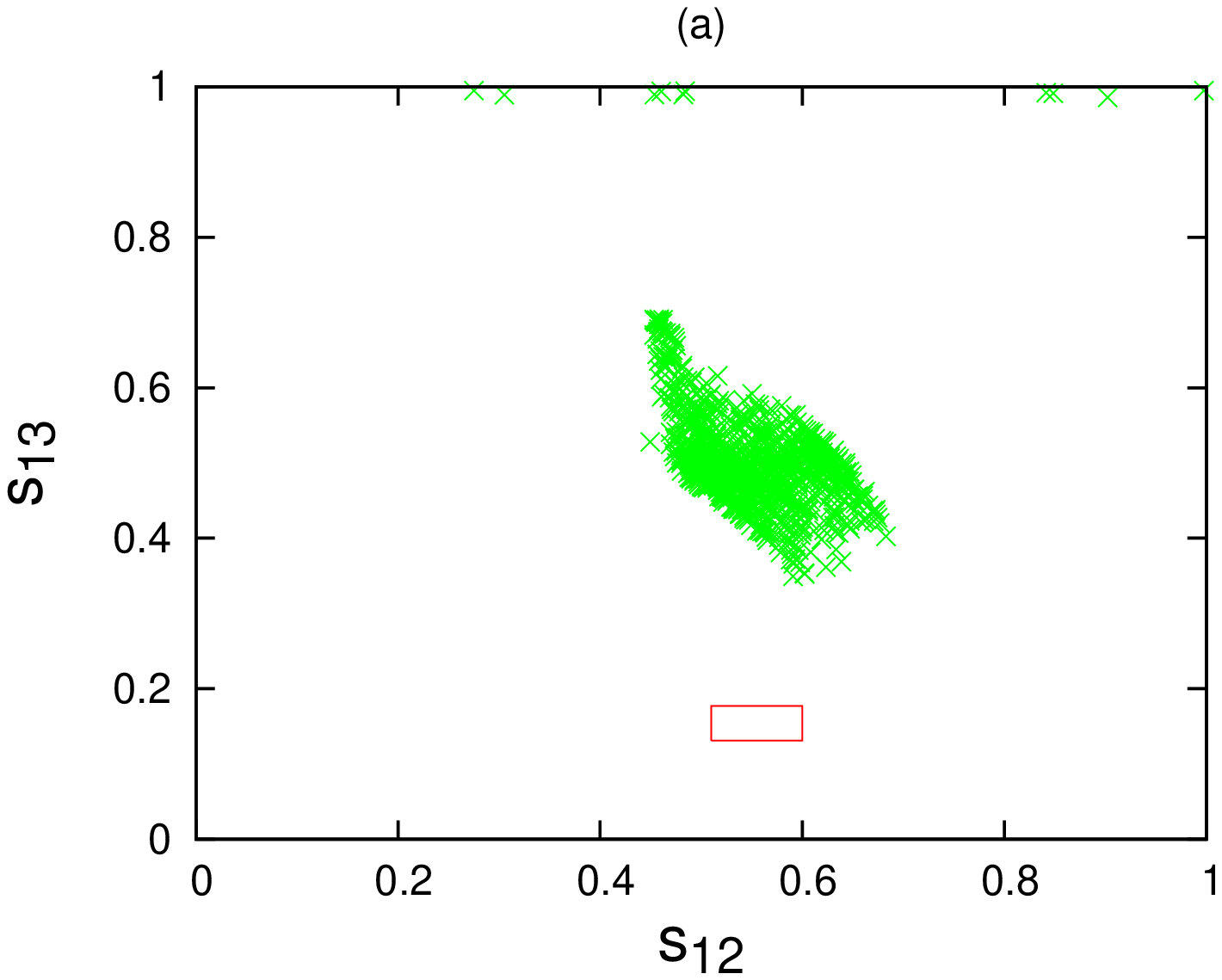}
    \end{minipage} \hspace{0.5cm}
\begin{minipage} {0.45\linewidth} \centering
\includegraphics[width=2.in,angle=-90]{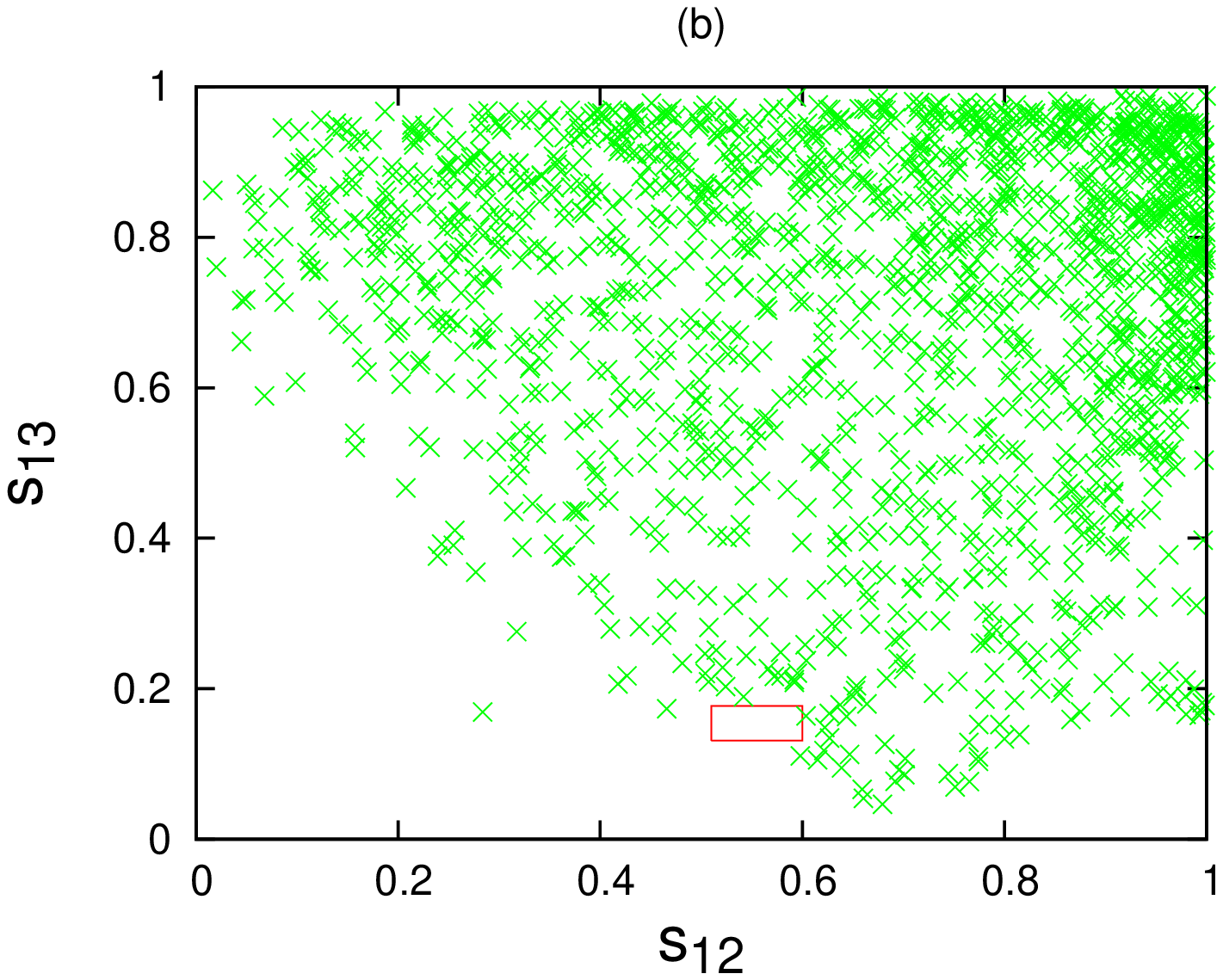}
\end{minipage}
\caption{Plots showing the parameter space for the mixing angles $s_{12}$ and $s_{13}$ when $s_{23}$ is constrained by
its  $3 \sigma$ range for Class II ansatz of texture five zero mass matrices pertaining 
to the possibilty (a)  $D_l = 0$ and $D_\nu \neq 0$ (b) $D_l \neq 0$ and $D_\nu = 0$ (inverted hierarchy).}
\label{t5cl2ih1}
\end{figure}

 We study both these possibilities in detail for all the neutrino mass orderings. Firstly, we examine
 the compatibility of matrices given in equations (\ref{cl2t51}) and (\ref{cl2t52}) with the inverted hierarchy
of neutrino masses. For this purpose, in figures \ref{t5cl2ih1}(a) and
 \ref{t5cl2ih1}(b), we present the plots showing the parameter space allowed by this ansatz for the 
 mixing angles $s_{12}$ and $s_{13}$ for the cases $D_l =0$, $D_\nu\neq 0$ and $D_l \neq 0$, $D_\nu = 0$ respectively.
 While plotting these graphs, the mixing angle $s_{23}$
has been  constrained by its $3\sigma$ experimental bound for inverted hierarchy of neutrino masses.
The rectangular regions in these plots represent the $3\sigma$ experimental ranges for
$s_{12}$ and $s_{13}$ . A general 
look at these plots reveals that inverted hierarchy seems to be viable for  
for the $D_l \neq 0$ and $D_\nu=0$ case, however it is clearly ruled out 
for the other possibilty i.e. $D_l =0$ and $D_\nu\neq 0$.

\begin{figure}
\begin{tabular}{cc}
  \includegraphics[width=0.2\paperwidth,height=0.2\paperheight,angle=-90]{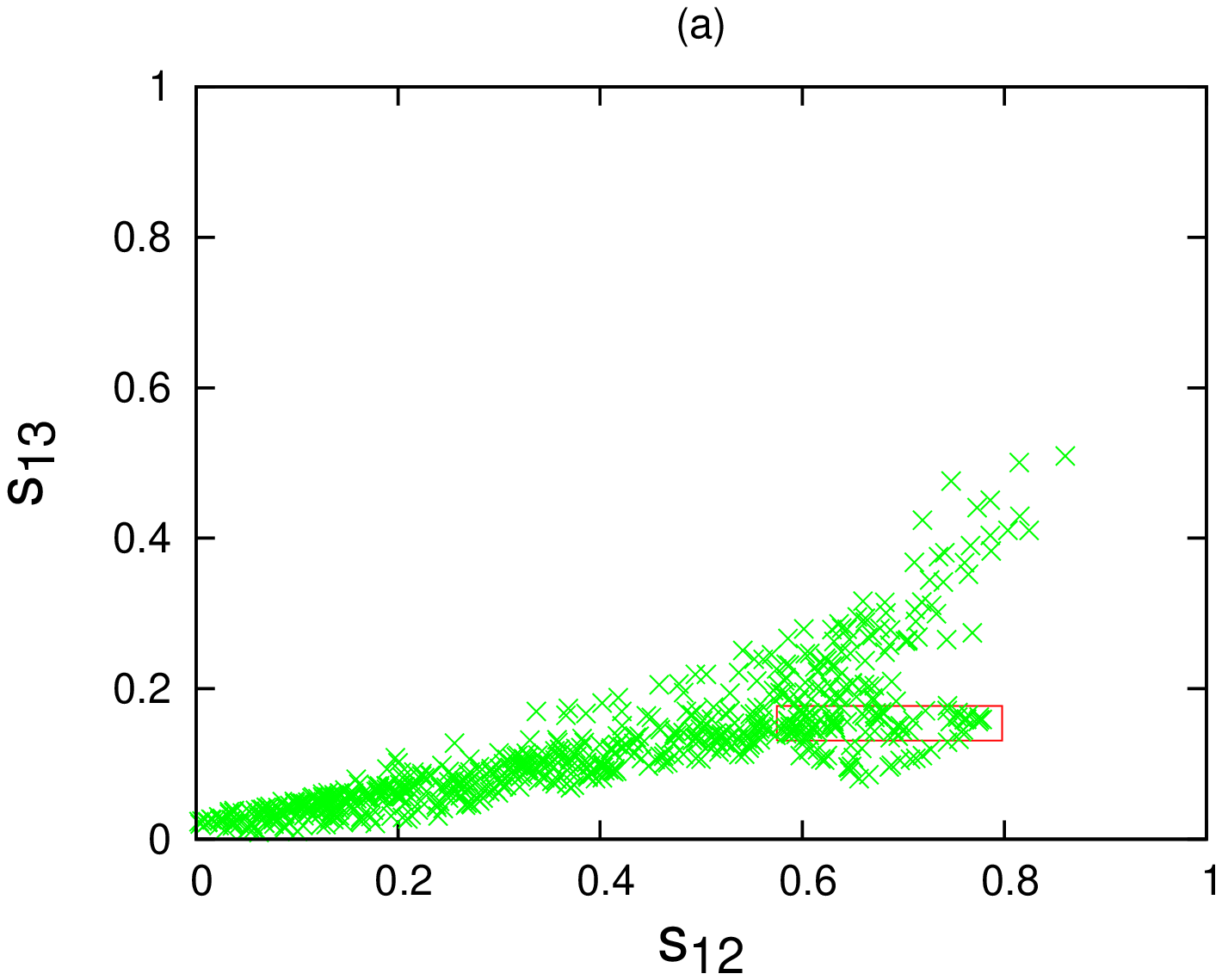}
  \includegraphics[width=0.2\paperwidth,height=0.2\paperheight,angle=-90]{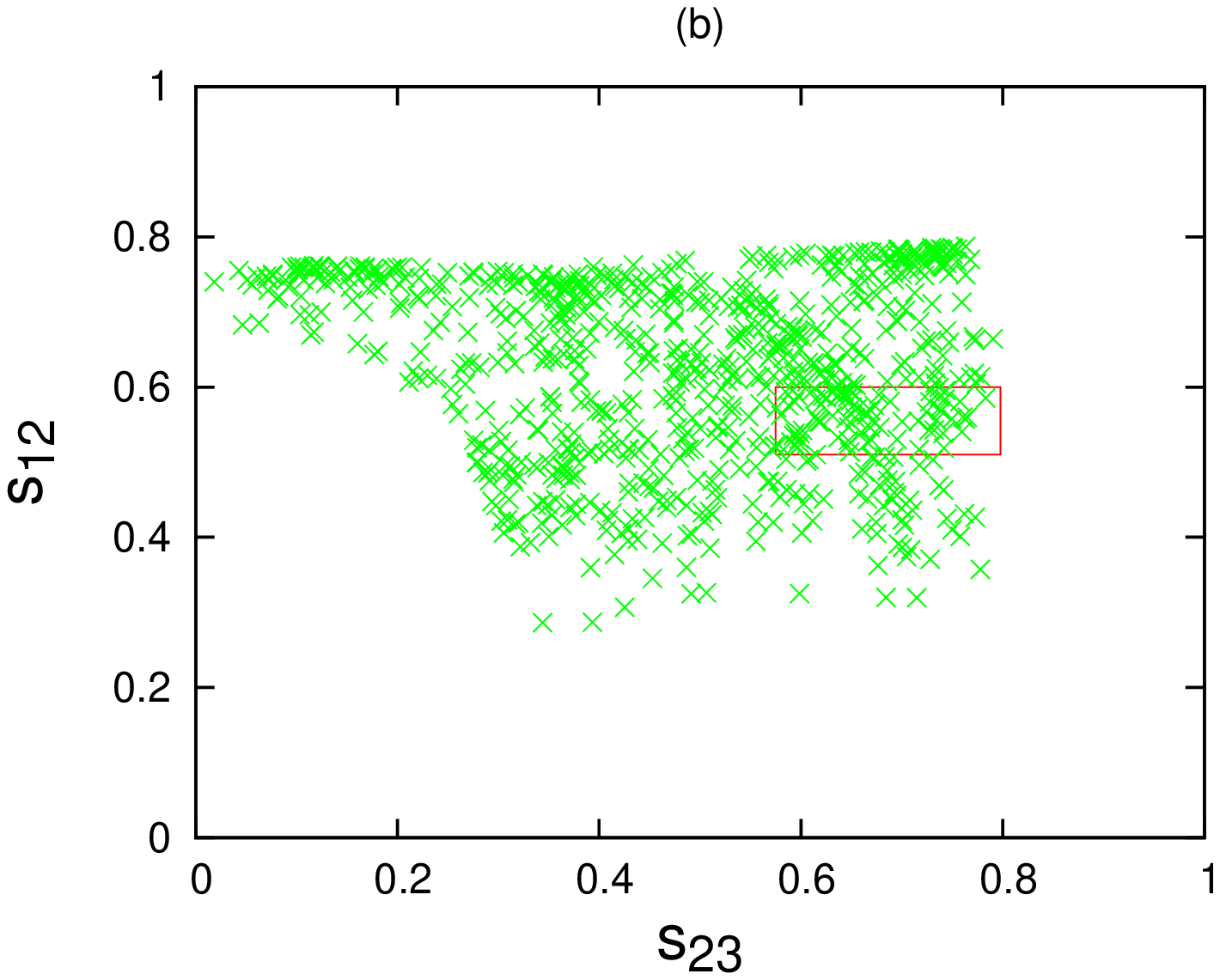}
  \includegraphics[width=0.2\paperwidth,height=0.2\paperheight,angle=-90]{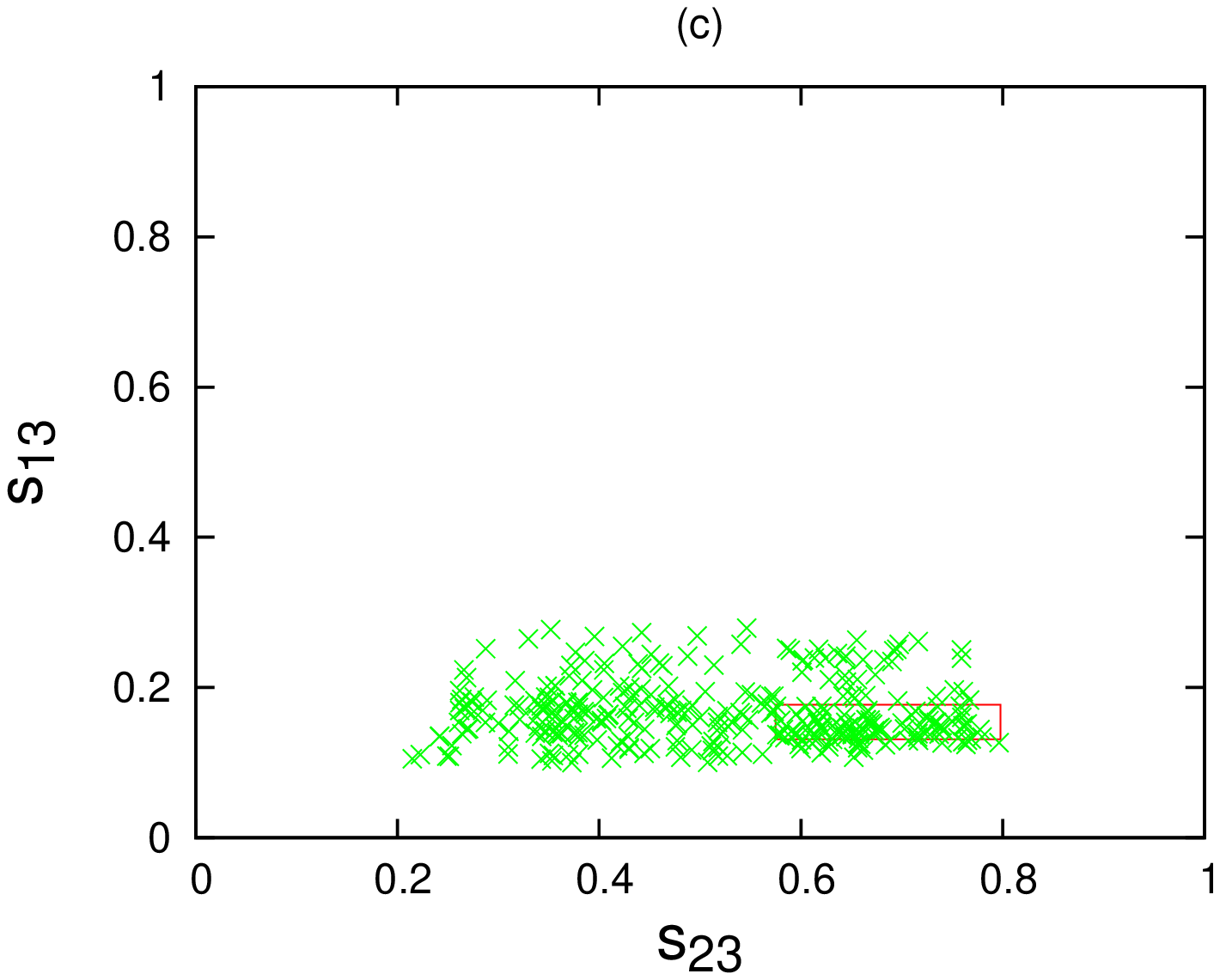}
\end{tabular}
\caption{Plots showing the parameter space for any two mixing angles when the third angle is constrained by
its  $1 \sigma$ range in the $D_l =0$ and $D_\nu\neq 0$ scenario 
for Class II ansatz of texture five zero mass matrices (normal hierarchy).}
\label{t5cl2nh1}
\end{figure}

\begin{figure}
\begin{tabular}{cc}
  \includegraphics[width=0.2\paperwidth,height=0.2\paperheight,angle=-90]{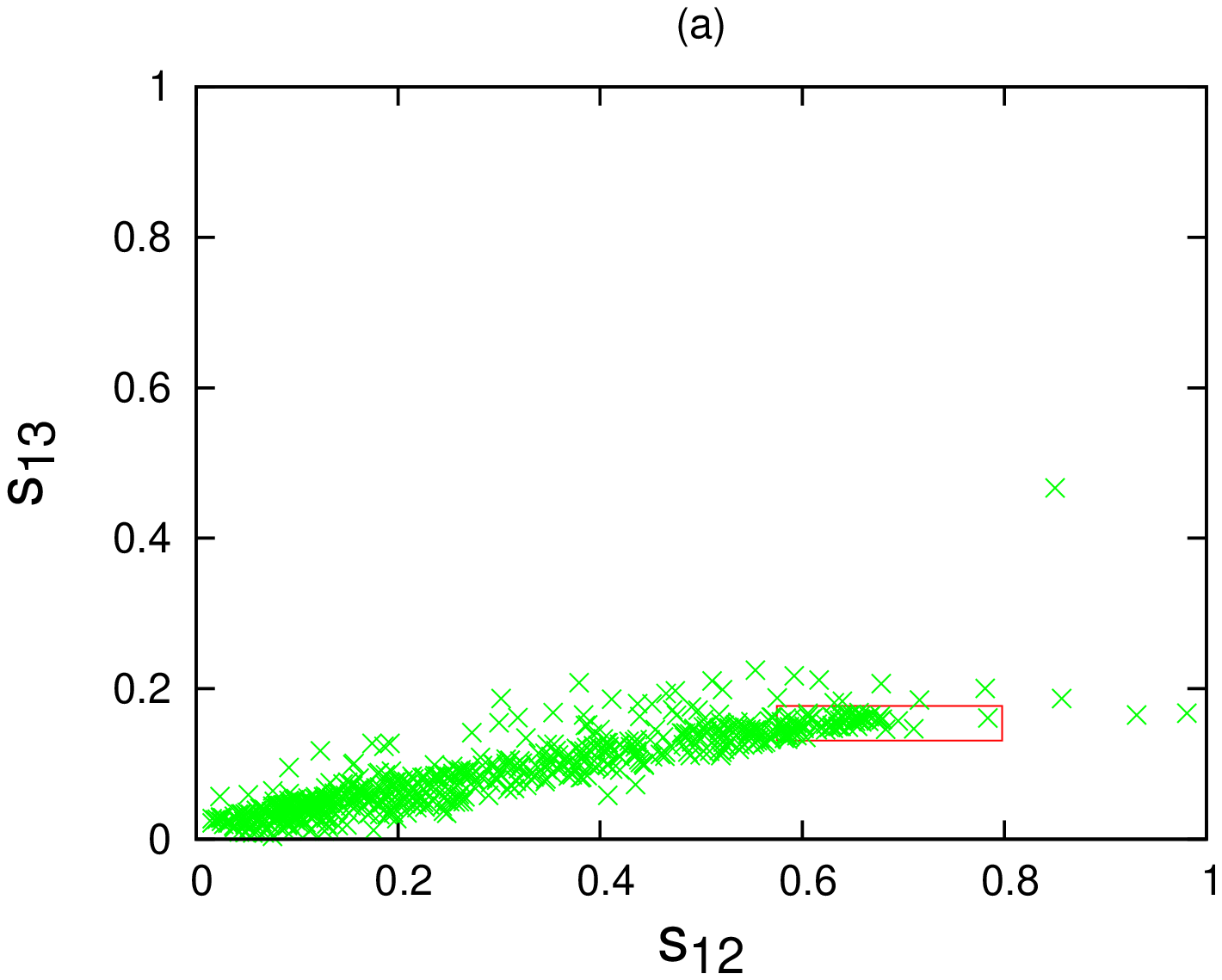}
  \includegraphics[width=0.2\paperwidth,height=0.2\paperheight,angle=-90]{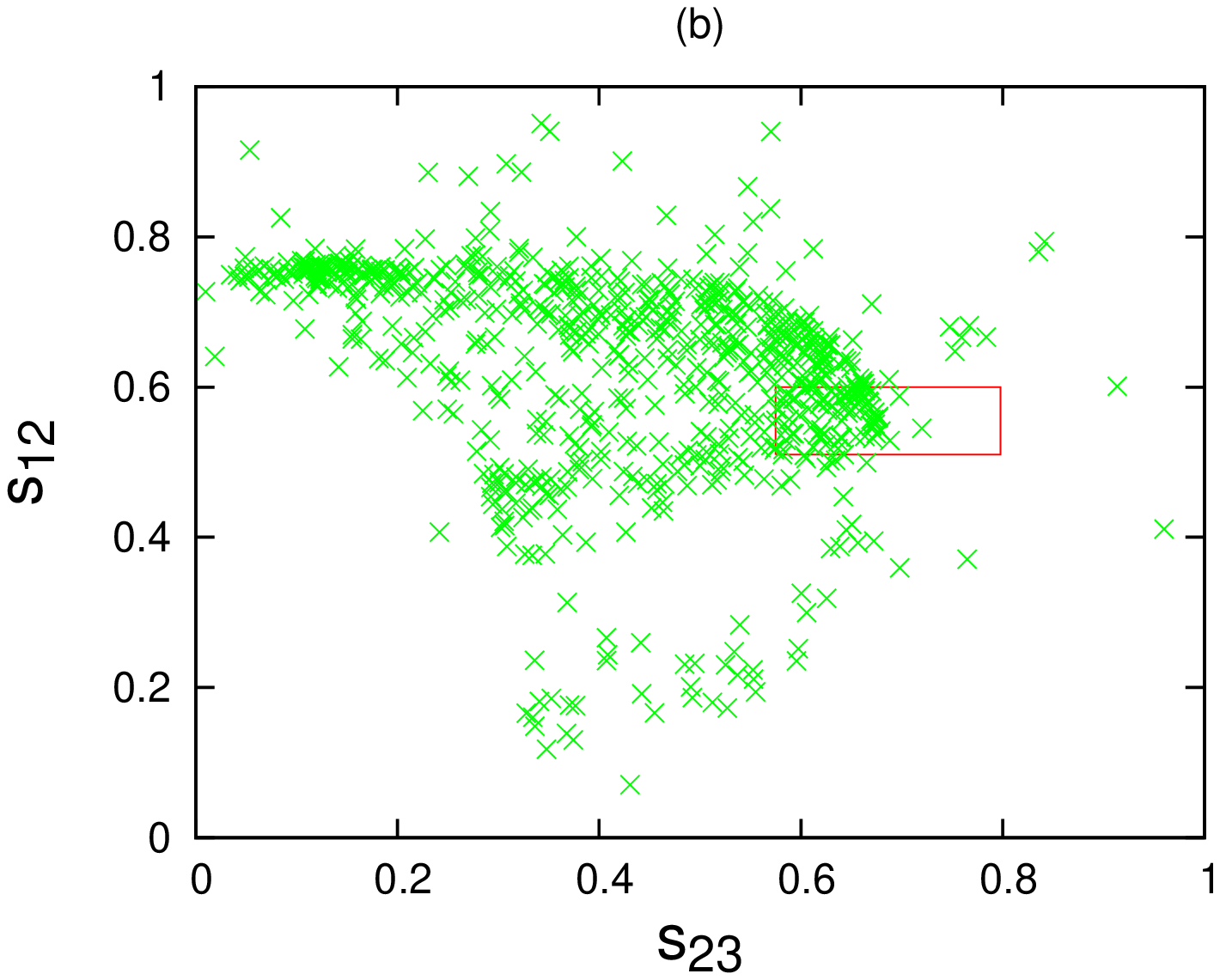}
  \includegraphics[width=0.2\paperwidth,height=0.2\paperheight,angle=-90]{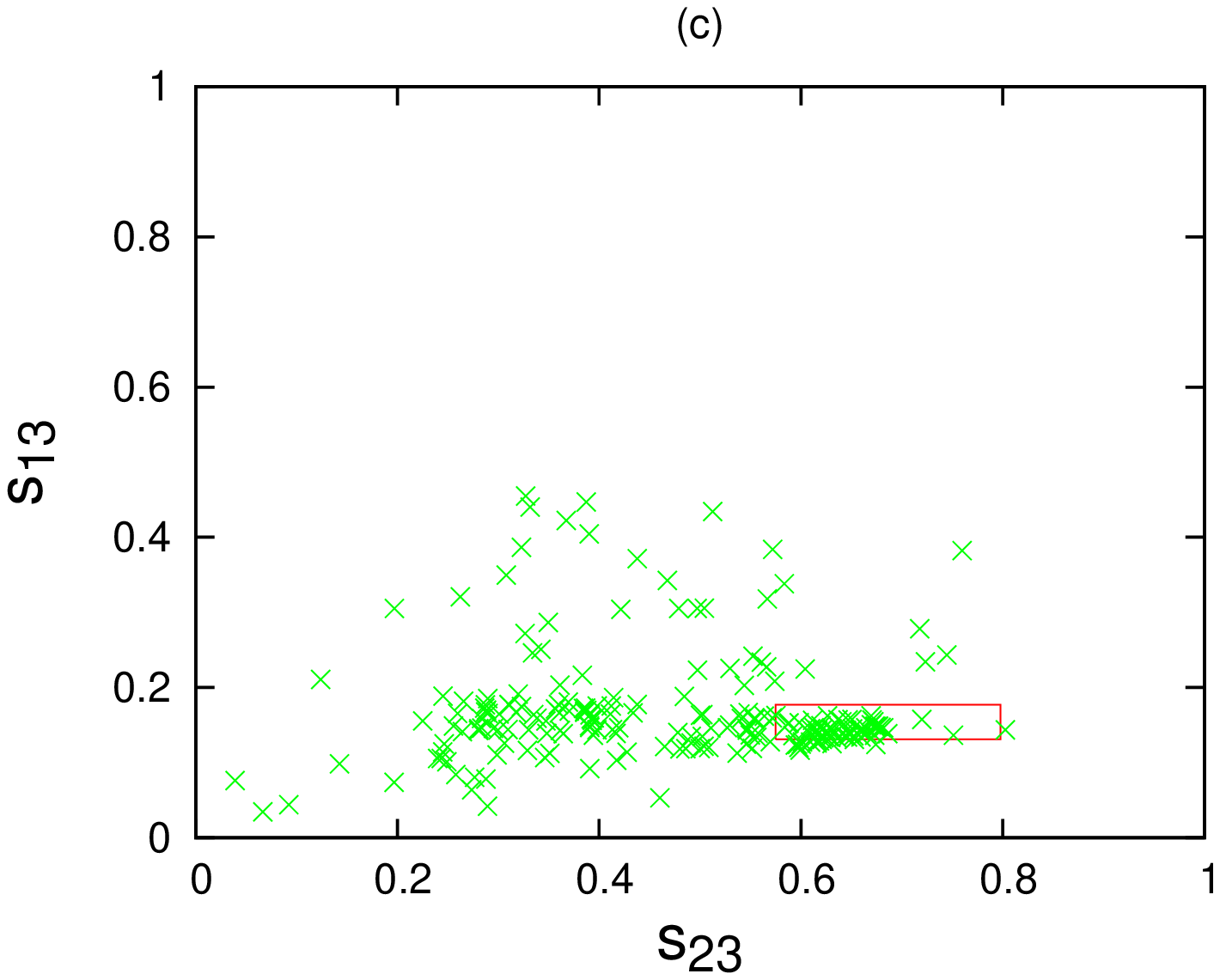}
\end{tabular}
\caption{Plots showing the parameter space for any two mixing angles when the third angle is constrained by
its  $1 \sigma$ range in $D_l \neq 0$ and $D_\nu = 0$ scenario for Class II ansatz of texture five zero mass
matrices (normal hierarchy).}
\label{t5cl2nh2}
\end{figure}

After examining the viability of  inverted hierarchy for both the cases 
of texture five zero mass matrices pertaining to class II, we 
now proceed to examine their compatibility with the normal hierarchy scenario. To this end, in figures (\ref{t5cl2nh1}) and
 (\ref{t5cl2nh2}), we present the plots showing the parameter space allowed by this ansatz for any two mixing angles wherein
the third one  is constrained by its $1\sigma$ experimental bound for normal hierarchy of neutrino masses.
The rectangular regions in these plots represent the $3\sigma$ ranges for the two mixing angles being considered. Normal hierarchy
seems to be viable for both the cases as can be seen from the significant overlap with the experimentally allowed region.

\par
\begin{figure}
\begin{minipage} {0.45\linewidth} \centering
\includegraphics[width=2.0in,angle=-90]{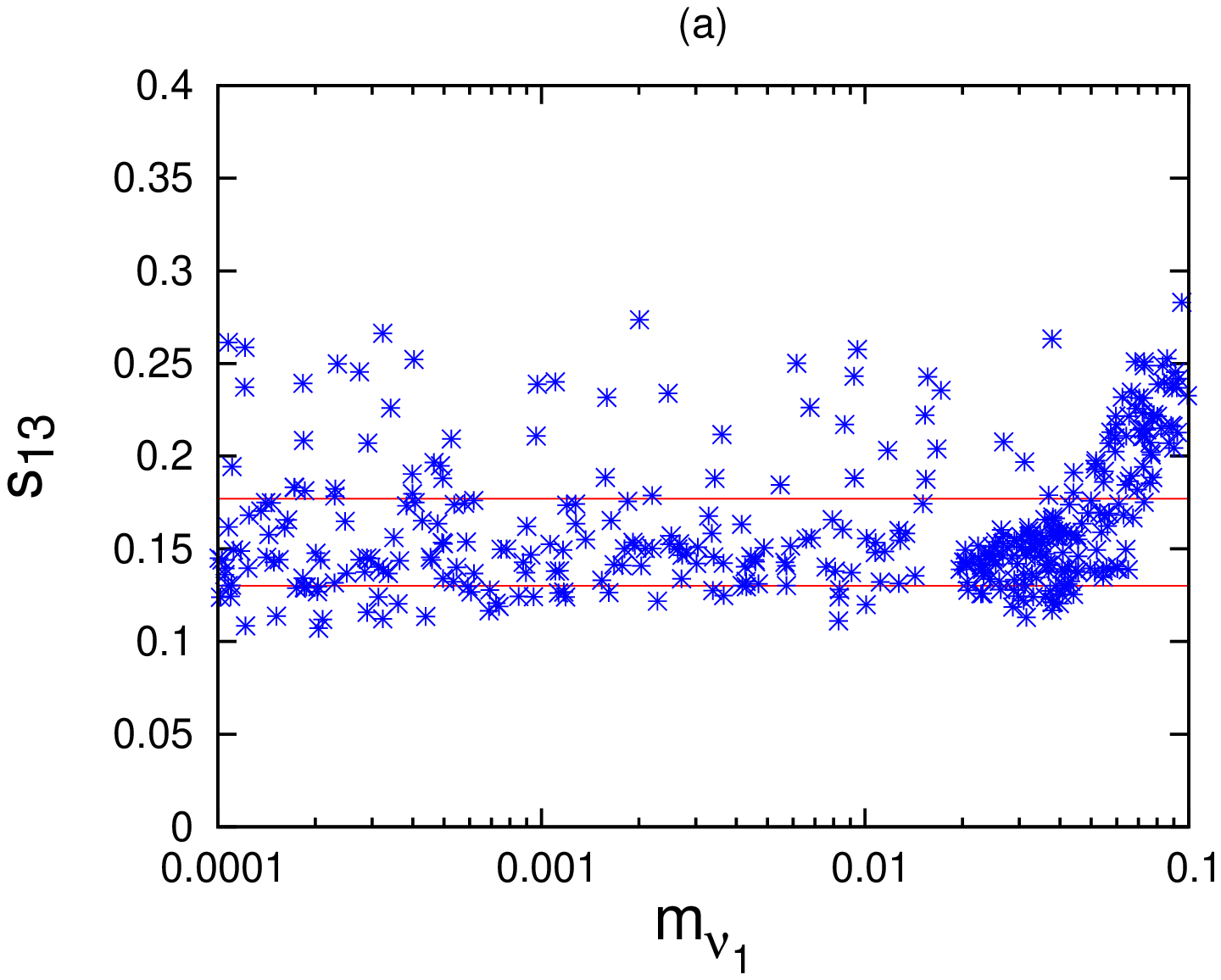}
 \end{minipage}
 \hspace{1.2cm}
 \begin{minipage} {0.45\linewidth} \centering
\includegraphics[width=2.0in,angle=-90]{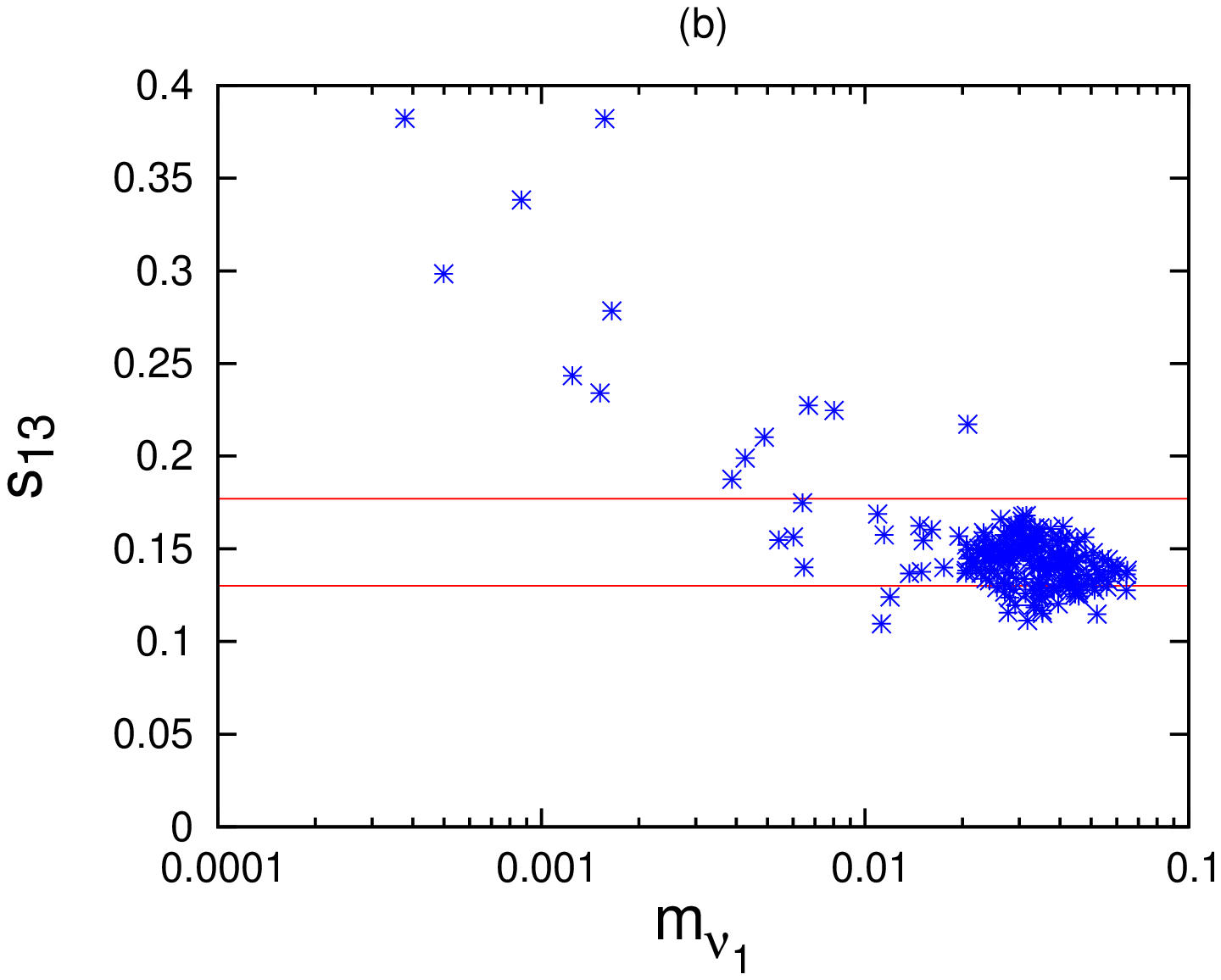}
 \end{minipage}
\caption{Plots showing the dependence of mixing angle $s_{13}$ on the lightest neutrino mass when the
other two angles are constrained by their $3 \sigma$ ranges  for Class II ansatz of texture five zero mass matrices pertaining 
to the possibilty (a)  $D_l = 0$ and $D_\nu \neq 0$ (b) $D_l \neq 0$ and $D_\nu = 0$ (normal hierarchy).}
\label{t5cl2nh3}
\end{figure}

\begin{figure}
\begin{minipage} {0.45\linewidth} \centering
\includegraphics[width=2.0in,angle=-90]{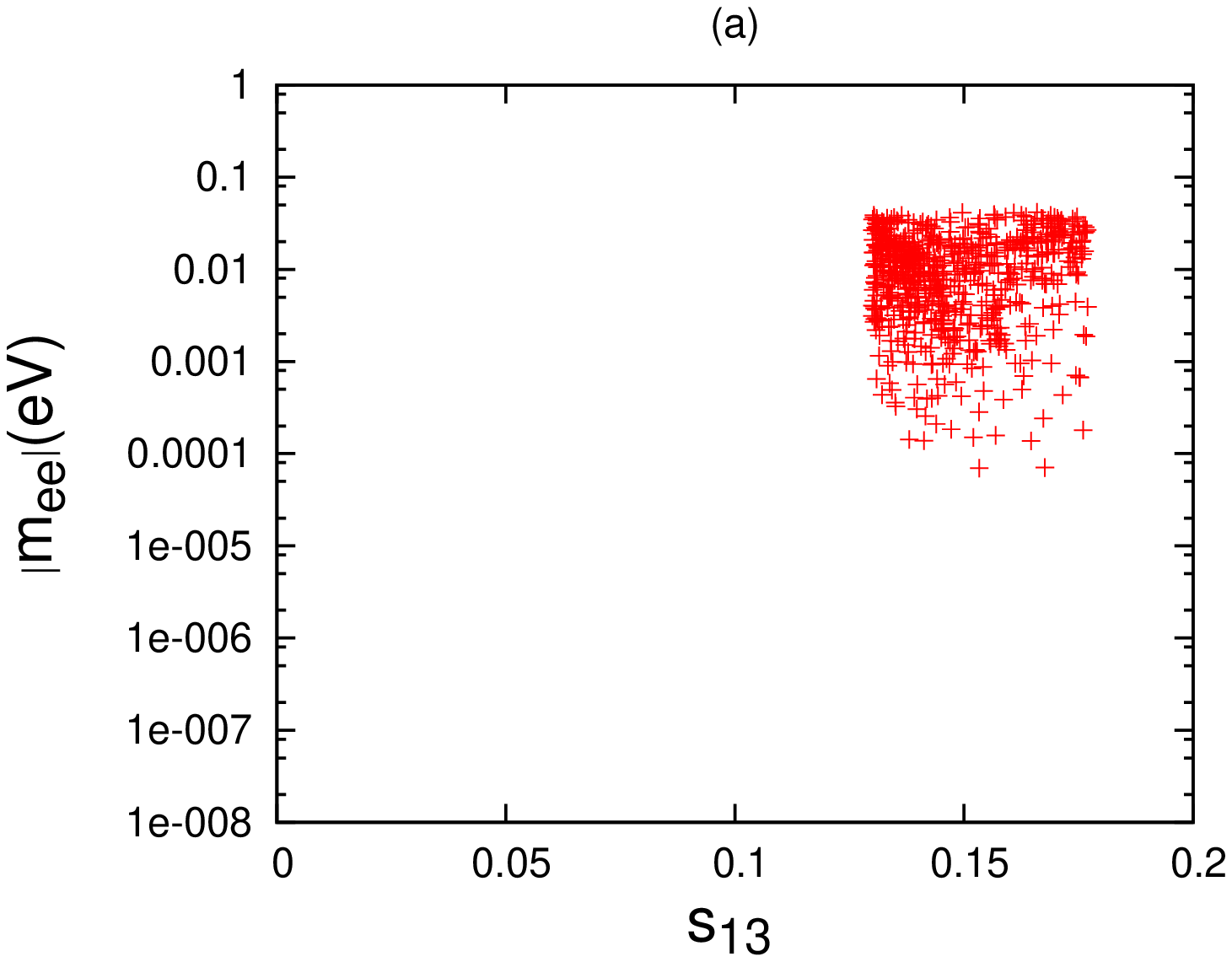}
 \end{minipage}
 \hspace{1.2cm}
 \begin{minipage} {0.45\linewidth} \centering
\includegraphics[width=2.0in,angle=-90]{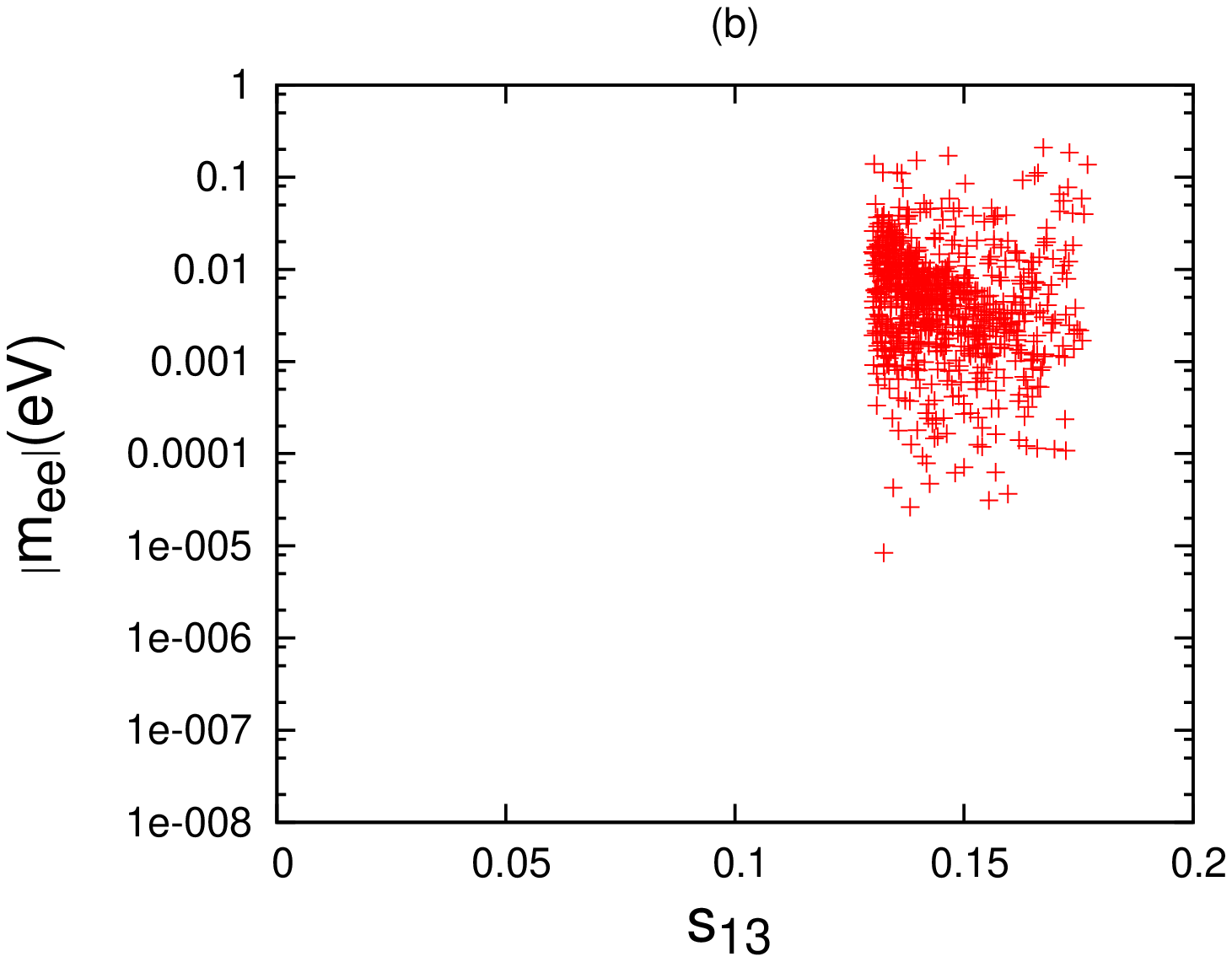}
 \end{minipage}
\caption{Plots showing the variation of the effective Majorana mass measured in 
neutrinoless double beta decay with $s_{13}$ for Class II ansatz of texture five zero mass matrices pertaining 
to the possibilty (a)  $D_l = 0$ and $D_\nu \neq 0$ (b) $D_l \neq 0$ and $D_\nu = 0$ (normal hierarchy)}
\label{t5cl2nh4}
\end{figure}

 Further, in figure (\ref{t5cl2nh3}) we present the graphs showing the variation
of the lightest neutrino mass with the mixing angle $s_{13}$ for both the cases $D_l =0$ and $D_\nu\neq 0$
as well as $D_l \neq 0$ and $D_\nu = 0$ pertaining to normal hierarchy of neutrino masses. While plotting
these graphs, the other two mixing angles have been constrained by their $3\sigma$ experimental bounds. The
parallel lines in each plot
show the $3\sigma$ range of the mixing angle $s_{13}$. Interestingly, taking a careful look at these graphs one finds that for 
the $D_l=0$ and $D_\nu \neq 0$ case the lightest neutrino mass is largely unrestricted, whereas for the $D_l\neq 0$ and $D_\nu = 0$
case a lower bound $\approx 0.005 eV$ can be obtained.
As a next step, in figures \ref{t5cl2nh4}(a) and \ref{t5cl2nh4}(b)  we study the variation of the 
effective Majorana mass with the $s_{13}$ for the cases  $D_l = 0$, $D_\nu \neq 0$ and $D_l \neq 0$ and $D_\nu = 0$
respectively texture five zero mass matrices in class II pertaining to normal hierarchy 
of neutrino masses. While plotting these 
figures, the mixing angle $s_{13}$ has been constrained by its $3\sigma$ experimental bound. A careful look 
at these figures reveals that the $3\sigma$ range for $s_{13}$ provides a lower bound $ \approx 10^{-4} eV$ and $ \approx 10^{-5} eV$
for $|m_{ee}|$ pertaining to the $D_l=0$, $D_\nu \neq 0$ and $D_l\neq0$, $D_\nu = 0$ cases respectively.

\subsubsection{Class III ansatz}

\begin{figure}
\begin{minipage} {0.45\linewidth} \centering
\includegraphics[width=2.0in,angle=-90]{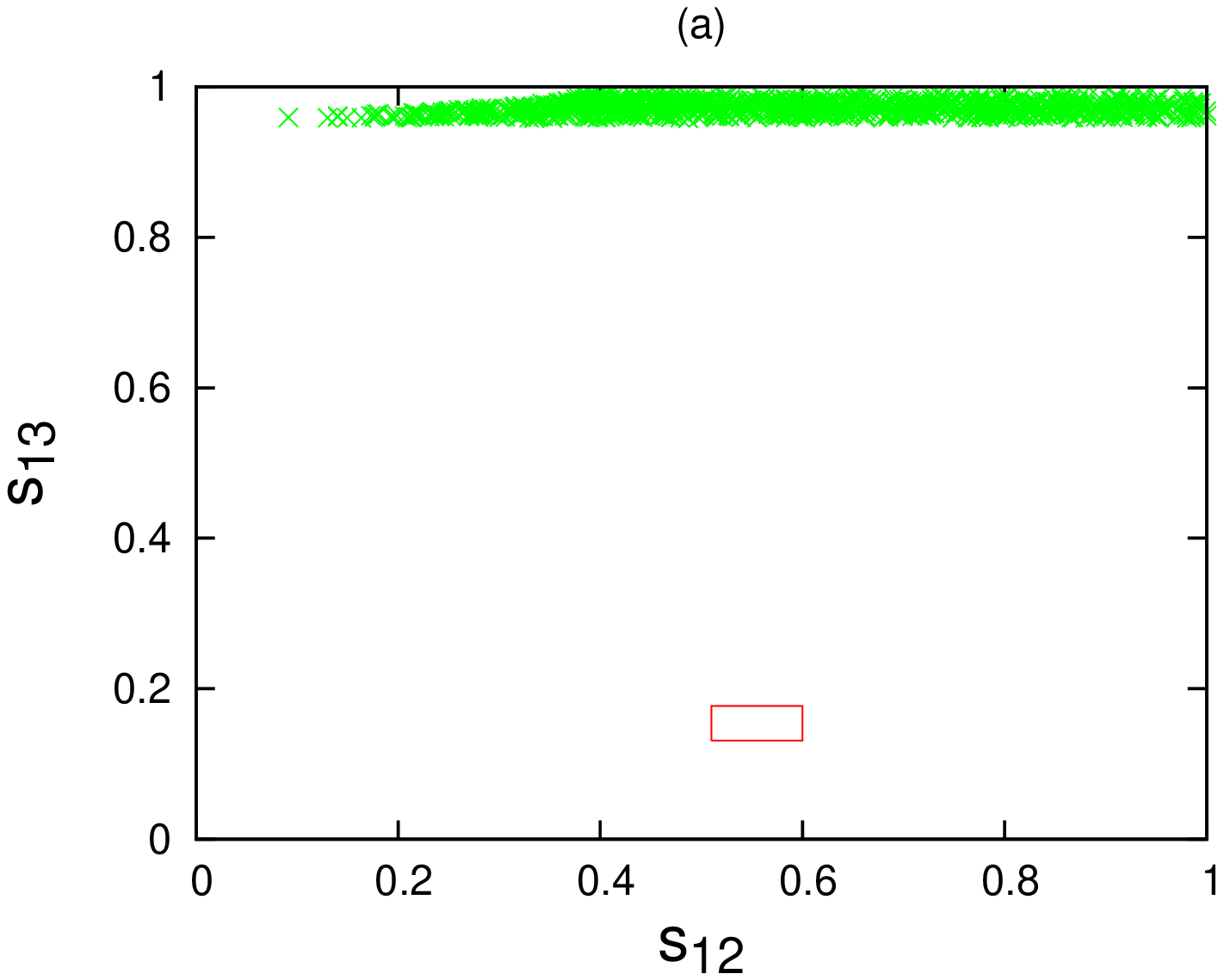}
 \end{minipage}
 \hspace{1.2cm}
 \begin{minipage} {0.45\linewidth} \centering
\includegraphics[width=2.0in,angle=-90]{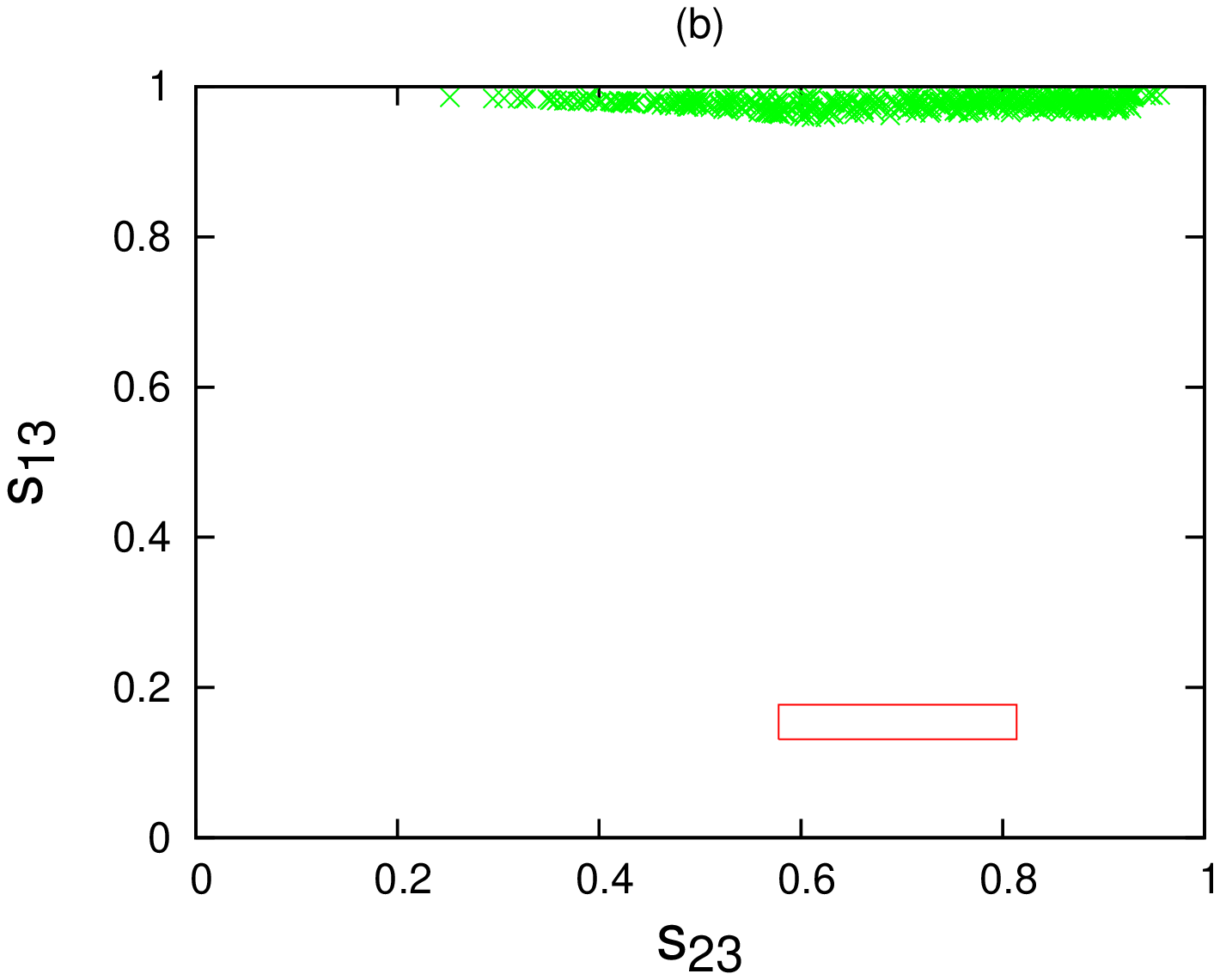}
 \end{minipage}
\caption{Plots showing the parameter space for (a) $s_{12}$ and $s_{13}$ (b)$s_{23}$ and $s_{13}$
 in the $D_l =0$ and $D_\nu\neq 0$ scenario for Class III ansatz of texture 
five zero mass matrices (inverted hierarchy).}
\label{t5cl3ih1}
\end{figure}

\begin{figure}
\begin{tabular}{cc}
  \includegraphics[width=0.2\paperwidth,height=0.2\paperheight,angle=-90]{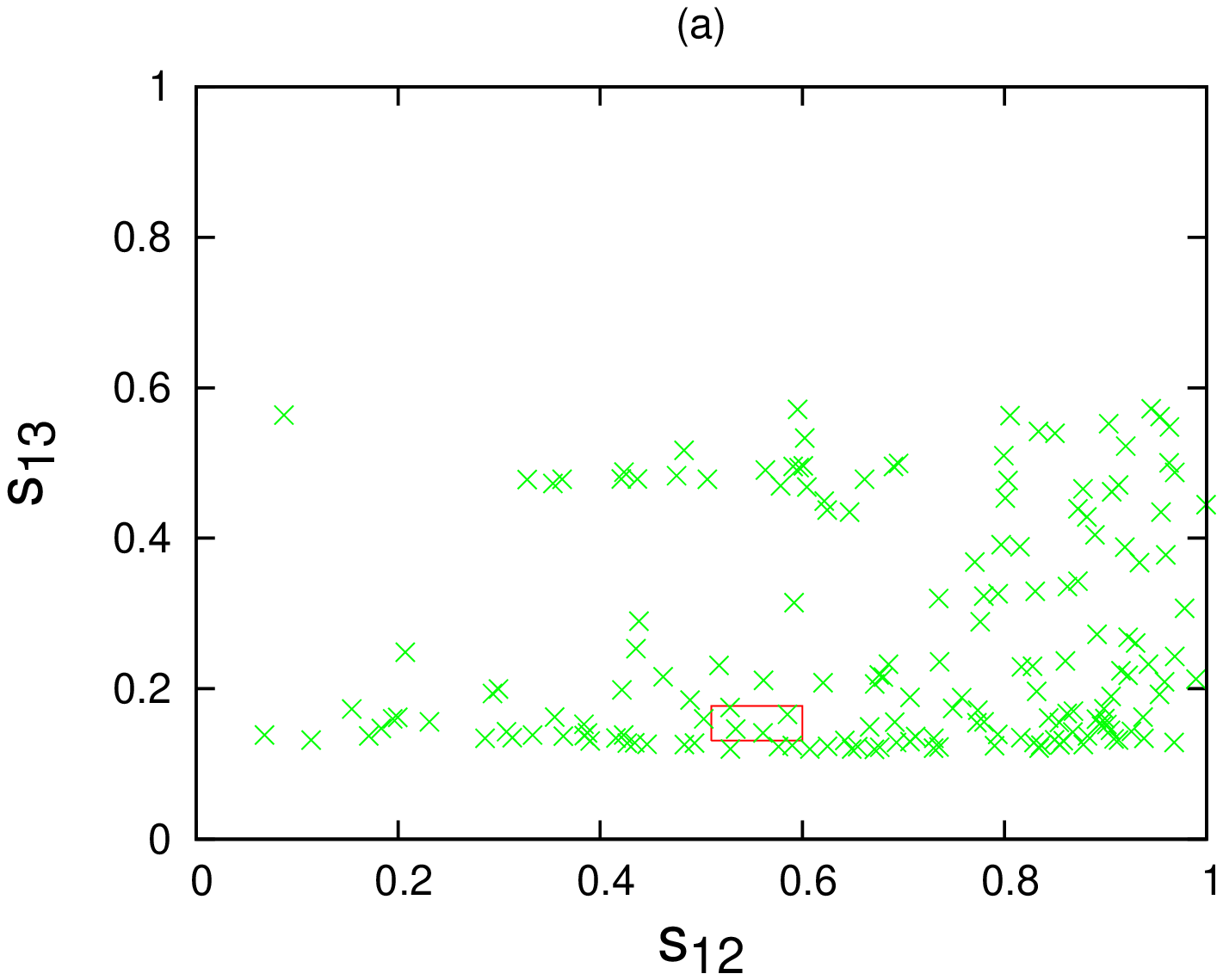}
  \includegraphics[width=0.2\paperwidth,height=0.2\paperheight,angle=-90]{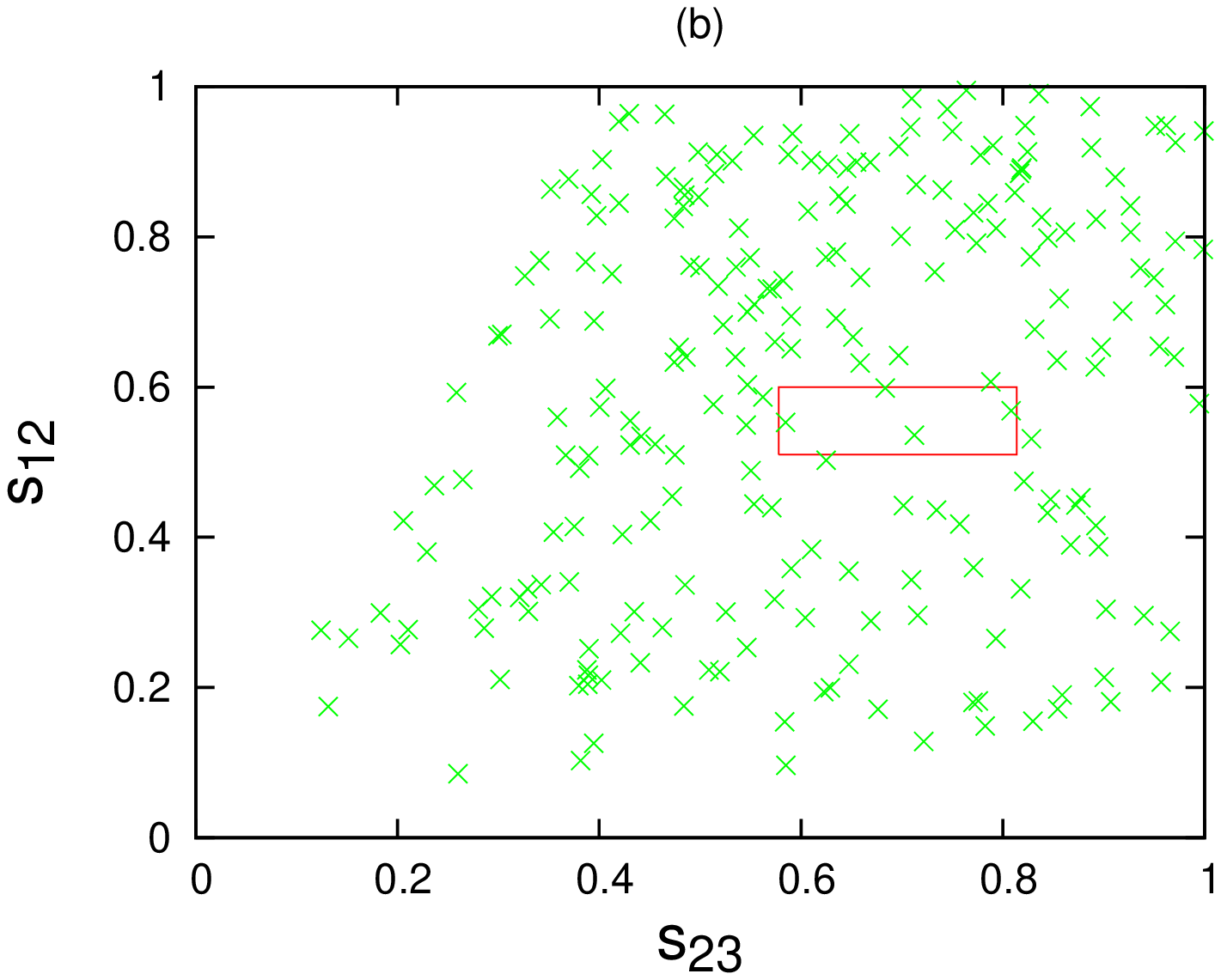}
  \includegraphics[width=0.2\paperwidth,height=0.2\paperheight,angle=-90]{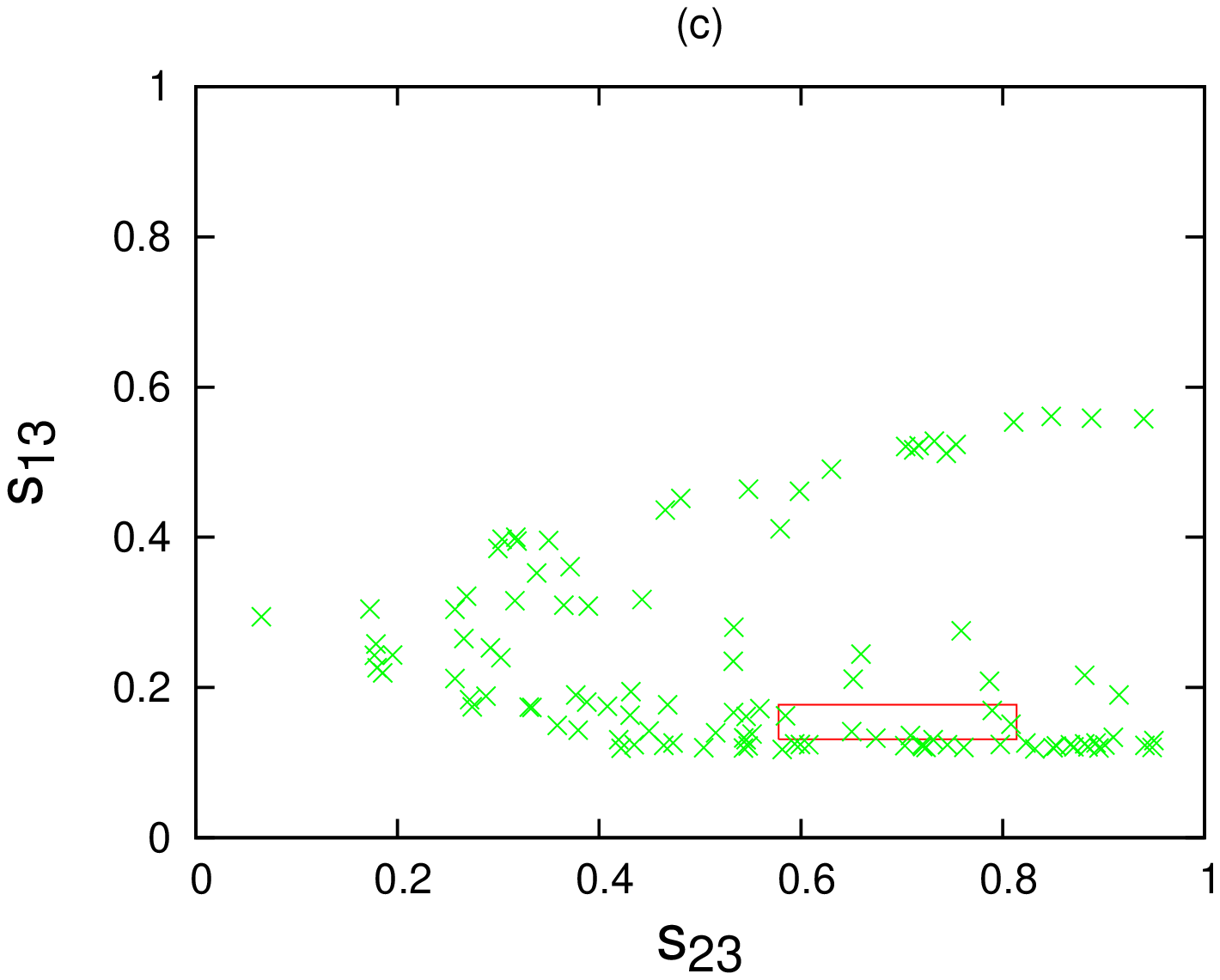}
\end{tabular}
\caption{Plots showing the parameter space for any two mixing angles when the third angle is constrained by
its  $3 \sigma$ range in the $D_l \neq 0$ and $D_\nu = 0$ scenario 
for Class III ansatz of texture five zero mass matrices (inverted hierarchy).}
\label{t5cl3ih2}
\end{figure}

The two possibilities for texture five zero lepton mass matrices for this class can be given as,

\be
 M_{l}=\left( \ba{ccc}
0 & A _{l}e^{i\alpha_l} & B_{l}     \\
A_{l}e^{-i\alpha_l} & 0 &  0     \\
 B_{l} & 0 &  E_{l} \ea \right), \qquad
M_{\nu}=\left( \ba{ccc}
0 & A _{\nu}e^{i\alpha_\nu} & B_{\nu}     \\
A_{\nu}e^{-i\alpha_\nu} & 0 &   D_{\nu}e^{i\beta_\nu}      \\
 B_{\nu}& D_{\nu}e^{-i\beta_\nu}  &  E_{\nu} \ea \right),
\label{cl3t51}
\ee
or \be
 M_{l}=\left( \ba{ccc}
0 & A _{l}e^{i\alpha_l} & B_{l}    \\
A_{l}e^{-i\alpha_l} & 0 &  D_{l}e^{i\beta_l}    \\
 B_{l} & D_{l}e^{-i\beta_l}  &  E_{l} \ea \right), \qquad
M_{\nu}=\left( \ba{ccc}
0 & A _{\nu}e^{i\alpha_\nu} & B_{\nu}     \\
A_{\nu}e^{-i\alpha_\nu} & 0 & 0        \\
 B_{\nu} & 0 &  E_{\nu} \ea \right),
\label{cl3t52}
\ee
We study both these possibilities in detail for all the neutrino mass orderings. Firstly, we examine
 the compatibility of matrices given in equations (\ref{cl3t51}) and (\ref{cl3t52}) with the inverted hierarchy
of neutrino masses. 
For this purpose, in figures (\ref{t5cl3ih1}) and
 (\ref{t5cl3ih2}), we present the plots showing the parameter space allowed by this ansatz, corresponding to the 
 cases $D_l =0$, $D_\nu\neq 0$ and $D_l \neq 0$, $D_\nu= 0$ respectively, for any two mixing angles wherein
the third one  is constrained by its $3\sigma$ experimental bound for inverted hierarchy of neutrino masses.
The rectangular regions in these plots represent the $3\sigma$ ranges for the two mixing angles being considered.
Interestingly, one finds that for the case $D_l =0$ and $D_\nu\neq 0$ of texture five zero lepton mass matrices
inverted hierarchy is ruled out, whereas for the case $D_l \neq 0$ and $D_\nu=0$ of texture five zero lepton mass matrices
inverted hierarchy scenario seems to be viable.

\par
\begin{figure}
\begin{minipage} {0.45\linewidth} \centering
\includegraphics[width=2.0in,angle=-90]{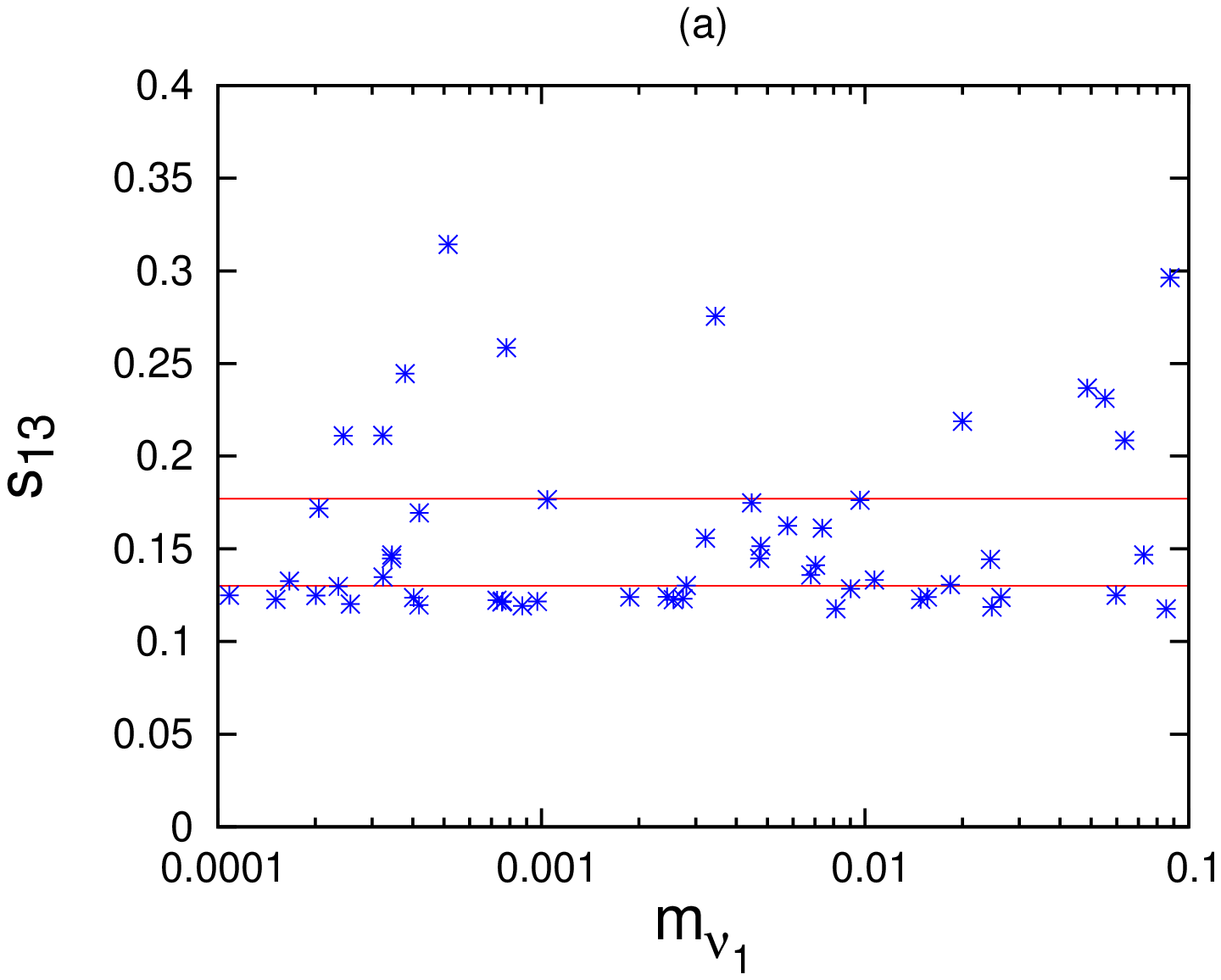}
 \end{minipage}
 \hspace{1.2cm}
 \begin{minipage} {0.45\linewidth} \centering
\includegraphics[width=2.0in,angle=-90]{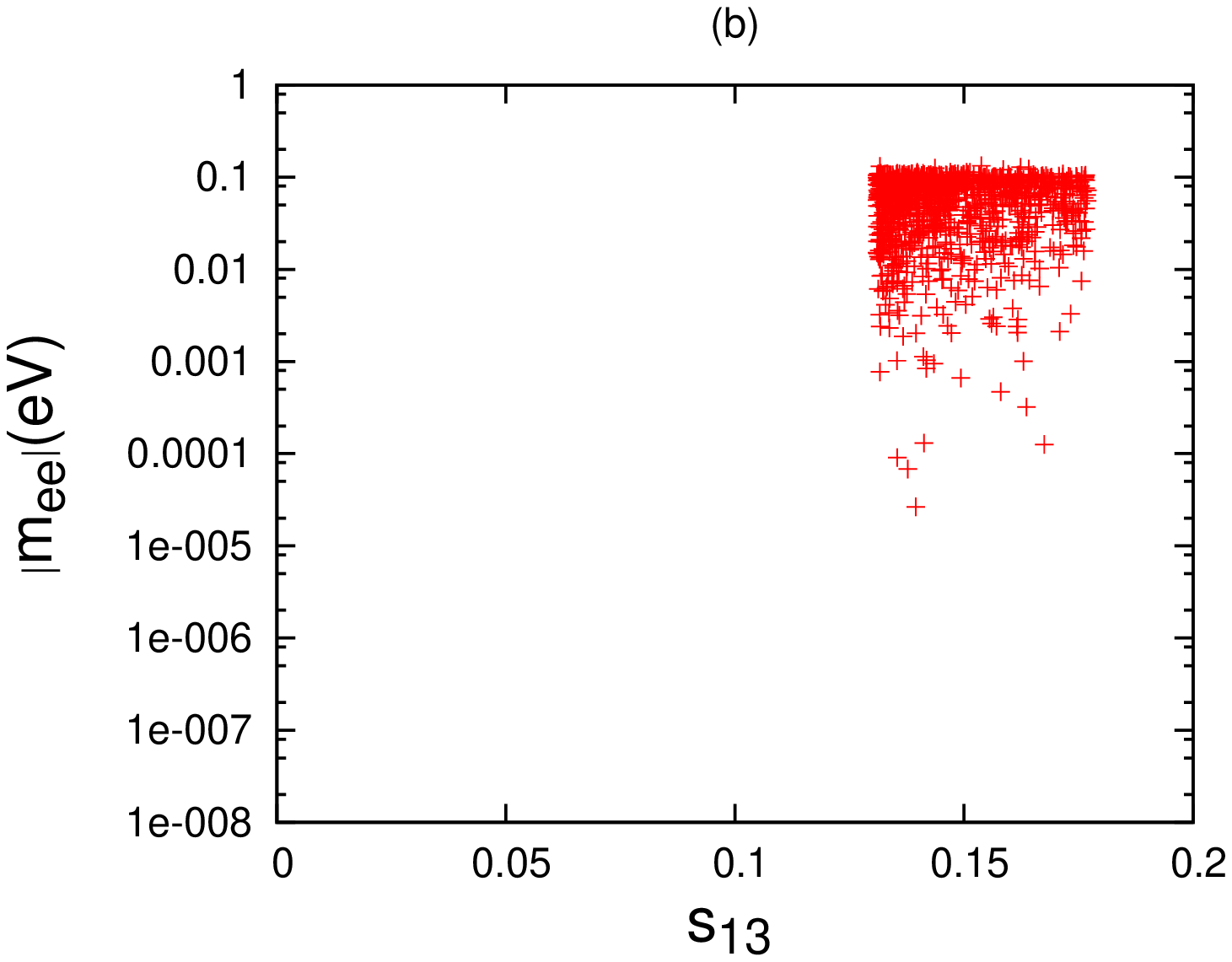}
 \end{minipage}
\caption{Plots showing the dependence of (a) the lightest neutrino mass (b) effective Majorana mass on the 
mixing angle $s_{13}$ in the $D_l\neq 0$ and $D_\nu = 0$ scenario for Class III ansatz of texture five
zero  mass matrices (inverted hierarchy).}
\label{t5cl3ih3}
\end{figure}

 For the $D_l \neq 0$ and $D_\nu = 0$ case of lepton mass matrices, wherein inverted hierarchy is shown 
to be viable, in figures \ref{t5cl3ih3}(a) and \ref{t5cl3ih3}(b) we present the plots showing the 
 dependence of the lightest neutrino mass and the effective Majorana mass respectively on the leptonic mixing 
 angle $s_{13}$. Interestingly, one finds that the lightest neutrino mass is unrestricted for this structure, whereas  
a lower bound $\approx 10^{-4}eV$ can be obtained 
for the effective Majorana mass $|m_{ee}|$.

\begin{figure}
\begin{tabular}{cc}
  \includegraphics[width=0.2\paperwidth,height=0.2\paperheight,angle=-90]{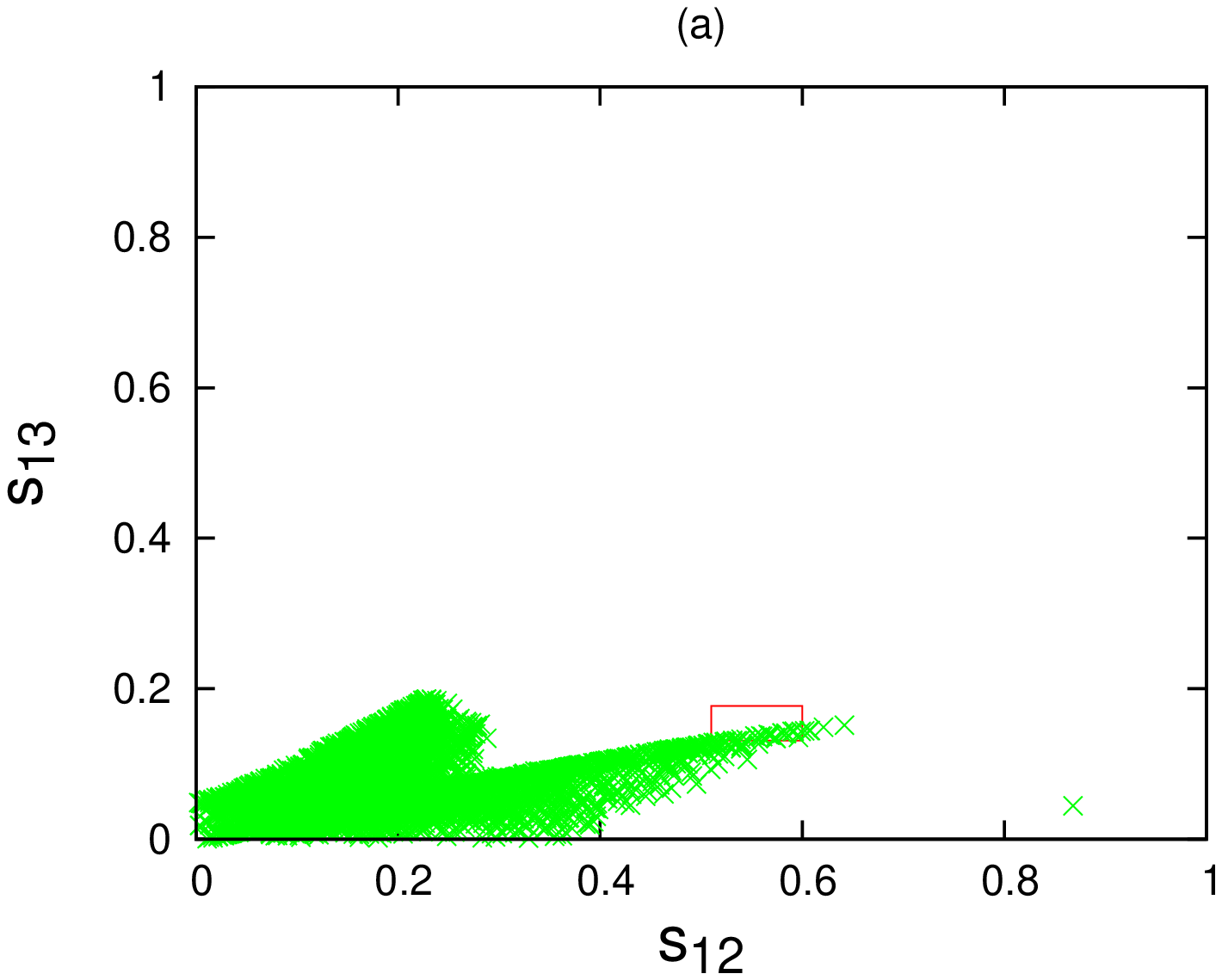}
  \includegraphics[width=0.2\paperwidth,height=0.2\paperheight,angle=-90]{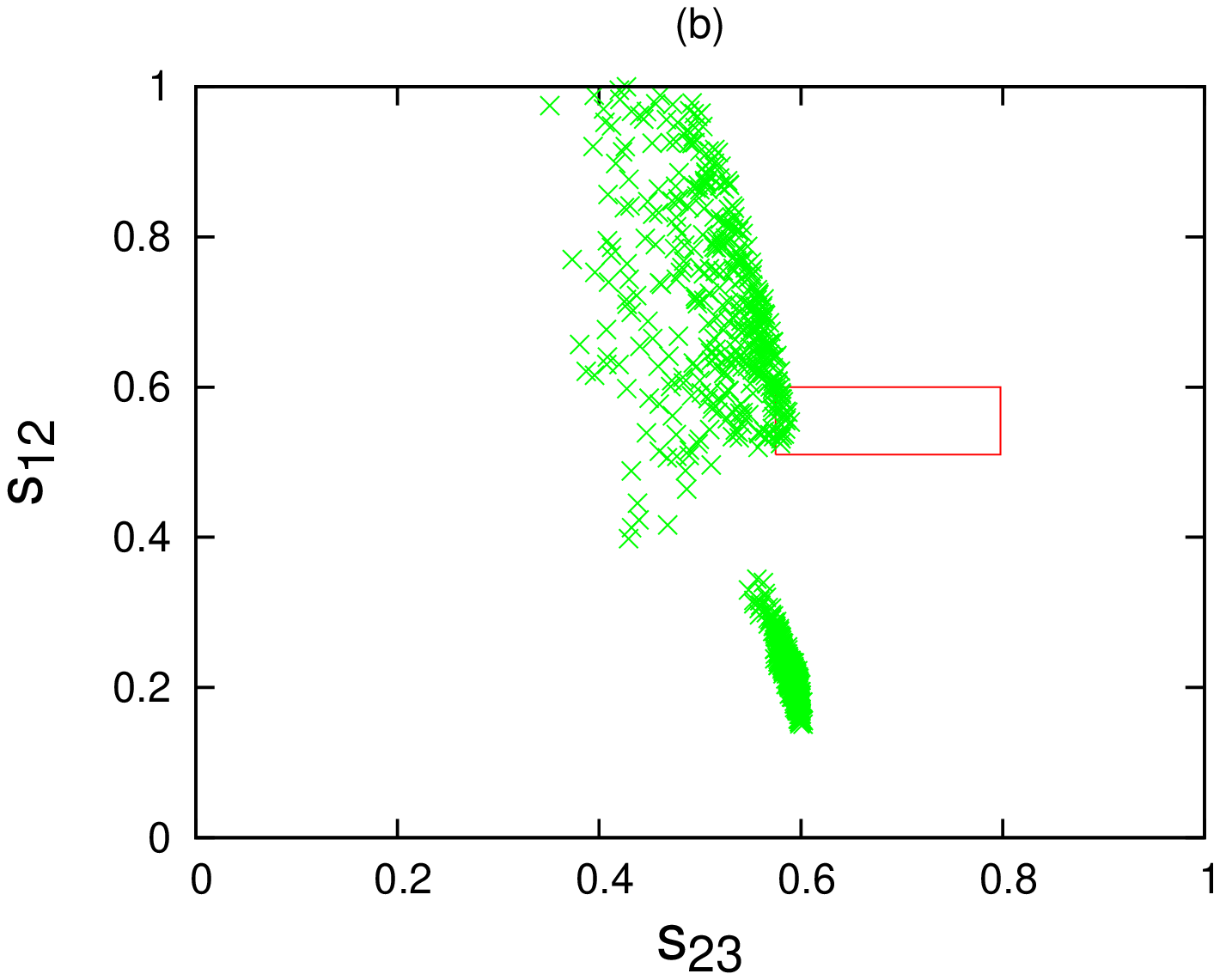}
  \includegraphics[width=0.2\paperwidth,height=0.2\paperheight,angle=-90]{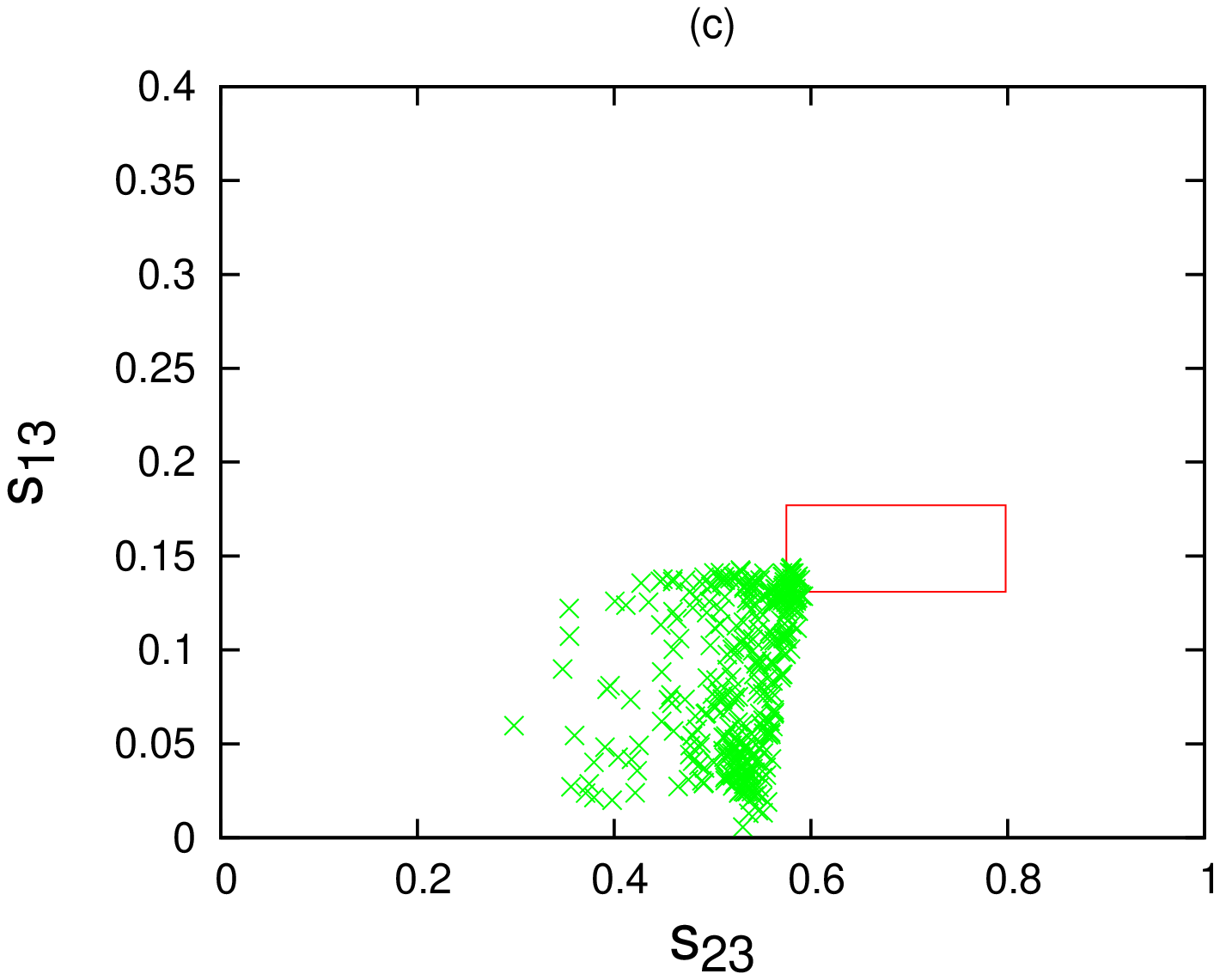}
\end{tabular}
\caption{Plots showing the parameter space for any two mixing angles when the third angle is constrained by
its  $3 \sigma$ range for normal
hierarchy scenario in the $D_l = 0$ and $D_\nu \neq 0$ scenario for Class III ansatz of texture five zero mass matrices.}
\label{t5cl3nh1}
\end{figure}

\begin{figure}
\begin{minipage} {0.45\linewidth} \centering
\includegraphics[width=2.0in,angle=-90]{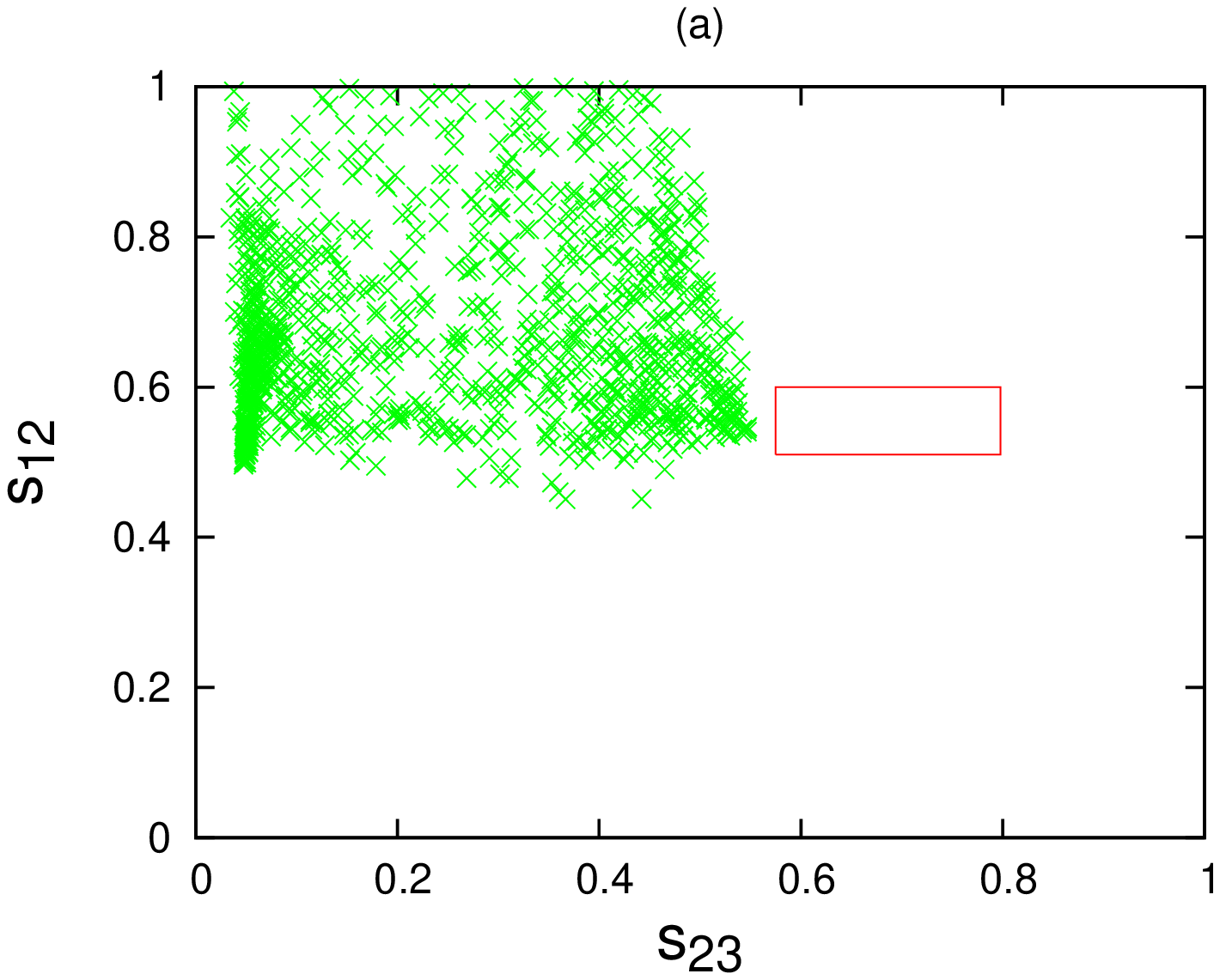}
 \end{minipage}
 \hspace{1.2cm}
 \begin{minipage} {0.45\linewidth} \centering
\includegraphics[width=2.0in,angle=-90]{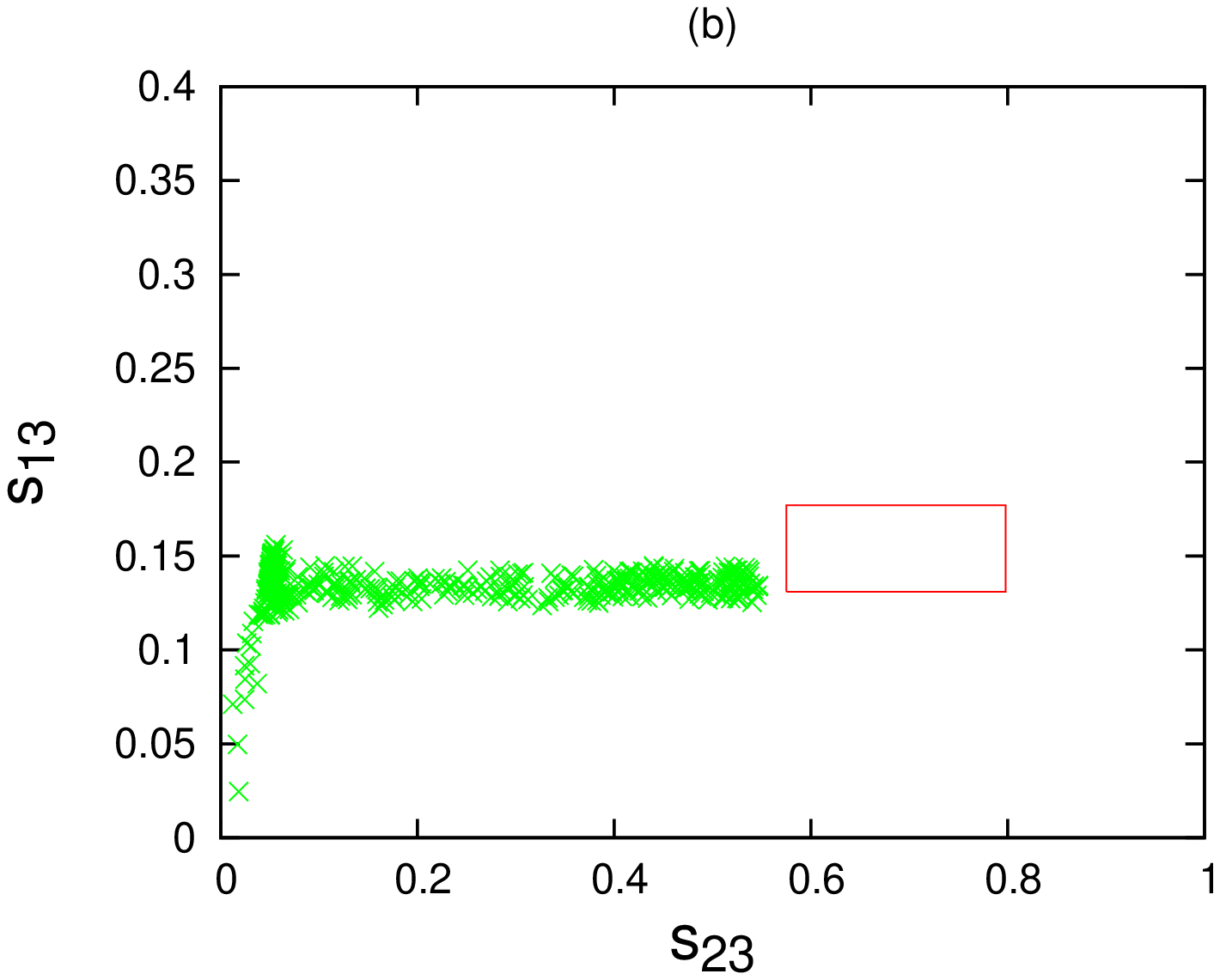}
 \end{minipage}
\caption{Plots showing the parameter space for any two mixing angles when the third angle is constrained by
its  $3\sigma$ range for normal
hierarchy scenario in the $D_l\neq0$ and $D_\nu= 0$ scenario for Class III ansatz of texture five zero mass matrices.}
\label{t5cl3nh2}
\end{figure}

After studying both the cases for texture five zero mass matrices for inverted hierarchy pertaining to class III ansatz,
we now carry out a similar
analysis pertaining to normal hierarchy. To this
end, in figures (\ref{t5cl3nh1}) and (\ref{t5cl3nh2}), we present the plots showing the parameter space corresponding
to any two mixing angles wherein the third one is constrained by its $3\sigma$ range. Interestingly, normal
hierarchy seems to be ruled out for the case $D_l \neq 0$ and $D_\nu=0$, whereas for the case $D_l=0$ and $D_\nu \neq 0$ 
of texture five zero lepton mass matrices normal hierarchy
seems to be viable.

\par
\begin{figure}
\begin{minipage} {0.45\linewidth} \centering
\includegraphics[width=2.0in,angle=-90]{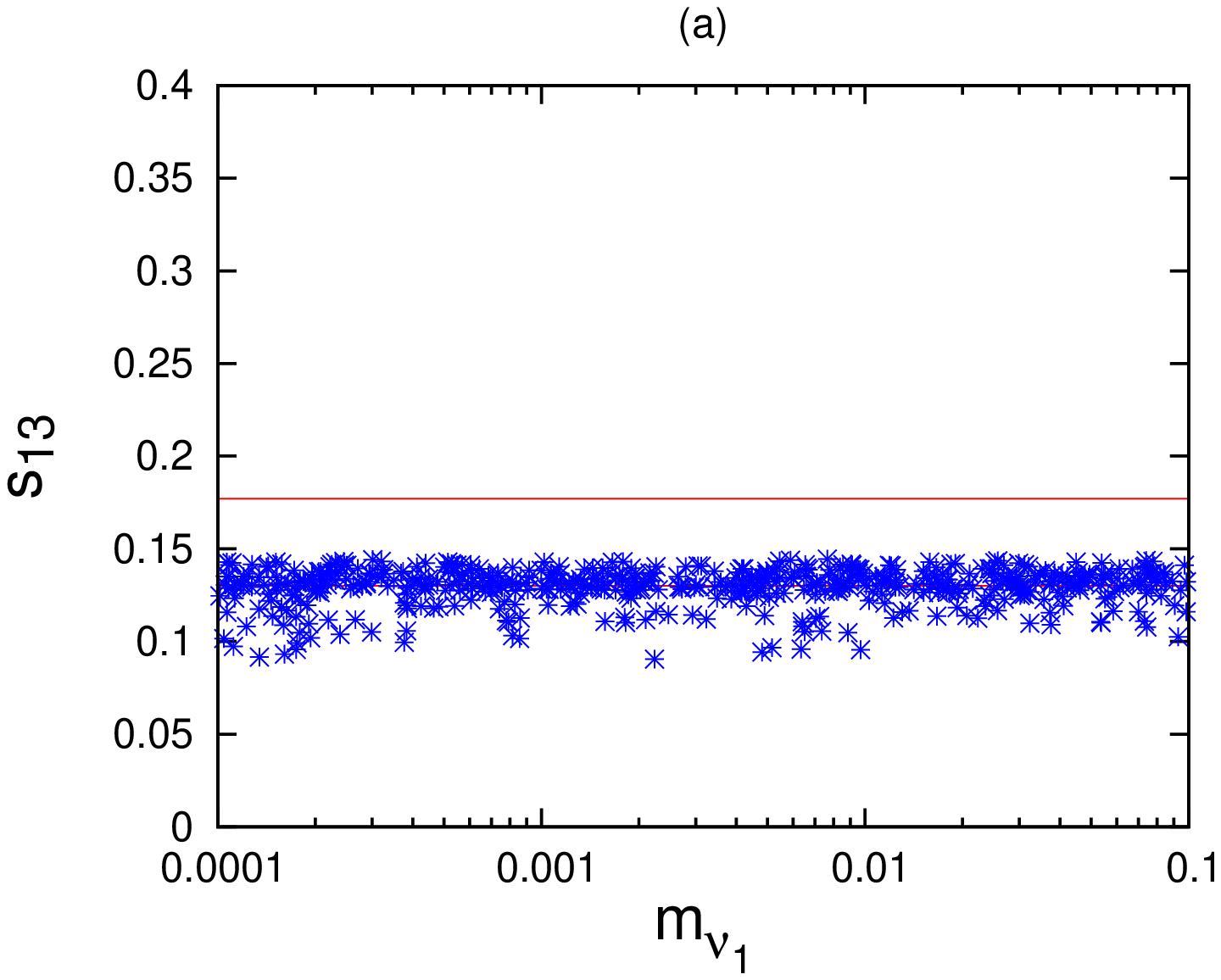}
 \end{minipage}
 \hspace{1.2cm}
 \begin{minipage} {0.45\linewidth} \centering
\includegraphics[width=2.0in,angle=-90]{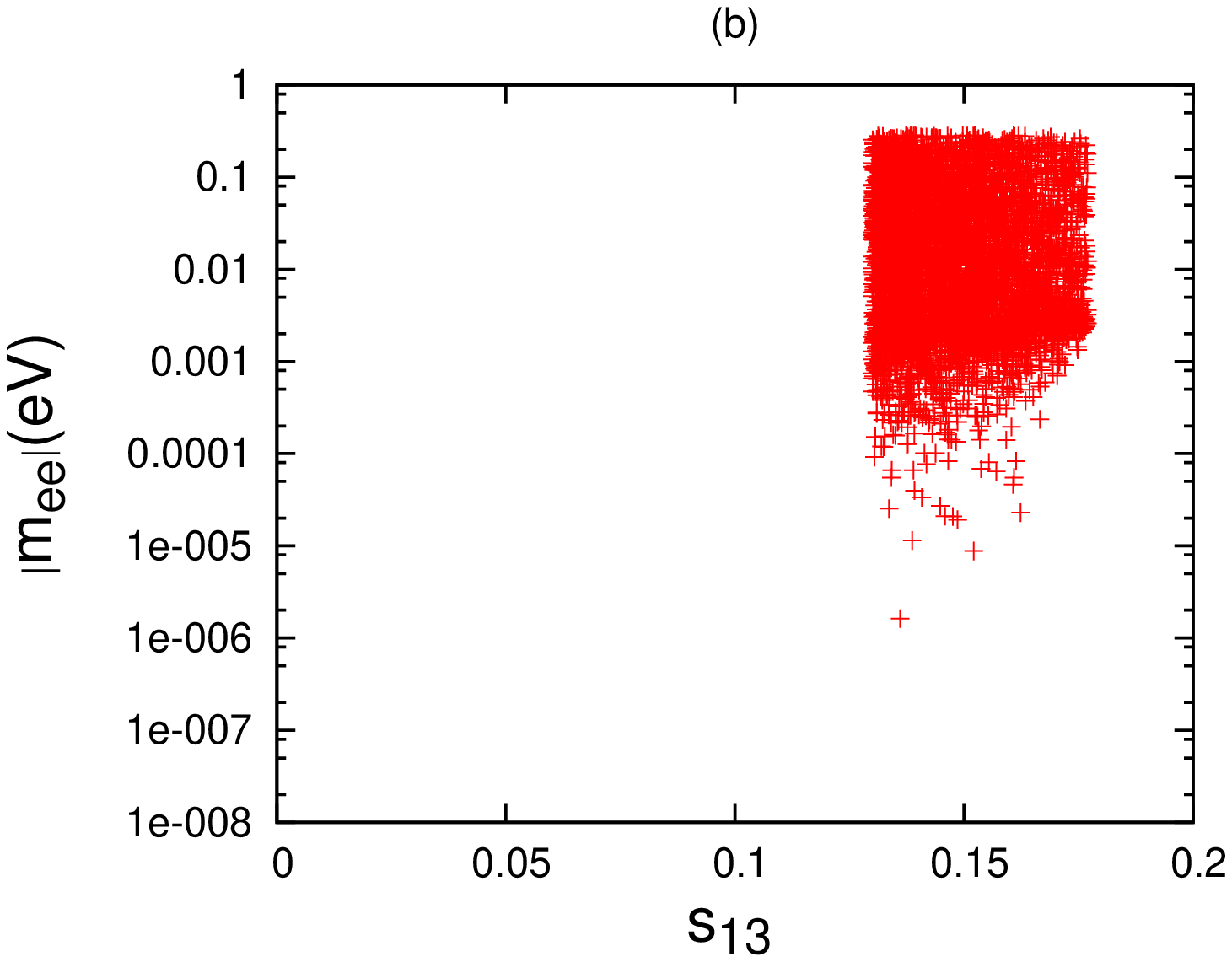}
 \end{minipage}
\caption{Plots showing the dependence of (a) the lightest neutrino mass (b) effective Majorana mass on the 
mixing angle $s_{13}$ in the $D_l=0$ and $D_\nu \neq 0$ scenario for Class III ansatz of texture five
zero  mass matrices (normal hierarchy)	.}
\label{t5cl3nh3}
\end{figure}

 For the $D_l = 0$ and $D_\nu \neq 0$ case of lepton mass matrices, wherein normal hierarchy is shown 
to be viable, in figures \ref{t5cl3nh3}(a) and \ref{t5cl3nh3}(b) we present the plots showing the 
 dependence of the lightest neutrino mass and the effective Majorana mass respectively on the leptonic mixing 
 angle $s_{13}$. Interestingly, one finds that the lightest neutrino mass is unrestricted for this structure, whereas  
a lower bound $\approx 10^{-6}eV$ can be obtained 
for the effective Majorana mass $|m_{ee}|$.

\section{Summary and conclusions}
To summarize, for Majorana neutrinos we have carried out detailed calculations pertaining to non minimal
textures characterized by texture two zero Fritzsch-like structure as well as all possibilities for 
texture four zero and five zero lepton mass matrices. Corresponding to these, we have considered all the
three possibilities for neutrino masses i.e. normal, inverted as well as degenerate scenarios. The 
compatibility of these texture specific mass matrices has been examined by plotting the parameter space 
corresponding to any two of the leptonic mixing angles. Further, for all the structures which seem to be 
compatible with the recent lepton mixing data, the implications of the mixing angles on the lightest 
neutrino mass as well as the effective Majorana  mass measured in neutrinoless double beta decay
have also been studied. 
\par The analysis reveals that the Fritzsch like texture two zero lepton mass matrices
are compatible with the recent lepton mixing data pertaining to normal as well as inverted
hierarchy of neutrino masses. Interestingly, one finds that both the normal as well
as inverted neutrino mass hierarchies are compatible with  texture four zero mass matrices pertaining to 
class II and III contrary to the case for texture four zero mass matrices pertaining to class I wherein inverted
hierarchy seems to be ruled out. Degenerate scenario pertaining to inverted hierarchy is clearly ruled out for texture
four zero mass matrices pertaining to class I, whereas for normal hierarchy this scenario can not be ruled out. For 
texture four zero mass matrices in class II and III, none of the possibilities
for degenerate neutrino mass scenario can be ruled out. Mass matrices in class IV are phenomenologically excluded.

\par For texture five zero lepton mass matrices, we analyse both the cases, viz. $D_l=0, D_\nu \neq 0 $ as well as
$D_l \neq 0, D_\nu =0$ for class II and class III ansatz. A detailed analysis for texture
five zero matrices pertaining to
class I has already been carried out in \cite{ourplb}. 
For class II, normal hierarchy is viable for both the cases while the inverted hierarchy seems to be 
ruled out for the case, $D_l=0, D_\nu \neq 0$. Finally, for texture five 
zero mass matrices pertaining to class III we find that inverted hierarchy 
is viable for the case $D_l \neq 0, D_\nu =0$, while the
normal hierarchy is compatible with the $D_l=0, D_\nu \neq 0$ case. Refinements in the 
measurements of the lightest neutrino mass and $<m_{ee}>$ are expected to have important implications 
for the texture specific mass matrices considered in the present work.

\vskip 0.5cm
{\bf Acknowledgements} \\S.S. would like to acknowledge UGC, Govt. of India, for financial support.
G.A. would like to acknowledge DST,
Government of India (Grant No: SR/FTP/PS-017/2012) for financial
support. S.S., P.F., G.A. acknowledge the Chairperson, Department
of Physics, P.U., for providing facilities to work.

\end{document}